\documentclass[hyper]{JHEP3}

\usepackage{epsfig}
\usepackage{latexsym}
\usepackage{amsfonts}
\usepackage{amsmath}
\usepackage{amsthm}
\usepackage{amssymb}
\usepackage{amsbsy}
\usepackage{multirow}
\usepackage{mathrsfs}
\usepackage[vcentermath, enableskew]{youngtab}
\usepackage{xspace}

\def\calb         {{\cal B}}

\def\calh         {{\cal H}}

\def\call         {{\cal L}}

\def\caln         {{\cal N}}
\def\calo         {{\cal O}}
\def\calp         {{\cal P}}
\def\calq         {{\cal Q}}
\def\calr         {{\cal R}}
\def\cals         {{\cal S}}

\def\calz         {{\cal Z}}

\def\Li           {\text{Li}}

\newsavebox{\uuunit}
\sbox{\uuunit}
    {\setlength{\unitlength}{0.825em}
     \begin{picture}(0.6,0.7)
        \thinlines
        \put(0,0){\line(1,0){0.5}}
        \put(0.15,0){\line(0,1){0.7}}
        \put(0.35,0){\line(0,1){0.8}}
       \multiput(0.3,0.8)(-0.04,-0.02){12}{\rule{0.5pt}{0.5pt}}
     \end {picture}}

\def\be{\begin{equation}}
\def\ee{\end{equation}}
\def\bea{\begin{eqnarray}}
\def\eea{\end{eqnarray}}

\def\a{\alpha}
\def\b{\beta}
\def\bn{\bar{n}}
\def\bm{\bar{m}}
\def\bd{\bar{d}}
\def\bh{\bar{h}}
\def\g{\gamma}

\def\d{\delta}

\def\D{\Delta}

\def\l{\lambda}
\def\ra{\rightarrow}

\def\lag{\langle}
\def\rag{\rangle}
\def\k{\kappa}

\def\m{\mu}
\def\n{\nu}

\def\on{\overline{N}}

\def\oy{\overline{\psi}}

\def\p{\pi}
\def\pt{\partial}

\def\s{\sigma}

\def\q{\theta}

\def\tb{\tilde{\beta}}

\def\tdim{\text{dim}}
\def\tnum{\text{num}}
\def\tmin{\text{Min}}
\def\tmax{\text{Max}}

\def\ty{\tilde{\psi}}

\def\tm{\tilde{m}}

\def\tN{\tilde{N}}

\def\ty{\tilde{\psi}}

\def\vf{\varphi}

\def\c{\chi}

\def\y{\psi}
\def\z{\zeta}

\title{A Universal Behavior of Half-BPS Probes in the Superstar Ensemble}

\author{Ilies Messamah$^1$ \\
\\
$^1$ National Institute for Theoretical Physics,\\
School of Physics and Centre for Theoretical Physics,\\
University of the Witwatersrand, Wits 2050, South Africa
 \\

\vspace*{2mm}
{\tt e-mail: \email{Ilies.Messamah@wits.ac.za}}
\\}

\abstract{In this paper, we probe the typical states of the superstar ensemble
of \cite{Balasubramanian:2005mg} using half-BPS states of type-IIB string
theory on AdS$_5 \times$ S$^5$. We find a very simple universal result that has
the structure $\log\, \lag\lag \y \; \y \rag\rag_\calo \approx \a\, h \, \log
N$, where $h$ is the conformal weight of the probe $\y$, and $\a$ is a constant
that depends mainly of the shape of the probe $\y$. A complete understanding of some properties of this leading term from the dual effective superstar geometry point of view is still lacking.}

\preprint{WITS-CTP-103}
\begin{document}
%
%
%
\section{Introduction}
%
%
%
The thermodynamical nature of black holes was revealed by studying the response
of black holes to both perturbations in their defining parameters (mass, angular
momentum, charge) \cite{Bardeen:1973gs}, as well as their response to the
presence of quantum fields in the bulk of their geometry \cite{Hawking:1974sw,
Hawking:1976de}. This unexpected nature of black holes led to the famous black
hole puzzles\footnote{Another puzzle, the singularity puzzle, is usually added
to the list. This puzzle questions the validity of general relativity near the
singularity and has nothing to do with the thermodynamical nature of black
holes. That is why we did not include it in our list.}: The black hole entropy
puzzle, and the information loss puzzle. A lot of effort was spent in the the
last four decades or so to solve these puzzles. Among the ideas that emerged
during this investigation is the fuzzball proposal, first advanced by Mathur and
collaborators \cite{Lunin:2001jy, Mathur:2002ie, Mathur:2005zp}. By now there
are many nice reviews on the subject, see for example \cite{Mathur:2005zp,
Mathur:2005ai, Bena:2007kg, Skenderis:2008qn, Mathur:2008nj,
Balasubramanian:2008da, Simon:2011zza}. The idea of the fuzzball proposal is
that the black hole geometry is an effective description of an underlying
exponentially large system of microstates. It is widely believed that some of
these microstates manifest themselves as smooth geometries on the gravity side.
However, depending on the kind of black holes under consideration, not all of
them can have a gravity description \cite{deBoer:2009un}. Another point of view
that was advanced in \cite{Sen:2009bm}, is that the black hole geometry is the
effective description of a set of microstates that do not have a smooth geometry
description. The situation is far from being conclusive and we will not be
concerned with these issues in the present paper. For a further discussion on
these issues see e.g. \cite{Simon:2009mf, Bena:2012hf}.

One of the papers that went beyond comparing the macroscopic and microscopic
entropies in checking the fuzzball proposal is \cite{Balasubramanian:2005mg}
(see also \cite{Shepard:2005zc} for a different approach). They studied a
specific ensemble of heavy half-BPS states of type-IIB string theory on
asymptotic AdS$_5 \times$ S$^5$, called the superstar ensemble, and
constructed its effective dual geometry (see \cite{Alday:2006nd} for the
application of similar ideas to the case of the D1-D5 system). It turned out
that this geometry is the same as the one of the superstar of
\cite{Myers:2001aq}. This led them to conjecture that the superstar is an
effective description of the superstar ensemble in line with the
fuzzball proposal. They supplemented this claim with further checks using some
correlation functions. We initiate in this paper a further check of this
proposal by studying the effect of light half-BPS probes on the heavy
states of the superstar ensemble. We find that at leading order, the final
answer is universal and does not depend on the details of the typical states of
the superstar ensemble.

This paper is organized as follows. In the second section we quickly review the
superstar ensemble and discuss some important properties of its typical states
that will be useful later on. This material is not new and can be found in
\cite{Balasubramanian:2005mg}. After that, we introduce the different classes
of probes that we will be dealing with in this paper, and discuss some of their
properties. In the third section, we introduce the two point function which is
the main quantity we will be evaluating in this paper. Essentially, half-BPS
states of type-IIB string theory on asymptotic AdS$_5 \times$ S$^5$ can be
described using Young diagrams (YDs) which are treated as irreducible
representations of the unitary group\footnote{Whether the actual group
is SU$(N)$ or U$(N)$ is still a matter of debate. We expect that our leading
large $N$ result will not be modified if we use SU$(N)$ group instead. See
\cite{deMelloKoch:2004ws, deMelloKoch:2005rq} for details on the use of the
SU$(N)$ group and modifications to be brought to the U$(N)$ formulas.} U$(N)$,
where $N$ is the flux of the background geometry \cite{Corley:2001zk,
Berenstein:2004kk}. When evaluating the two point function, we we face the
problem of decomposing the tensor product of two U$(N)$ representations into
irreducible ones. This is the topic of the fourth section. Although it is
impossible to completely carry out this decomposition, we manage to extract
enough information about it to be able to evaluate the leading order of the two
point function. In the fifth section, we calculate the leading order term of the
log of the two point function for the different classes of probes. We close the
paper by discussing these results, and pointing out some further directions of
research. Part of the conclusion is devoted to discussing a mysterious non-differentiability of the leading term of the two point function at the point $h \sim N$. A full understanding of this might have deep implications on the physics of black holes. We left some details to the appendices. Among them, let us mention
appendix-\ref{conventions} which includes a summary of our notations and YD
terminology that is heavily used in this paper. We advise the reader to read it
before reading the main part of the paper (sections \ref{tensor-prod-section}
and \ref{two-pt-funct-full-section}).
%
%
%
\section{Backgrounds and probes} \label{intro-ensemble}
%
%
%
Our central aim in this paper is to probe a class of ``heavy'' half-BPS
states of type-IIB string theory on asymptotically AdS$_5 \times$ S$^5$
spacetimes, using ``light'' half-BPS states of the same theory. By heavy we mean
states with energy/conformal weight that scales as $N^2$, whereas by light we
mean states with energy/conformal weight that scales slower than $N^2$. Since on
the gravity side, the heavy half-BPS states backreact on spacetime and generate
the bulk geometry, an LLM geometry \cite{Lin:2004nb}, we will call them the
background states. The light half-BPS states on the other hand probe these
geometries, hence we name them probes.

Our probe analysis will take place entirely in the dual conformal field theory,
the $\caln =4$ SU$(N)$ super Yang-Mills theory \cite{Maldacena:1997re}. As is
well known, the half-BPS states of this theory can be described using YDs
with at most $N$ rows \cite{Corley:2001zk, Berenstein:2004kk}. The conformal
weight of the state is the total number of the boxes of its corresponding YD. As
a result, we will heavily use the YDs technology in this paper. For the needed
notions, properties, as well as conventions used in this paper we refer
the reader to appendix-\ref{conventions}.

In this section, we will review the background states that we are interested in.
We will discuss some of their main properties that will be crucial later on.
After that, we will discuss the type of probes we will be using and some of
their most important properties which will play a prominent role in sections
\ref{tensor-prod-section} and \ref{two-pt-funct-full-section}.
%
%
\subsection{The superstar ensemble}\label{superstar-ensemble}
%
%
In the following, we review the superstar ensemble discussed in
\cite[section 3.3]{Balasubramanian:2005mg} in some detail. Our main interest is
the main structure of the typical YDs of this ensemble. Hence, we will neither
discuss the mapping between the LLM geometries \cite{Lin:2004nb} and the YDs of
this ensemble, nor its effective description in terms of the superstar of
\cite{Myers:2001aq}. We refer the interested reader to the paper
\cite{Balasubramanian:2005mg}. We will start by describing the ensemble and its
``average'' YD. Then, we will discuss some general properties of the typical
states that will be of interest to us in the bulk of the paper. We will be
following closely \cite{Balasubramanian:2005mg}.
%
%
\subsubsection{The ensemble and its limiting shape YD}
%
%
The superstar ensemble is the set of YDs with fixed number of columns $D$, fixed
number of rows $N$ and fixed number of boxes $\D$, such that:
\begin{equation}
 D \sim N \;, \qquad \D = \frac{1}{2} \, N \, D \sim N^2 \;,
\label{sustar-def}
\end{equation}
which are weighed equally. Let $r_i$ be the length of row $i$, and let $c_j$
be the number of columns of length $j$ \footnote{Notice that we have a different
convention for numbering the YD rows than the one used in
\cite{Balasubramanian:2005mg}. We start the numbering from top to bottom,
whereas they number the way around, from bottom to top. This is the reason we
have different expressions for $r_i$ and $c_i$ than them.}. In our conventions
(see appendix-\ref{conventions}), we have the relations:
\begin{equation}
  c_N = r_N \;, \qquad c_i = r_i - r_{i+1} \;; \quad 1 \leq i \leq N -1 \;.
\label{column-row-relation}
\end{equation}
We can describe the superstar ensemble using a canonical ensemble. The
associated partition function is given by:
\begin{align}
 \calz = \sum_{c_1 \,,\, c_2 \,,\, \ldots \,,\, c_N =1}^\infty e^{-\b \,
\sum_{j=1}^N j \, c_j - \l \, \sum_{j=1}^N c_j} = \prod_{j=1}^N \frac{1}{1
- p \, q^j} \;, \label{part-funct}
\end{align}
where $q = e^{-\b}$, $p = e^{-\l}$, $\b$ and $\l$ are some positive parameters
that will be fixed later on. Since we are dealing with the canonical ensemble
instead of the microcanonical one, we need to fix the average of the number of
boxes $ \D $, as well as the average of the number of columns $ D $ such that:
\begin{align}
 \D &= \lag \sum_{j=1}^N j \, c_j \rag = q \, \pt_q \, \log \calz = \sum_{j=1}^N
\frac{j \,p\, q^j}{1- p \, q^j} = \frac{1}{2} \,N \, D \;,
\label{beta-lambda-const-1} \\
 D  &= \lag \sum_{j=1}^N c_j \rag = p \, \pt_p \, \log \calz = \sum_{j=1}^N
\frac{p\,q^j}{1- p \, q^j}\;.
\label{beta-lambda-const-2}
\end{align}
We restrict ourselves to spelling out the results here, leaving the details to
appendix-\ref{app-thermo-dyn}. We find that by fixing $\b$ and $p$ as:
\begin{equation}
  p = \frac{1-q^D}{1-q^{D+N}} \approx \frac{D}{D+N} \;, \qquad \b \sim
\frac{1}{N} \;,
\end{equation}
we satisfy the constraints above. The scaling of $\log q$ with $N$ is fixed by
evaluating the entropy of this ensemble and using that $\log q = (\pt\,
S/\pt\, \D)$.

When dealing with very large YDs that come with certain probability/weight, it
is usually beneficial to construct the limit shape YD. This YD can be thought
of as the ``average'' YD, as it is constructed by finding a relation between the
average length of a row and its position. In our present case, if we denote by
$y(x)$ the length of a row whose position is given by its number $x$, the limit
shape YD turns out to be a triangle with the diagonal given by the equation:
\begin{equation}
  y (x) = D \, \left( 1 - \frac{x}{N} \right) \;.
\end{equation}
According to the usual intuition from statistical mechanics, most of the YDs in
the superstar ensemble will be close to this limit shape YD. How close are they
will be the subject of the next subsection.
%
%
\subsubsection{Typical states}
%
%
In the following, we will summarize some of the main properties of typical YDs
in the superstar ensemble that will be of importance to us. More precisely, we
want to know the deviation between a random typical YD $\calo$ and the limit
shape YD $\calo_0$. Since the number of boxes, columns and rows is held fixed in
the superstar ensemble, the only thing that can happen is for boxes near the
diagonal of $\calo_0$ to move from one row to another.

Let us first worry about boxes in the same row. A good estimate for the number
of moved boxes is given by the variance $\s (D)$. The reason being that for the
limit shape YD, there is a linear relation between $D$ and the length of a row
$i$, $\calo_i = D \, (1- i/N)$. For a more precise treatment of $\s (\calo_i)$
see \cite{Balasubramanian:2005mg}. We have:
\begin{align}
 \s^2 (D) = \sum_{i=1}^N \left(\lag c_i^2 \rag - \lag c_i \rag^2 \right) = (p \,
\pt_p )^2 \, \log \calz \approx \frac{D}{N} \, (D+N) \sim N \;.
\end{align}
Hence the fluctuation in the length of a row is of order $\sqrt{N}$. This is
the usual thermal fluctuation since the length of almost all of the rows of the
limit shape YD is of order $N$.

The other quantity that will be crucial to us is the total number of migrating
boxes i.e. the total number of boxes that are moved around when comparing a
typical YD in the superstar ensemble with its limit shape YD. A good estimate of
that is the fluctuation in $\D$. We have:
\begin{equation}
 \s^2 (\D) = (q \, \pt_q)^2 \, \log \calz \approx \frac{1}{2} \, D \, (D+N)
\sim N^2 \;.
\end{equation}
Hence, the fluctuation in $\D$ is of order $N$. Once again, we find the usual
thermal fluctuation.

Before moving on, let us estimate the number of corners in a typical YD of the
superstar ensemble. This quantity will play an important role later on, see
appendix-\ref{app-inclusion-yd-antisym-rules}. First of all, due to the nature
of the limit shape YD, and the fact that both the number of columns and the
number of rows are of order $N$, one easily concludes that we have order $N$
corners in the limit shape YD $\calo_0$. Next, using the fact that when
comparing $\calo_0$ to a typical YD $\calo$, the total number of moved boxes is
of order $N$, we arrive at the conclusion that typical YDs of the superstar
ensemble have also order $N$ corners. Let us check this claim in the simple case
of the superstar ensemble of YDs with the same number of columns and rows, $N$.
In this case, the limit shape YD $\calo_0$ has $N$ corners. Let us suppose for
a moment that the number of corners of a typical YD $\calo$ is of order $N^b$,
where $b \leq 1$. We need to prove that $b=1$. In order for a typical
YD $\calo$ to have order $N^b$ corners, this YD $\calo$ needs to have
$N^b$ sets of equal length rows. Suppose that there are $n_i$ rows in each set,
$i=1 \,,\,2\,,\, \ldots \,,\, N^b$. Then, the total number of moved boxes $(\d\,
\D)$ in this case is:
\begin{align*}
 2\,(\d\, \D) \approx \sum_{i=1}^{N^b} n_i^2 = \sum_{i=1}^{N^b} \left[ \left(n_i
- N^{1-b}\right) + N^{1-b} \right]^2 = \sum_{i=1}^{N^b} \left(n_i -
N^{1-b}\right)^2 + N^{2-b} \geq N^{2-b} \;,
\end{align*}
where we used the fact that $\sum_i n_i = N$. But we know from our discussion
above that $(\d \, \D) \sim N$ at worst. Hence $b =1$. We can repeat the same
arguments for other superstar ensembles, which confirms our previous claim.

Let us summarize the main properties of typical YDs here. If we pick a random
typical YD $\calo$ from the superstar ensemble, almost all of its rows have a
length which is of order\footnote{This is a straightforward result of the
combined fact that almost all the rows of the limit shape YD have a
length which is of order $N$, and that the difference in length of the same row
in the limit shape YD and a typical YD is at worst of order $\sqrt{N}$.} $N$.
Furthermore, this YD $\calo$ has order $N$ corners.
%
%
\subsection{The probes and their SU$(N)$ and $S_h$ dimensions}
\label{probes-section}
%
%
After we discussed the most important properties of our background YDs, we turn
our attention to the probe YDs. Their main property is that the total number of
their boxes $h$ is much less than $N^2$ i.e. $h \ll N^2$. Another property of
these YDs is that the total number of their rows $n$ equals $N$ at most. This
is because we will be treating them as irreducible representations of U$(N)$
when evaluating their two point function in our backgrounds, see section
\ref{two-point-function-section} for more details.

Among all possible probe YDs, we will limit our discussion in this paper to the
{\it homogeneous} YDs. These are YDs where the ratio of the numbers of rows
(columns) whose length does not scale with $N$ in the same way as the number of
columns $d$ (respectively rows $n$) to the total number of rows (respectively
columns) tends to zero in the limit $N \ra \infty$. If $d$ denotes the total
number of columns, $n$ the total number of rows then we have:
\begin{equation}
  h \sim n \, d \;. \label{homogenuous-YD-boxes-number}
\end{equation}
For the other kinds of probes, we should think of them as the result of a
tensor product of two or more homogeneous probes. In a sense, the homogeneous
YDs are our building blocks that generate all the other YDs by the means of
taking the tensor product between them. Form now on, whenever we talk about a
probe YD we mean a homogeneous one.

Before going on, let us fix the notation once and for all\footnote{The
conventions to be used throughout this paper are collected in
appendix-\ref{conventions}.}. $\y$ will stand for a probe YD, $\y_i$ the length
of its $i^{\text{th}}$ row, $d$ the total number of its columns, $n$ the total
number of its rows, $\y_0 = \tmax \, \{d \,,\, n\}$, and $h$ the total number of
its boxes.

In the remaining of this subsection, we will study the leading behavior of
$\tdim_N \, \y$ the dimension of a probe YD $\y$ as an irreducible
representation of SU$(N)$, as well as $\tdim_h \, \y$ its dimension as an
irreducible representation of the permutation group $S_h$. These quantities will
play an important role in this paper as they are intimately connected to the
decomposition of the tensor product $\calo \otimes \y$ (see section
\ref{tensor-prod-section} for more details). The latter will play a role in the
evaluation of the two point function (\ref{two-pt-bc}) whose leading term we are
after. Let us first start by giving two different formulas for the dimension of
an SU$(N)$ irreducible representation specified by a YD $\y$ (see for example
\cite{Fulton, B-Sagan}). The first one is in terms of the difference in the
lengths of different rows and reads:
\begin{equation}
 \tdim_N \, \y  = \prod_{k=1}^{N-1} \; \prod_{i=1}^{N-k} \left(1 +
\frac{\y_i - \y_{i+k}}{k}\right) \;. \label{dim-step}
\end{equation}
The second formula is in terms of the hook lengths. It reads:
\begin{equation}
 \tdim_N \, \y = \frac{\prod_{i=1}^N \; \prod_{j=1}^{\y_i}
(N-i+j)}{\calh_\y} \;, \label{dim-hooks}
\end{equation}
where $ \calh_\y = \prod_{i,j} h_{(i,j)}$, $h_{(i\,,\,j)}$ is the hook length
associated to the box $(i\,,\, j)$,  see appendix-\ref{conventions} for its
definition. Although we will heavily use the first expression, we will still
need the second expression since it has a similar form as the dimension of $\y$
as a representation of the permutation group $S_h$. The latter reads (see for
example \cite{Fulton, B-Sagan}):
\begin{equation}
 \tdim_h \, \y = \frac{h!}{\calh_\y} \;. \label{dim-perm}
\end{equation}
Since we are working with exponentially large quantities, it is much useful to
evaluate the log of these dimensions given that we are interested in the large
$N$ limit. Taking into account that the number of rows of $\y$ is given by
$n \leq N$, we find using equation (\ref{dim-step}):
\begin{align}
 \log \, \tdim_N \, \y = \sum_{i=1}^{n-1} \sum_{j=1}^{n-i} \log \left(1 +
\frac{\y_i - \y_{i+j}}{j} \right) + \sum_{i=1}^n \sum_{j=n+1-i}^{N-i} \log
\left( 1+ \frac{\y_i}{j} \right) \;. \label{dim-SUN-n-small}
\end{align}
It is clear that the leading term of the expression above will depend on how
$\y_i$ scales with $N$. Such behavior leads us to classify the probes into
three classes which are the following:
\begin{itemize}
  \item {\underline{\bf Generic probes class:}} In this case, both $n$ and $d$
are very small compared to $N$ i.e. $n$, $d \ll N$. The reason we call them
generic is that they exist for all regimes of $h$ of interest to us ($h \ll
N^2$).
  \item {\underline{\bf Linear probes class:}} In this case, either $n$ or $d$
but not both scales as $N$. The reason we coin them the name linear is that the
leading behavior of the log of their SU$(N)$ dimension is linear in $h$ as we
will see below. This class of probes is associated to the following regime of
$h$: $N \lesssim h \ll N^2$.
  \item {\underline{\bf Long probes class:}} In this case $d \gg N$. Notice that
this class of probes exists only in the regime $h \gg N$.
\end{itemize}
We will also use the nomenclature ``{\it non-generic probes class}'' to
collectively denote the linear and long probes classes.

When evaluating the leading order of the log of the SU$(N)$ and S$_h$
dimensions of $\y$ in the following, we will discuss each class of probes on its
own. For reasons that will be clear later on (see section
\ref{tensor-prod-section}), the leading behavior of $\tdim_h \, \y$ will be of
interest to us only if $h \ll N$. We will keep the discussion general and leave
the treatment of a concrete example to appendix-\ref{sample-YD}. But before
continuing with the discussion of the dimensions of the different probe classes,
let us pause for a moment and discuss a curious duality of YDs\footnote{This
duality looks like a manifestation of the hole/particle symmetry
\cite{Berenstein:2004hw, Ghodsi:2005ks} but not quite. This is because $\tdim_N
\, \y$ does not have a clear physical interpretation. However, through its
connection with the decomposition of the tensor product $\calo \otimes \y$ (see
section \ref{tensor-prod-section}), it has a connection with the aforementioned
symmetry.} that will be useful below.
%
\subsubsection{An approximate duality of Young diagrams} \label{approx-dual}
%
%
We know that two YDs that are related by the flip row $\leftrightarrow$
column have the same $S_h$ dimension. This is easily understood from equation
(\ref{dim-perm}) since the number of boxes as well as the hook length remain
the same under such a flip. We will show in this section that the leading term
of $\log \,\tdim_N \, \y$  exhibits such an invariance for\footnote{This is
because $n \leq N$ as a result of the YD being an U$(N)$ representation.} $d
\leq N$.

We will assume in the following that $d \ll n$. This assumption is not
restrictive since we know that such YDs are one end of this duality. The only
non-covered case is when $ d \sim n$, but as we will see later on, this case
fits nicely in the manipulations used for $d \ll n$. A property that will be
used below, which will be argued for in subsections \ref{gen-probes-section} and
\ref{linear-probes-section} below and further checked in a concrete example in
appendix-\ref{sample-YD}, is that the leading behavior of $\log \, \tdim_N \y$
is either of order $h \, \log N$ or at worst of order $h$ for the cases of
interest to us. We will take these properties as granted for now.

Our starting point is the relation (\ref{dim-hooks}) in the case $d \ll
n$. First, we rewrite it as:
\begin{align}
  \log \, \tdim_N \, \y = \sum_{i=1}^{N-1} \sum_{j=1}^{\y_i} \log (N - \y_i + i
+ j) - \sum_{i=1}^{N-1}\; \sum_{j=1}^{\y_i} \log h_{(i,j)} \;.
\label{dim-hook-modif}
\end{align}
The duality in this formula is a statement about the exchange $i
\leftrightarrow j$. The part that depends only on the hook length is trivially
invariant under such a change. The quantity $(i+j)$ is also invariant. The only
problematic part is $\y_i$. However, its contribution is much smaller than $h$,
and hence, subleading given our claim above. To prove this claim, we use that
$\y_i \ll N$ to expand the $\log$ term and get:
\begin{align*}
 \d\, \log \, \tdim_N \, \y \approx \sum_{i=1}^{N-1} \sum_{j=1}^{\y_i}
\frac{\y_i}{N+i+j} < \sum_{i=1}^{N-1} \frac{\y_i^2}{N+i} < d^2 \ll h \;,
\end{align*}
where we used that $h \sim n\, d$. Notice that these manipulations do not work
if $ d \gg n$. But this is not a problem since this regime will be at the other
end of the duality for the case $n \gg d$.

As already advanced in the beginning, the same manipulations lead to the same
conclusion in the case $ n \sim d$. The reason being that, in this case both $n$
and $d \ll N$ because we are interested in cases where $h \ll N^2$. Hence, these
kind of YDs belong to the generic class. Although the correction that we get
form the terms depending on $\y_i$ will be of order $h$, the leading behavior of
$\log \, \tdim_N \, \y$ is of order $h \, \log N$ (see the subsection below, and
also appendix-\ref{sample-YD}). Hence, the duality survives in this case as
well.
%
\subsubsection{The generic probes class}\label{gen-probes-section}
%
In this case, we have $n \ll N$ and $\y_i \ll N$. In turns out that we need to
distinguish between two cases: $n \ll d$, and $ d \ll n$. The case $ d \sim n$
is a trivial consequence of the previous two cases. The easiest case to deal
with is when we have $n \ll d$. Let us discuss it first, then turn to the second
case $d \ll n$. In the case $ n \ll d$ (or $n^2 \ll h$), we have from equation
(\ref{dim-SUN-n-small}) the following upper bound:
\begin{align}
 \log \, \tdim_N \, \y &\leq \sum_{i=1}^n \sum_{j=1}^{N-i} \left[\log \left( j +
\y_i \right) - \log j \right] \nonumber \\
                       &\lesssim \sum_{i=1}^n \left[ (N-i + \y_i) \, \log (N-i +
\y_i)  - (N-i) \, \log (N-i) - \y_i \, \log \y_i \right] \nonumber\\
                       &\lesssim \sum_{i=1}^n \left[ \y_i \, \log N + \y_i -
\y_i \, \log \y_i \right] \approx h \, \log \left( \frac{N}{d} \right) \;,
\label{leading-behav-up-lim}
\end{align}
where in the first line we used that $\left(\y_i - \y_{i+j}\right) < \y_i$ for
$1 \leq i <n$. To move form the first to the second line, we used equation
(\ref{gen-sum-m-0}) to evaluate the sum over $j$, and to get the third line we
used that $\y_i \ll N$ and $n \ll N$. The last result is a consequence of:
\begin{equation}
 \sum_{i=1}^n \y_i \, \log \y_i - h \, \log d = \sum_{i}^n \y_i \, \log
\left( \frac{\y_i}{d} \right) \sim \sum_{i=1}^n \y_i \sim h \;,
\end{equation}
where we used the fact that the ratio $(\y_i/d)$ is independent of $N$ for
almost all of the rows.

For the lower bound of (\ref{dim-SUN-n-small}), we get:
\begin{align*}
 \log \, \tdim_N \, \y & \geq \sum_{i=1}^n \sum_{j=n+1-i}^{N-i} \left[\log
\left( j + \y_i \right) - \log j \right] \\
                       & \gtrsim h \, \log N + h - \sum_{i=1}^n \left[
(n-i+\y_i) \, \log (n-i +1 + \y_i) - (n-i) \, \log (n-i+1) \right] \;,
\end{align*}
where in the first line we use that the first sum in (\ref{dim-SUN-n-small}) is
a positive number. To get the last line, we used the same steps as above. We
need to deal with the last sum. We find using that $\y_i \sim d \gg n$ the
following approximate value:
\begin{align*}
 \sum_{i=1}^n \left[ (n-i+\y_i) \, \log (n-i +1 + \y_i) - (n-1) \, \log
(n-i+1) \right] \approx \sum_{i=1}^n \y_i \, \log \, \y_i \approx h \, \log d
\;,
\end{align*}
which we discussed before. Plugging this result in the expression for $\log \,
\tdim_N\, \y$ above, we find:
\begin{equation}
 \log \, \tdim_N \, \y \gtrsim h \, \log \left( \frac{N}{d} \right) \;.
\nonumber\\
\end{equation}
Combining this lower limit with the upper limit in
(\ref{leading-behav-up-lim}), we conclude that the leading behavior of $\log
\,\tdim_N \, \y$ in the case $ N \gg d \gg n$ is given by:
\begin{equation*}
 \log \, \tdim_N \, \y \approx h \, \log \left( \frac{N}{d} \right) \;.
\end{equation*}

What about the other case $ n \gg d$? For this case, we take advantage of the
duality discussed in section-\ref{approx-dual} to swap the rows and columns of
$\y$, which brings us to the previous case where $n$ here plays the role of $d$
there and vice versa. Hence, in the case $n \gg d$, the leading behavior of
$\log \, \tdim_N \, \y$ is given by:
\begin{equation}
 \log \, \tdim_N \, \y \approx h \, \log \left( \frac{N}{n} \right) \;.
\nonumber
\end{equation}
As a conclusion, we find that the leading behavior of $\log \, \tdim_N \, \y$
in the large $N$ limit in the case where $n$ and $d \ll N$ is given by:
\begin{equation}
 \log \, \tdim_N \, \y \approx h \, \log \left( \frac{N}{\y_0} \right) \;,
\label{leading-behav-d-n-ll-N}
\end{equation}
where $\y_0 = \tmax \{d \,,\, n\}$.

The next quantity we want to find is the leading term of $\log \, \tdim_h
\, \y$ in the case $ h \ll N$. The only non-trivial term in the dimension
relation (\ref{dim-perm}) is $\calh_\y$. Its leading behavior can be derived
using the expression (\ref{dim-hooks}) together with the leading behavior
(\ref{leading-behav-d-n-ll-N}). First, we need to deal with the numerator of
(\ref{dim-hooks}), we have:
\begin{align}
 \log \, \tnum_N \, \y &= \sum_{i=1}^n \sum_{j=1}^{\y_i} \log (N +j-i) \nonumber
\\
                       &\approx \sum_{i=1}^n \left[ (N-i + \y_i) \, \log (N-i
+\y_i) - (N-i) \, \log (N-i) - \y_i \right] \nonumber \\
                       &\approx \sum_{i=1}^n \y_i \, \log (N-i) \approx h \,
\log N \;, \label{dim-num-leading-n-d-ll-N}
\end{align}
where we used the approximation (\ref{gen-sum-m-0}) to get the second line, then
we used that $\y_i \ll N$ and $n \ll N$ to approximate the sums in the second
and third line respectively. Next, we plug the leading behavior of $\log \,
\tdim_N \, \y$ given in equation (\ref{leading-behav-d-n-ll-N}) and the leading
behavior of the numerator of (\ref{dim-hooks}) that is given in
(\ref{dim-num-leading-n-d-ll-N}) above, in the equation (\ref{dim-hooks}), to
get the following leading behavior of the product over the hook lengths:
\begin{equation}
 \log \, \calh_\y \approx h \, \log \y_0 \;, \nonumber
\end{equation}
where $\y_0 = \tmax \, \{d\,,\,n\}$. As a result, the expression
(\ref{dim-perm}) for the dimension of $\y$ as a representation of $S_h$ leads
to the following leading behavior:
\begin{equation}
 \log \, \tdim_h \, \y \approx h \, \log \left( \frac{h}{\y_0} \right) \;,
\label{perm-dim-leading-bahav-n-d-ll-N}
\end{equation}
where we used that $\log \, h! \approx h \, \log \,h$. Notice that in the case
where $h \sim \y_0$, the leading behavior of $\log \, \tdim_h \, \y$ will be
proportional to $h$. We will use the expression above for the leading
behavior of the $S_h$ dimension for all cases of this class of probes. This is
because the final result of the leading term of the two point function will not
change due to continuity, see subsection \ref{two-pt-funct-full-gen-h-ll-N}.

This is the only class of probes where we will need the leading behavior of
their dimension as a representation of $S_h$. In the remaining of this section
we will look for the leading behavior of the SU$(N)$ dimension of YD in the
other two classes. Although a universal exact expression for the leading term is
not always possible, we will derive its leading behavior in the worst
situations.
%
\subsubsection{The linear probes class}\label{linear-probes-section}
%
In this class of probes, we have either $d \sim N$ or $n \sim N$. Following the
same route as above, we first discuss the case $ d \sim N$, then use the duality
discussed in section-\ref{approx-dual} for the case $n \sim N$. The reason we
can use this duality has to do with the leading term of the dimension $\tdim_N
\, \y$, which will be evaluated below.

Our starting point is once again the relation (\ref{dim-SUN-n-small}). We have
in the case $d \sim N \gg n$ the following upper bound:
\begin{align}
 \log \, \tdim_N \, \y &\leq \sum_{i=1}^n \sum_{j=1}^{N-i} \left[ \log
(j + \y_i) - \log j \right] \nonumber \\
                       &\lesssim \sum_{i=1}^n \left[ (N-i+\y_i) \, \log
(N+\y_i) - (N-i) \, \log N - \y_i \, \log \y_i\right] \nonumber \\
                       &\lesssim \sum_{i=1}^n \left[(N + \y_i) \, \log (1+
\oy_i) - \y_i \, \log \oy_i \right] \sim h \;, \nonumber
\end{align}
where we used the approximation (\ref{gen-sum-m-0}) to evaluate the sum over
$j$ in the first line and that $n \ll N$ to get the second line. In the last
line, we introduced the quantity $\oy_i = \y_i/N$, which is independent of $N$.
At the end we used that $\y_i \sim d \sim N$ and $h \sim n\,d \sim n N$.
The reason we did not try to get an exact result here is that knowing
that the upper bound is of order $h$ is more than enough for our purposes, see
subsection \ref{ngen-probes-two-pt-funct-section} for more details.

To complete the circle of thoughts, although not needed, let us look for a
lower bound on the leading behavior of $\log \, \tdim_N \, \y$. We have:
\begin{align}
 \log \, \tdim_N \, \y &\geq \sum_{i=1}^n \sum_{j=n+1-i}^{N-i} \left[ \log
(j+\y_i) - \log j \right] \nonumber \\
                       &\gtrsim \sum_{i=1}^n \left[(N-i + \y_i) \, \log
(N+\y_i) - (N-i) \, \log N \right] \nonumber \\
                       & \quad - \sum_{i=1}^n \left[ (n-i+\y_i) \, \log
\y_i - (n-i) \, \log (n-i)\right] \nonumber \\
                       &\gtrsim \sum_{i=1}^n \left[ (N + \y_i) \, \log
(1+\oy_i) - \y_i \, \log \oy_i\right] \sim h \;, \nonumber
\end{align}
Where we used the same steps as in the derivation above. Notice that we get the
same expression as for the upper bound. Hence, we conclude that in the case
$d \sim N$, we have the following leading behavior of $\log \, \tdim_N \, \y$:
\begin{equation*}
 \log \, \tdim_N \, \y \approx \sum_{i=1}^n \left[ (N + \y_i) \, \log
(1+\oy_i) - \y_i \, \log \oy_i\right] \sim h \;,
\end{equation*}
which is of order $h$ as argued above.

This was for the case $d \sim N$, what about the other possibility $n \sim N$.
Once again, we take advantage of the duality discussed in
section-\ref{approx-dual} to safely conclude that the leading behavior
of the dimension in this case ($n \sim N$), is given by:
\begin{equation*}
 \log \, \tdim_N \, \y \approx \sum_{i=1}^n \left[ (N + n_i) \, \log
(1+\bn_i) - n_i \, \log \bn_i\right] \sim h \;,
\end{equation*}
where $n_i$ stands for the length of the columns of $\y$ and $\bn_i = n_i/N$.

All in all, we conclude that the leading behavior of the dimension of the
linear probes is such that:
\begin{align}
 \log \, \tdim_N \, \y \sim h \;, \label{dim-SUN-behav-giant}
\end{align}
where the actual value of the non-zero coefficient that multiplies $h$ is not
important to us.
%
\subsubsection{The long probes class} \label{long-probes-section}
%
The shape of these probes suggest that their dimension will have a slower
growth than the previous cases. This is because $\calh_\y$ in (\ref{dim-hooks})
will be of the same order as the numerator. Let us proceed and check this
intuitive guess. As usual, we are going to look for an upper and a lower bounds
of $\log \, \tdim_N \, \y$ using equation (\ref{dim-SUN-n-small}). For the upper
bound, we find:
\begin{align}
 \log \, \tdim_N \, \y &\leq \sum_{i=1}^n \sum_{j=1}^{N-i} \left[ \log (j+\y_i)
- \log j\right] \nonumber \\
                       &\lesssim \sum_{i=1}^n \left[ (N-i + \y_i) \, \log
(N-i+\y_i) - \y_i \, \log \y_i - (N-i) \, \log (N-i) \right] \nonumber \\
                       &\lesssim \sum_{i=1}^n \left[ N \, \log \y_i - N \, \log
N \right] \approx n\, N \, \log \left( \frac{d}{N} \right) \;,
\nonumber
\end{align}
where we used as usual (\ref{gen-sum-m-0}) to evaluate the sum over $j$ to get
the second line. To get the third line, we used that $\y_i \gg N \gg n$. To
arrive at the last result, we took advantage of the fact that:
$$ n \, \log d - \sum_{i=1}^n \log \y_i = \sum_{i=1}^n \log \, \frac{d}{\y_i}
\sim n \;, $$
since the ratio $\left(\y_i/d\right)$ is $N$-independent for almost all
$\y_i$'s.

What about the lower bound? Using the same manipulations as above, we easily
find:
\begin{align}
 \log \, \tdim_N \, \y &\geq \sum_{i=1}^n \sum_{j=n+1-i}^{N-i} \left[ \log (\y_i
+j) - \log j \right] \nonumber \\
                       &\gtrsim \sum_{i=1}^n \left[ (N-i+\y_i) \, \log
(N-i+\y_i) - (n-i+\y_i) \, \log (n-i+\y_i) \right] \nonumber \\
                       & \quad - \sum_{i=1}^n \left[(N-i) \, \log (N-i) - (n-i)
\, \log (n-i)\right] \nonumber \\
                       & \gtrsim \sum_{i=1}^n \left[ N \, \log \y_i - N \, \log
N \right] \approx n\, N \, \log \left( \frac{d}{N} \right) \;.
\nonumber
\end{align}
Combining this result with the upper bound above, we conclude that the leading
behavior of the SU$(N)$ dimension of these probes is given by:
\begin{equation}
 \log \, \tdim_N \, \y \approx n \, N \, \log \left( \frac{d}{N} \right) \;.
\label{dim-SUN-leading-ngeom}
\end{equation}
Notice that by exchanging the roles of $N$ and $d$, one can map this leading
term to the corresponding one in the case of generic probes given in
(\ref{leading-behav-d-n-ll-N}), taking into account that $h \sim n\,d$.
%
%
%
\section{Probes in AdS$_5 \times$ S$^5$} \label{two-pt-funct-section}
%
%
%
The states we are dealing with in this paper are half-BPS states of type-IIB
string theory on asymptotic AdS$_5 \times$ S$^5$. These states can be described
on the dual field theory side using YDs, which are seen as irreducible
representation of U$(N)$, where $N$ is the number of fluxes in the background
geometry \cite{Corley:2001zk, Myers:2001aq}. We are mainly interested in probing
the backgrounds associated to the microstates of the superstar of
\cite{Myers:2001aq} according to the proposal advanced in
\cite{Balasubramanian:2005mg}. Remember that these microstates are characterized
by YDs whose number of columns $N_c$, number of rows $N_r$ and number of boxes
$\D$ are all held fixed as follows:
$$ N_r = N \;, \quad N_c = D \sim N \;, \quad \D = \frac{1}{2} \, N\, N_c \;, $$
see subsection \ref{superstar-ensemble} and reference
\cite{Balasubramanian:2005mg} for more details. In this section, we will spell
out the probing tool that we will be using in this paper in its full generality.
After that, we will warm up with a toy model that describes a background which
does not belong to the superstar ensemble for reasons that will be clear later
on.
%
%
\subsection{The two point function} \label{two-point-function-section}
%
%
The game we will be playing in this paper is as follows. We have a background
that is generated by a typical state $\calo$ of the superstar ensemble and we
add on top of that a light probe $\y$, then see what happens. According to
\cite{Corley:2001zk}, modulo an overall trivial space time dependence which is
completely fixed by conformal symmetry, everything boils down to evaluating the
two point function:
\begin{equation}
  \lag\lag \y \; \y \rag\rag_{\calo} = \frac{\lag (\calo \otimes \y) \;
(\calo \otimes \y) \rag}{\lag \calo \; \calo \rag} \;, \label{two-pt-bc}
\end{equation}
where $\y$ is our probe, $\calo$ is our background, and the vacuum two point
function $\lag \a \; \b \rag $ is given by:
\begin{equation}
    \lag \a \; \b \rag = \d_{\a \b} \; \prod_{i\,,\,j} (N-i+j) \;,
\label{two-pt-vc}
\end{equation}
where $\d_{\a \b}$ is a schematic notation that means that the two YDs $\a$
and $\b$ should be identical, and the product is over all the boxes of the YD
$\a$ where $i$ is the row number and $j$ is the column number. The expression
(\ref{two-pt-bc}) is evaluated as follows. We first decompose the tensor
product $\calo \otimes \y$ into irreducible representations of U$(N)$:
\begin{equation}
 \calo \otimes \y = \bigoplus_k \; d_k \, \vf_k \;,
\label{tensor-prod-decomp-schem}
\end{equation}
where $d_k$ is the degeneracy of the YD $\vf_k$, and $k$ is a summation index
that will not play any important role in the following. Next, we use that the
two-point $\lag \calo \; \calo' \rag $ is bilinear to find:
\begin{equation}
 \lag\lag \y \; \y \rag\rag_{\calo} = \sum_k d_k^2 \; \lag \vf_k \; \vf_k
\rag_\calo \;, \label{two-point-funct-explicit}
\end{equation}
where we used that the two point function (\ref{two-pt-vc}) is diagonal and
introduced the notation:
\begin{equation}
 \lag \bullet \; \bullet \rag_\calo = \frac{\lag \bullet \; \bullet
\rag}{\lag \calo \; \calo \rag} \;, \label{lag-rag-calo-notation}
\end{equation}
in order not to get a cluttered expression. We will be using this
simplified notation from now on. So, our task can be summarized into the
following steps:
\begin{enumerate}
  \item Get the needed information from the tensor product decomposition
(\ref{tensor-prod-decomp-schem}). These include the degeneracy $d_k$, the type
of YDs $\vf_k$, and their total number. From now on, we will refer to the tensor
product of two irreducible representations of U$(N)$ by the tensor product
between the associated YDs, in an abuse of language.
  \item Evaluate the term $\lag \vf_k \; \vf_k \rag_\calo$ in the expression
(\ref{two-point-funct-explicit}).
  \item Finally collect all the intermediate results to get the final answer.
\end{enumerate}
We will deal with each step on its own in the following sections. Some of the
details will be left to appendices that we will refer to at the right places.
But before doing so, let us discuss a simple toy model. What we will obtained
here will serve as a good reference point for results to be derived later on.
%
%
\subsection{A toy model: A background outside the superstar ensemble}
\label{toy-model-section}
%
%
The decomposition of the tensor product $\calo \otimes \y$ is very involved in
general (see subsection \ref{tensor-prod-rules}), but there are some simple
situations where things become straightforward. One of the simplest cases
correspond to a background YD $\calq$ where all of its $N$ rows have the same
length:
$$ \calq_i = \k \, N \;, $$
where $\k$ is an arbitrary constant. Although this background $\calq$ is not of
immediate importance to us since it is not part of the superstar ensemble, it is
both a good warm up exercise as well as a good reference point for the two
point function (\ref{two-pt-bc}) of the superstar ensemble.

The simplicity of the background $\calq$ resides in the fact that there is only
one YD $\vf_0$ when decomposing the tensor product $\calq \otimes \y$. If we
denote by $\d_i$ the number of boxes added to the $i^{\text{th}}$ row of
$\calq$ to form the YD $\vf_0$ then we have:
$$ \d_i = \y_i \;. $$
The two point function (\ref{two-pt-bc}) simplifies drastically in this case as
we need to only evaluate $\lag \vf_0 \; \vf_0 \rag_\calq$. The latter can be
easily evaluated using its defining equation (\ref{lag-rag-calo-notation})
together with the explicit expression of the vacuum two point function $\lag \a
\; \b \rag$ given by (\ref{two-pt-vc}). Using theses equations we find that:
\begin{equation}
 \log \, \lag \vf_0 \; \vf_0 \rag_\calq = \sum_{i=1}^N \sum_{j=1}^{\d_i}
\log (N + \calq_i + j -i) \;, \label{two-point-funct-summand-log-extra}
\end{equation}
where $\calq_i$ stands for the length of row $i$ of the YD $\calq$. We can
easily evaluate the sum over $j$ using equation (\ref{gen-sum-m-0}) to
get the following approximate expression:
\begin{align}
 \log \, \lag \vf_k \; \vf_k \rag_\calq \approx& - h + \sum_{i=1}^N (N +
\calq_i + \d_i -i) \, \log (N + \calq_i + \d_i -i) \nonumber \\
                                               & \quad \qquad - \sum_{i=1}^N (N
+ \calq_i -i) \, \log (N + \calq_i -i) \;, \label{two-point-funct-summ-1-extra}
\end{align}
where we used $\sum_{i=1}^k \d_i = h$. Let us now specify this formula to our
case where $\calq_i = \k \, N$ and $\d_i = \y_i$ is nonzero only for $i \leq n$.
We get:
\begin{align}
 \log \, \lag \vf_0 \; \vf_0 \rag_\calq \approx& - h + \sum_{i=1}^n [(1+\k) N +
\y_i -i] \, \log [(1+\k) N + \y_i -i] \nonumber \\
                                               & \quad \qquad - \sum_{i=1}^n
[(1+\k) N -i] \, \log [(1+\k) N -i]\;. \nonumber
\end{align}
To proceed further, we need to distinguish between different probe classes.
\paragraph{The generic probes class} In this class, we have both $\y_i \ll N$
and $n \ll N$. We find after expand the log:
\begin{equation}
 \log \, \lag\lag \y \; \y \rag\rag_\calq = \log \, \lag \vf_0 \; \vf_0
\rag_\calq \approx h \, \log N \;. \label{two-pt-funct-ntyp-gen}
\end{equation}
\paragraph{The linear probes class} In this class, we have either $d \sim N$ or
$n \sim N$. In the case $n \sim N$, we have $d \ll N$ and hence we can use the
same manipulations as before to get:
\begin{equation*}
 \log \, \lag\lag \y \; \y \rag\rag_\calq \approx h \, \log N \;.
\end{equation*}
In the opposite case $\y_i \sim N$, we have $n \ll N$ which allows us to expand
the log term around $i=0$. We get in this case the same result as above.
\paragraph{The long probes class} In this class, we have $d \gg N$ and $n\ll N$.
Expanding the log term appropriately, we get:
\begin{equation*}
 \log \, \lag\lag \y \; \y \rag\rag_\calq \approx h \, \log d \;.
\end{equation*}
Notice that this result does not depend explicitly on $N$. Actually, we can
combine the results for the linear and long probes in a single one that reads:
\begin{equation}
  \log \, \lag\lag \y \; \y \rag\rag_\calq \approx h \, \log \y_0 \;.
\label{two-pt-funct-ntyp-ngen}
\end{equation}
Notice that the leading term of the log of the two point function is continuous,
however it is not differentiable at the point $\y_0 \sim N$. This
non-differentiability can be related to the phase transition between gravitons
and (dual-)giant gravitons when the angular momentum of the probe becomes very
large \cite{McGreevy:2000cw, Hashimoto:2000zp}.

Although the results derived here are in a background that is not related to
the superstar ensemble, some of the results are valid for a wide range of
backgrounds including the superstar ones. We will discuss this observation in
subsection \ref{ngen-probes-two-pt-funct-section}. For the moment let us
continue with our investigation and discuss the tensor product decomposition of
$\calo \otimes \y$.
%
%
%
\section{Dealing with the tensor product $\calo \otimes \y$}
\label{tensor-prod-section}
%
%
%
Our starting point in evaluating the two point function (\ref{two-pt-bc}) is to
construct the tensor product decomposition (\ref{tensor-prod-decomp-schem}).
For our purposes, we do not need to get all the details of this decomposition
which is hopeless. We only need to get an estimate of the degeneracies $d_k$ of
the YDs $\vf_k$, an estimate of the total number of these YDs $d_t = \sum_k
d_k$, and a rough idea on the shape of the YDs $\vf_k$ so that we can evaluate
the quantities $\lag \vf_k \; \vf_k \rag_\calo $. In the following, we will
start by reviewing the construction of the YDs that appear in the decomposition
of the tensor product of two YDs of U$(N)$. By studying carefully the conditions
on these YDs that appear in such decomposition, we propose a map between
semi-standard Young tableaux (SSYTx) and these YDs. Next, using this map we
discuss the shape of ``dominant'' YDs $\vf_k$ and their degeneracies $d_k$
according to the scaling of $h$ with $N$. Some of the details will be left to
appendices \ref{Kostka-numbers-app} and \ref{app-inclusion-yd-antisym-rules}.
%
%
\subsection{The tensor product decomposition and the semi-standard Young
tableaux} \label{tensor-prod-ssyt-section}
%
%
The first question to answer is how to construct the YD $\vf_k$ starting from
the YDs $\calo$ and $\y$, when the tensor product $\calo \otimes \y$ is seen as
a tensor product between two irreducible representation of U$(N)$? Fortunately,
there is a well known recipe to construct the decomposition of the tensor
product of two YDs of U$(N)$. We will first review this recipe, then we will
repackage the information about such decomposition into labelings of $\y$ in the
case of our interest.
%
\subsubsection{Decomposing the tensor product of two representations}
\label{tensor-prod-rules}
%
Let $A$ and $B$ be two YDs associated to two irreducible representations of
U$(N)$. The YDs corresponding to the irreducible representations appearing in
the decomposition of the tensor product $A \otimes B$ are constructed by adding
all the boxes of one of the two YDs, say $B$, to the other YD, here $A$, in all
possible ways subject to the following rules (see for example
\cite[Chapter.9]{Predrag}):
\begin{itemize}
   \item First, we fill each box of the YD $B$ with a label $a_i$, where
$i$ stands for the number of the row the box belongs to.
   \item We start by adding the leftmost box\footnote{This is a matter of
choice as boxes belonging to the same row are treated on equal footing, hence
the reason they carry the same label. In most literature, one starts from the
rightmost box. We choose to start from the leftmost one for reasons to be
clear later on.} form the first row of $B$, which carries the label $a_1$, to
the YD $A$ in all possible positions such that we end up with a YD of U$(N)$
i.e. the length of rows is decreasing from top to bottom, and the length of each
column is at most $N$. We repeat the same process with the remaining boxes in
the first row of $B$ following their order from left to right, keeping in mind
that each new added box should be to the right or below the previous one, and
that, no two boxes among the added ones are in the same column. The latter two
conditions are solely for boxes that belong to the same row in $B$.
   \item We repeat exactly the same process for each row of the YD $B$ until we
   finish all of its boxes. The order in which the rows of $\y$ are dealt with,
   follows exactly their position in the YD $\y$. Explicitly and in our conventions,
   we start by the first row, then the second one, and so on and so forth.
   \item Finally, we keep only the YDs that satisfy the following rule. Let $C$
denote one of the resultant YDs. We start our journey at the upper rightmost box
of the YD $C$ going from the right to the left of the first row, then move down
to the second row and, once again, start from its rightmost box and move to the
left, and so on and so forth until we reach the lower leftmost box of $C$. At
each box in this journey, the number of the newly added boxes, encountered so far,
with label $a_i$ should not exceed the number of boxes with label $a_j$ if $j<i$.
\end{itemize}
At the end, we collect the resulting YDs according to their shape. Since the
decomposition of the tensor product will play a crucial role in this paper, it
is a good idea to name the rules in the construction above, instead of trying to
explain them each time. Essentially, we have three important rules which are as
follows. The {\it YD} rule, which refers to the condition that the resultant digram
should be a YD after each step of adding a new box. The {\it antisymmetry} rule, which
refers to the condition that two boxes originating from the same row of the YD $B$
should not belong to the same column in the resulting YDs. The {\it ordering} rule,
which refers to the combination of the order in which we add boxes originating from
the same row in the YD $B$ (explained in the first step above), which will be called
the {\it row} rule from now on, and the last condition above which will be called
the {\it column} rule from now on.

Before continuing, let us apply these rules in the case of the following tensor
product:

\begin{align}
 \yng(2,1) \;\otimes\; \young(aa,b) &= \left( \young(\cdot\cdot a,\cdot) \;
\oplus \; \young(\cdot\cdot,\cdot a) \;\oplus\;
\young(\cdot\cdot,\cdot,a)\right) \;\otimes\; \young(:a,b) \nonumber \\
                                                   &= \left( \young(\cdot\cdot
aa,\cdot) \;\oplus\; \young(\cdot\cdot a,\cdot a) \; \oplus \;
\young(\cdot\cdot a,\cdot,a) \; \oplus \; \young(\cdot\cdot,\cdot a,a)\right)
\; \otimes \young(b) \nonumber \\
                                                   &= \young(\cdot\cdot
aa,\cdot b) \;\oplus\; \young(\cdot\cdot aa,\cdot,b) \;\oplus\;
\young(\cdot\cdot a,\cdot ab) \; \oplus \; \young(\cdot\cdot
a,\cdot a,b) \;\oplus\; \young(\cdot\cdot a,\cdot b,a) \nonumber\\
                                                   & \quad \qquad \;\oplus\;
\young(\cdot\cdot a,\cdot,a,b) \;\oplus \; \young(\cdot\cdot,\cdot a,ab)
\;\oplus \; \young(\cdot\cdot,\cdot a,a,b) \nonumber\\
                                                   &=
\yng(4,2) \;\oplus\; \yng(4,1,1) \;\oplus\; \yng(3,3) \; \oplus \; 2\;\;
\yng(3,2,1) \nonumber\\
                                                   & \quad \qquad \;\oplus\;
\yng(3,1,1,1) \;\oplus \; \yng(2,2,2) \;\oplus \; \yng(2,2,1,1) \quad .
\label{tensor-prod-decomp-eg}
\end{align}
\\
In this example, we used letters to label the boxes, which is fine since we are
still following the rules. Notice that in this example we have a single YD
that appears twice with different order of the letters $a$, $b$, hence it is
doubly degenerate. Notice also that the box with the letter $b$ never
appears in the first row. This is a straightforward consequence of the ordering
rule. In general, boxes from row $i$ in $B$ cannot be attached to the row $j$ in
$A$ if $j<i$.

There is another immediate consequence of the ordering rule, which has to do
with the number of added boxes to a row $i$ in $A$. Let us denote by $n^j_i$
the number of boxes from row $j$ in $B$ that are added to row $i$ in $A$.
Suppose that $i$ is the row we are interested in, and suppose also that the
added boxes to this row belong to the rows $m \leq k \leq n$ from $B$. We have
the following conditions:
\begin{align*}
  n^k_i \leq \sum_{j=1}^{i-1} \left( n^{k-1}_j - n^k_j \right) \;; \qquad
\forall k \;, \quad m+1 \leq k \leq n \;.
\end{align*}
Summing these relations gives us:
\begin{equation}
 \sum_{k=m}^n n^k_i \leq n^m_i + \sum_{j=1}^{i-1} \left( n^{m}_j - n^n_j
\right) < b_m \;, \label{added-box-cond}
\end{equation}
where $b_m$ is the length of row $m$ in $B$. Hence, if the boxes added to
row $i$ of the YD $A$ used to be in rows of the YD $B$, with $m$ denoting the
number of the upper row among them, then, the number of added boxes to the row
$i$ is smaller or equal the number of boxes in the row $m$ of the YD $B$. These
two implications of the ordering rule will have a nice interpretation after
introducing our map between the YD $\vf_k$ and SSYTx $\y$. See
appendix-\ref{Kostka-numbers-app} for more details.

Our next target, after we learned how to decompose a tensor product into
irreducible representations (YDs), is to count the number of the resulting YDs
and their degeneracies. This will be the subject of the next few subsections. We
will first start by encoding the information about YDs that appear in the
tensor product decomposition of $\calo \otimes \y$ into labelings of $\y$. Then,
using this new way of characterizing the tensor product decomposition, we
proceed to deal with our main target, estimating the number of YD $\vf_k$ and
their degeneracies $d_k$.
%
%
\subsubsection{Repackaging the information about the tensor product
decomposition} \label{tensor-prod-repackage}
%
Our aim in this paper is to study the two point function (\ref{two-pt-bc}) as
we vary the probe $\y$. Hence, it will be very effective and fruitful to use the
YD $\y$ in packaging the information about the tensor product $\calo \otimes
\y$. It is clear from the way we construct the decomposition of the tensor
product of two YDs, that it is much easier in the present case $\calo \otimes
\y$ to distribute the boxes of $\y$ on $\calo$. A naive guess for packaging the
information about the YDs $\vf_k$ that will appear in the decomposition of
$\calo \otimes \y$, is to associate to each box of $\y$ the number of the row
of the YD $\calo$ it is attached to. As a result, we get for each $\vf_k$ a
certain labeling of $\y$. To get more information about the type of these
labelings, we need to answer the following question: What are the implications
of the tensor product decomposition rules, that we discussed previously, on the
labelings of $\y$?

The easiest condition to apply on the possible labelings of $\y$, is that the
number of rows of $\vf_k$ should not exceed $N$. Since the YD $\calo$ has
already $N$ rows, see section \ref{superstar-ensemble}, the labels of $\y$
should be among the numbers $\call = \{1 \,,\, 2 \,,\, \ldots \,,\, N\}$. For a
specific $\vf_k$, the set of numbers appearing in associated labeling of
$\y$ will in general be a subset of $\call$. This observation will play an
important role later on. What about the impact of the three rules: the YD rule,
the antisymmetry rule, and the ordering rule? It is clear that the last rule,
the ordering rule, is universal in the sense that its outcome does not depend on
the YD $\calo$, in contrast to the other two rules whose outcomes depend highly
on which background $\calo$ we are studying. Hence, it is a good idea to
look first for the implications of the ordering rule on the possible labelings
of $\y$, then look for possible corrections due to the other two rules. In the
remaining of this subsection, we will only discuss the impact of the ordering
rule on the labelings of $\y$ leaving the inclusion of the other two rules
mainly to appendix-\ref{app-inclusion-yd-antisym-rules}.

The ordering rule as defined previously (section \ref{tensor-prod-rules}) has
two parts to it. The first part, called the row rule, has to do with boxes that
belong to the same row of $\y$ whereas the second one, called the column rule,
has to do with boxes that belong to different rows of $\y$. It is easy to see
that the row rule is equivalent to the requirement that, in the associated
labeling of $\y$, the numbers in the same row should be weakly increasing from
left to right. What about the column rule? To understand its consequence, let us
assume for a moment that $\y$ is a YD with a single column. To simplify the
discussion below, we will also number the boxes of $\y$ according to their row
i.e. box $i$ means a box in row $i$ of $\y$. It is easy to see that in this
case, the column rule is equivalent to the condition that the box $i$ should not
be added to the rows of $\calo$ that are before the one that has box $j$ added
to it if $j <i$. Hence, the acceptable labelings of $\y$ in this case should be
such that, the numbers labeling boxes of the single column are strictly
increasing from top to bottom. What about the case where $\y$ has two columns?
Let us concentrate on the four boxes belonging to rows $i$ and $j$ where $j>i$.
The column rule implies that, the row to which the box $j$ from the first column
of $\y$ is attached to, should be below the row to which the box $i$ from the
first column of $\y$ is attached to. The same condition should be satisfied by
the boxes $i$ and $j$ from the second column of $\y$. Hence, in this case, the
column rule requires that the numbers labeling boxes in the same column of $\y$
to be strictly increasing from top to bottom for an acceptable labeling of $\y$.
It is not hard to see that this is the manifestation of the column rule for
acceptable labelings of a generic $\y$. This can be easily seen by iterating
the previous discussion for each row of $\y$.

In our previous example of the tensor product decomposition
(\ref{tensor-prod-decomp-eg}), the labelings of the resultant YDs are:

\begin{align}
 \young(11,2) \;,\; \young(11,3) \;,\; \young(12,2) \;,\; \left(\young(12,3)
\;,\; \young(13,2) \right) \;,\; \young(13,4) \;,\; \young(23,3) \;,\;
\young(23,4)
\end{align}
\\
which clearly satisfies the two conditions: numbers in the same row are weakly
increasing from left to right, and numbers in the same column are strictly increasing
from top to bottom. The two YTx between bracket correspond to the degenerate YD.

To summarize, the shape of the YD $\calo$, the fact that we are dealing with the
U$(N)$ group, together with the ordering rule, imply that the acceptable
labelings of $\y$ that are associated to a YD $\vf_k$ in the decomposition of
$\calo \otimes \y$ should be such that:
\begin{itemize}
 \item The labels belong the set of numbers $\call = \{1 \,,\, 2 \,,\, \ldots
\,,\, N\}$.
 \item Numbers labeling boxes in the same row, are weakly increasing from left to right.
 \item Numbers labeling boxes in the same column, are strictly increasing from
top to bottom.
\end{itemize}
These last two conditions define the so called SSYT, see
appendix-\ref{conventions}. If we take into account the first condition, the
number of these SSYTx $\y$ is the dimension of $\y$ as a representation of
SU$(N)$, see for example \cite{Fulton}. Since we are not taking into account
neither the YD rule nor the antisymmetry rule, $\tdim_N \, \y$ is an upper
bound on the total number of the YDs $\vf_k$ i.e. $d_t = \sum_k d_k \leq
\tdim_N \, \y$.

Parameterizing the decomposition of the tensor product $\calo \otimes \y$ in
terms of SSYTx $\y$, gave us an idea about the total number of the YDs $\vf_k$.
But, what we are really after is the value of the degeneracies $d_k$ and the
range of the index $k$. Let us pick a YD $\vf_k$. This YD is fixed once we know
the $N$-tuple $\b = (n_1 \,,\, n_2 \,,\, \ldots \,,\, n_N)$, where $n_i$ stands
for the added boxes to row $i$ of the YD $\calo$. Its degeneracy $d_k$ on the
other hand, corresponds to the total number of the possible different rows
origin\footnote{Remember that boxes in the same row are treated symmetrically.}
in the YD $\y$ of these added boxes. In our interpretation of the tensor product
decomposition as associating fillings to the YD $\y$ to get a SSYT, the
$n_i$ in the $N$-tuple $\b$ counts the number of times the number $i$ appears in
the SSYT. As a result, the $N$-tuple $\b$ is precisely a filling, see
appendix-\ref{conventions}. So, the degeneracy $d_k$ is bounded from
above\footnote{Since we are not taking into account the modifications due to the
YD and antisymmetry rules.} by the number of different SSYTx $\y$ that are
associated to the same filling $\b$. Such a number is called a Kostka number,
and is denoted by $K_{\y \,,\, \b}$, see
appendix-\ref{conventions}. Actually, there is a precise relationship between
Kostka numbers and degeneracies of YDs appearing in the decomposition of a
tensor product, see for example \cite{ettienne}. However, we will not need this
exact relation in what we are trying to do in this paper.

Before continuing, let us clarify a point that will play an important
role in extracting the leading behavior of $d_k$ in the following.
We concluded that we can repackage the information about the tensor product
$\calo \otimes \y$ in terms of SSYTx, where the labels take values in $ \call =
\{ 1 \,,\, 2 \,,\, \dots \,,\, N \}$. We, as well, mentioned above that the number
of SSYTx $\y$ with a certain filling $\b$ is given by the Kostka numbers
$K_{\y \,,\,\b}$. However, this is not completely exact. Essentially, the Kostka
number $K_{\y \,,\, \b}$ encodes only the information about $\y$ and $\b$ being
partitions of the same integer $h$, but not the information about the range of
labels, see for example \cite{stanley}. Remember that $\b$ is a collection of
positive integers $\b_i$, some of them can be zero, which are ordered according
to their index $i$. We know on the other hand that, a partition of an integer is
a collection of strictly positive ordered integers that sum to that integer. So,
we seem to have a mismatch between the number of SSYTx of interest to us and
Kostka numbers: We can have zero entries in the filling $\b$, as well as ordering
that depends on the index of the entry $\b_i$ and not its value. We will discuss
the issue of the zeros here, and leave the ordering issue to
appendix-\ref{Kostka-numbers-app}. This appendix contains also some further useful
properties of Kostka numbers.

Remember that in order for a YT to be SSYT, the labels should have a certain
order along the rows and the columns of the associated YD (see
appendix-\ref{conventions}). This means that if $\b$ has some zero entries, the
number of SSYTx is the same as the one for a filling $\tb$, which is constructed
from $\b$ by omitting the zero terms, then relabeling its remaining entries
$\tb_j$, keeping the order of their index untouched. Take as an example $\b =
(\b_1 \,,\, 0 \,,\, 0 \,,\, \b_4 \,,\, \b_5)$. Then, $\tb = (\tb_1 \,,\, \tb_2
\,, \, \tb_3)$, where $\tb_1 = \b_1$, $\tb_2 = \b_4$ and $\tb_3 = \b_5$. Since
the map $\c$ between the indices $i$ and $j$ of the two fillings preserves the
order i.e. $(i_1 \geq i_2) \Longleftrightarrow \c (i_1) \geq \c (i_2)$, the
conditions on the order of labels along the rows and column of the YD remain
intact. Hence, the number of SSYTx for both fillings is the same. Notice that
$\tb_j$ counts how many times the index $j$ appears in the labeling of the YD.
%
%
\subsection{Non-generic probes: The toy model reloaded}
\label{ngen-probes-two-pt-funct-section}
%
%
Before continuing with our discussion of the tensor product decomposition, let
us briefly discuss the claim raised at the end of subsection
\ref{toy-model-section} regarding the universality of some of the values of
the leading term of the two point function derived there. To this end, we will
review the most important properties that were used in the derivations in
subsection \ref{toy-model-section} and, at the same time, discuss possible
modifications when dealing with general backgrounds.

The first important ingredient in the derivations of subsection
\ref{toy-model-section} is the YD $\vf_0$. Notice that this YD (with appropriate
modifications) always exists for any background YD $\calo$ as it satisfies all
the rules of the tensor product decomposition reviewed in the beginning of this
section. On top of this, the YD $\vf_0$ enjoys another equally important property:
it is the YD that maximizes the quantity $ \lag \vf_k \; \vf_k \rag_\calo $ as can
be easily deduced starting from equation
(\ref{two-point-funct-summ-1-extra}). To see this, let us look for the values of
$\d_i$ such that we maximize the aforementioned expression. From its explicit
form given by equation (\ref{two-point-funct-summ-1-extra}), it is easy to see
that we reach the maximum if we minimize the possible values of the summation
index\footnote{This is a straightforward result of the fact that the function
$f(x) = (B-x) \, \log (B-x)- (A-x) \, \log (A-x)$, with $B > A > x >0$, is a
decreasing function.} $i$ and maximize the values of $\d_i$. Taking into account
the tensor product decomposition rules, the solution to these conditions is: $\d_i
= \y_i$ i.e. the YD $\vf_0$. To arrive at this conclusion, we start by the smallest
value of $i$ which is $1$. Due to the ordering rule we can only add boxes from the
first row of $\y$. Maximizing this number leads us to the equality $\d_1 = \y_1$.
Since we are left with boxes from the YD $\y$, let us move to the next smallest
value of $i$ which is $2$. The ordering rule tells us to only add boxes from the
first two rows of $\y$. Since we are not left with any boxes from the first row
of $\y$, maximizing $\d_2$ leads us to the equality $\d_2 = \y_2$. Repeating the
same arguments for the following values of $i$ leads us to our claim that $\vf_0$
maximizes the quantity $ \lag \vf_k \; \vf_k \rag_\calo $.

The next important ingredient used in subsection \ref{toy-model-section} is the
length of the rows of $\calq$, which is the same and equals to $\k \,N$. This
enters essentially in the evaluation of the quantity $\lag \vf_0 \; \vf_0
\rag_\calq$. What happens in the case of a generic background $\calo$? Looking
back at the construction of $\vf_0$, it is easy to see that the leading term of
$\lag \vf_0 \; \vf_0 \rag_\calo$ will be the same as the one of $\lag \vf_0 \;
\vf_0 \rag_\calq$, if the length of almost all\footnote{Actually, it is enough
that the first $n$ rows of $\calo$ satisfy this requirement. But since we allow
$n$ to be of order $N$ (linear probes class), we end up with our condition above.}
of the rows of $\calo$ is of order $N$, even if they are not equal to each other.

Building on the observations above, can we say something on the full two point
function (\ref{two-pt-bc})? Taking into account that the YD $\vf_0$ is
non-degenerate ($d_0 =1$), there is a term associated to the DY $\vf_0$ in the
explicit expression of the two point function given in equation
(\ref{two-point-funct-explicit}). The log of the leading term contribution of
this quantity ($\lag \vf_0 \; \vf_0 \rag_\calo $) is given by either equation
(\ref{two-pt-funct-ntyp-gen}) or (\ref{two-pt-funct-ntyp-ngen}) depending on
the type of the probe $\y$. What happens when we take into account the tensor
product decomposition? Let us denote by $d_t$ the total number of the YDs $\vf_k$
i.e. $d_t = \sum_k d_k$. We proved in the previous subsection that $d_t \leq \tdim_N
\, \y$, in the case where $\calo$ is a member of the superstar ensemble. Actually,
this inequality is valid for any YD that is treated as a representation of the
group U$(N)$. This is because the biggest allowed number of rows of these YDs is
$N$. Hence, the two point function (\ref{two-pt-bc}) satisfies the following
inequality:
$$ \lag \vf_0 \; \vf_0 \rag_\calo \leq \lag \lag \y \; \y \rag \rag_\calo \leq
\left( \tdim_N \, \y \right)^2 \, \lag \vf_0 \; \vf_0 \rag_\calo \;, $$
which is a consequence of the fact that $\vf_0$ maximizes the quantity $\lag
\vf_k \; \vf_k \rag_\calo$. Given the leading behavior of $\log \,\lag \vf_0 \;
\vf_0 \rag_\calo $ (see equations (\ref{two-pt-funct-ntyp-gen}) and
(\ref{two-pt-funct-ntyp-ngen})), it is clear that in the case where:
$$ \log \,\tdim_N \, \y \ll h \, \log N \;,$$
the leading term of the two point function (\ref{two-pt-bc}) is completely
fixed by the YD $\vf_0$ according to the equation:
$$ \log \, \lag \lag \y \; \y \rag \rag_\calo \approx \log \, \lag \vf_0 \;
\vf_0 \rag_\calo \;. $$
In the case where almost all the rows of $\calo$ have a length of order $N$,
the value of the leading term above will be the same as the one associated to
the background $\calq$. In our case of interest, the non-generic probes class
satisfy the above inequality on their SU$(N)$ dimension, see equations
(\ref{dim-SUN-behav-giant}) and (\ref{dim-SUN-leading-ngeom}). Hence, in the
case of the non-generic probes $\y$, the leading term of the full two point
function (\ref{two-pt-bc}) reads:
\begin{equation}
 \log \, \lag \lag \y \; \y \rag \rag_\calo \approx h \, \log \, \y_0 \;,
\label{two-pt-funct-ngen-univ}
\end{equation}
for any YD $\calo$ whose almost all of its rows have a length of order $N$.
This set of YDs includes almost all of the YDs of the superstar ensemble and
many more. This universal result has a nice intuitive interpretation from the
dual gravity perspective. The dual objects of this class of probes are a bound
state of dual giant gravitons \cite{McGreevy:2000cw, Hashimoto:2000zp} with a
large angular momenta. Since the radius of a dual giants is proportional to its
angular momentum, these giants probe the outer region of the geometry which is
completely fixed by its asymptotic AdS$_5 \times$ S$^5$ and the background
flux. Given the result above, the only class of probes that we need to worry
about in the remaining of this paper are the generic probes that satisfy $\y_0
\ll N$. The study of the tensor product decomposition of $\calo \otimes \y$ in
this class of probes will be the main topic of the remaining of this
section.
%
%
\subsection{The tensor product $\calo \otimes \y$: I-The case $h \ll N$}
\label{tensor-prod-h-ll-N}
%
%
We are ready to tackle our main question of estimating the maximal possible
degeneracy of a YD in the decomposition of $\calo \otimes \y$ through the
corresponding Kostka number. Our idea is to use $\tdim_N \, \y$ since it is
related to the number of possible SSYTx and hence to the Kostka numbers.
Remember that $\tdim_N \, \y$ is the number of all possible SSYTx with labels
among the set $\{ 1 \,,\, 2\,,\, \ldots \,,\, N\}$. This number has two
contributions: The first one is the number of possible choices of labels, and
the second one is the associated Kostka number $K_{\a \,,\, \b}$. Using that
$K_{\a\,,\,\b} = 0$ if $|\b| < n$, where $n$ is the number of rows of $\y$, and
$|\b|$ is the number of nonzero entries of the filling $\b$ (see
appendix-\ref{Kostka-numbers-app}), it is easy to see that the dimension of $\y$
as a representation of SU$(N)$ written in terms of Kostka numbers reads:
\begin{equation}
  \tdim_N \, \y = \sum_{\b \,; \; n \leq |\b|\leq \tmin\,\{N\,,\,h\}}
C_N^{|\b|} \, K_{\y\,,\,\b} \;, \label{suN-dim-kostka}
\end{equation}
where $n$ is the number of rows of $\y$, $|\b|$ stands for the number of
nonzero elements in the filling $\b$, $h$ is the number of boxes of $\y$,
$C_n^m$ is the usual binomial coefficient:
$$ C_n^m = \frac{n!}{m!\, (n-m)!} \;, $$
and $K_{\y \,,\, \b}$ is the Kostka number associated to the filling $\b$. The
expression (\ref{suN-dim-kostka}) above reflects the fact that when labeling the
YD $\y$, we first need to choose the labels to use, among all the possibilities
$\call = \{ 1 \,,\, 2 \,,\, \ldots \,,\, N\}$ which gives rise to the binomial
coefficient, then we need to decide on the possible multiplicities of these
labels (i.e. choose a filling) each of which gives rise
to its associated Kostka numbers.

Notice that when fixing the value of $|\b|$, not all fillings $\b$ contribute
to the sum (\ref{suN-dim-kostka}) above as the corresponding Kostka number might
vanish. Our approach will be to include all of these fillings $\b$ in the sum
keeping in mind that some of the associated Kostka numbers can vanish. How many
$\b$'s are there? For a fixed $|\b|$, the number of possible fillings $\g$ such
that $|\g|=|\b|$ is the same as the number of partitions of the integer $h$ into
$|\b|$ strictly positive integers $\g_i$ that are ordered according to
their index $i$ and not their value. This is similar to the problem discussed in
section-\ref{counting-formula} of appendix-\ref{useful-form}. The difference
resides in the fact that we want to only count partitions that do not have zero
entries here. It is easy to adapt the counting there to this case by putting from
the start one ball in each box, reducing the number we want to partition to $h -
|\b|$. Hence, we get the number:
\begin{equation}
 \caln_{|\b|} = C_{h-1}^{|\b| -1} \;. \label{nconstrained-filling-fixed-length}
\end{equation}
It is clear that this is an over-estimate for the actual number of fillings
$\g$ with fixed $|\g| = |\b|$, in the sense that not all the associated Kostka
numbers are different from zero, but this will be enough for our arguments in
this and the next subsection. Using the estimate above, the total number of
fillings $\b$ is bounded from above by:
\begin{equation}
 \caln_\b = \sum_{|\b|=1}^h \caln_{|\b|} = 2^{h-1} \;.
\label{total-number-filling}
\end{equation}
It is an upper bound since we allowed for fillings $\b$ with $|\b| < n$ in this
sum for which we know that their associated Kostka number vanishes. We have
also allowed the range of summation to run all the way to $h$ even if $h>N$.
Notice that we are still missing the contribution of the binomial coefficient
$C_N^{|\b|}$ to get the actual number of fillings, however we will not do so
here and keep calling $\caln_\b$ above the total number of fillings in an abuse
of language.

As is clear from the form of the expression (\ref{suN-dim-kostka}), our
search for the ``dominant'' SSYTx $\y$ will depend on whether $h$ or $N$ is
bigger. We will concentrate on the case $h \ll N$ here, and leave the cases $h
\sim N$ and $ h \gg N$ to the next subsection. Before continuing with our
discussion, let us be more precise about what we are dealing with. First of all,
what we really mean by comparing $h$ and $N$ is comparing the leading behavior
of $h$ with $N$. Hence, $h = N/2 - \sqrt{N}$ is part of the cases $h \sim N$.
The next point we need to keep in mind is that, we are only interested in the
generic probes class, see the previous subsection for details. These probes
are such that $\y_0 = \tmax\, \{d \,,\, n\} \ll N$.

The idea in the case $h \ll N$ is to maximize both the binomial coefficient
$C_N^{|\b|}$ and the Kostka number $K_{\y \,,\,\b}$ independently. Notice that
the sum in this case is over the range $n \leq |\b| \leq h$. Hence, the maximum
of the binomial coefficient $C_N^{|\b|}$ is reached for $|\b|=h$. This is
satisfied only for the filling $\b_0 = (1)$ \footnote{The notation $\b_0 = (1)$
means that all the entries in the filling $\b_0$ equal $1$.}, which gives rise
to standard Young tableaux (SYTx). The reason we have only one filling $\b=(1)$
for $|\b|=h$ is that the number of fillings in a family with fixed $|\b|$ is the
number of $|\b|$-tuples $(n_1\,,\, n_2 \,,\, \ldots \,,\, n_{|\b|})$ such that:
$$ \forall \; i\,; \quad n_i >0 \;, \quad \sum_{i=1}^{|\b|} n_i = h \;.$$
It is clear from this condition that the only solution in the case $|\b| = h$
is $n_i =1 \;, \; \forall \; i$, and hence the claim that $\b_0 = (1)$ is the
only filling such that $|\b|=h$. What about the Kostka number $K_{\y \,,\, \b}$?
A moment thought reveals that this number is also maximized by the filling $\b_0
= (1)$. This is because in this filling all the labels used to label the boxes
of the YD $\y$ are different from each other. Combining this observation with
the fact that for SSYTx, the labels along the same column should be strictly
increasing from top to bottom leads us the aforementioned claim. As a result,
the filling with the maximum contribution to the sum in
(\ref{suN-dim-kostka}) is $\b_0 =(1)$. Notice that maximum here does not mean
that the contribution coming from the filling $\b_0 =(1)$ dominates over the
other contributions to the sum (\ref{suN-dim-kostka}). It will be nice to find
out if there is a condition on the shape of the YD $\y$ such that this is
true\footnote{The author is very grateful to Robert de Mello Koch for pointing
out a mistake in an argument related to this point in a previous version of this
paper.}. What is the value of the Kostka number $K_{\y \,,\, (1)}$? Remember
that this is the same as the number of SYTx associated to the YD $\y$, which in
turn is the dimension of the YD $\y$ when seen as an irreducible representation
of the permutation group $S_h$. The latter is given by (see equation
(\ref{dim-perm})):
$$ K_{\a\,,\,(1)} = \tdim_h \, \y = \frac{h!}{\calh_\y} \;. $$
As a side note, one can get this formula by rescaling $N$ as $(\l \, N)$ in both
the formulas (\ref{suN-dim-kostka}) and (\ref{dim-hooks}) of $\tdim_N \, \y$
without touching the YD $\y$, then picking the leading contribution to both
of them in the limit $\l \ra \infty$ and equating them. This scaling argument
may be useful in answering the question raised above about the condition
on $\y$ so that the dominant contribution to (\ref{suN-dim-kostka}) comes form
SYTx.

Let us continue our investigation of the SYTx associated to $\y$ in the present
situation i.e. the labels of the fillings of $\y$ are subsets of $\call = \{1
\,,\,2 \,,\, \ldots N\}$. Their total number, taking into account the
contribution coming form the possible different choices of labels, is given by
$\caln_{SYT} = \caln_{lab} \times \caln_{perm}$, where:
\begin{equation}
 \log \, \caln_{lab} = \log \, C_N^h \approx h \, \log \frac{N}{h} \;, \quad
\log \, \caln_{per} = \log \, \tdim_h \, \y \approx h \, \log \frac{h}{\y_0}
\;, \label{h-ll-N-info}
\end{equation}
where $\caln_{lab}$ is the number of possible labelings and $\caln_{perm}$ the
number of SYTx $\y$ given fixed labels. Notice that:
$$ \log \, \tdim_N \, \y \approx \log \, \caln_{lab} + \log \, \caln_{perm} \;.
$$
This is a very interesting observation as it suggests that we can restrict
ourselves to the SYTx $\y$ when studying the SSYTx $\y$ if we are only
interested in leading order quantities. For example, the contribution from the
other fillings to (\ref{suN-dim-kostka}) is bounded from above by:
$$ \d \, \tdim_N \, \y \approx 2^{h-1} \, C_N^h \, K_{\y \,,\, (1)} \;. $$
As a result, the correction to $\log \, \tdim_N \, \y$ due to SSYTx that are
not SYTx, after taking the contribution of the latter into account, is subleading
as one expects based on the observation above.

Although we managed to get ample information about the SSYTx $\y$ in the case
$h \ll N$, we still lack the information we need: $d_k$ the degeneracy of the
YDs $\vf_k$ in (\ref{tensor-prod-decomp-schem}) and their type. Using the
discussion of the Kostka numbers above as a guide, together with the map between
SSYTx $\y$ and $\vf_k$, one expects that one can also take as representatives
of the YDs $\vf_k$ appearing in the decomposition of $\calo \otimes \y$, the
ones that we get by adding at most one box to each row of $\calo$. These are the
``duals'' of the SYTx $\y$ according to our map between SSYTx and YDs $\vf_k$,
and we will denote them from now on by $\vf_k^0$. The reasons that we can do so
can be summarized as follows. First of all, the leading terms of log of their
degeneracy $d_k^0$ and their number $\caln_0$ which are given by:
\begin{equation}
 \log \, d_k^0 \approx h \, \log \frac{h}{\y_0} \;, \quad \log \caln_0 \approx
h \, \log \frac{N}{h} \;, \label{tensor-prod-info-h-ll-N}
\end{equation}
do no change once we take into account the the YD and the antisymmetry rules.
See section \ref{h-ll-n-discussion} of
appendix-\ref{app-inclusion-yd-antisym-rules} for more details. Secondly, the
leading term of the log of $\lag \vf_k \; \vf_k \rag_\calo$, that appears in the
explicit form (\ref{two-point-funct-explicit}) of the two point function
(\ref{two-pt-bc}), is the same for almost all the YDs $\vf_k$, see section
\ref{two-pt-funct-full-section} for more details. Lastly, the degeneracies
$d_k$ are maximized for this kind of YDs. This is because $d_k$ are related to
Kostka numbers and these YDs are associated to the filling $\b = (1)$ which
maximizes the Kostka number. We will come back to these issues and others when
we discuss the full two point function of these probes in subsection
\ref{two-pt-funct-full-gen-h-ll-N}.
%
%
\subsection{The tensor product $\calo \otimes \y$: II-The cases $h \sim N$ and
$h \ll N$} \label{tensor-prod-h-sim-gg-N}
%
%
As previously mentioned, when studying the decomposition of the tensor product
$\calo \otimes \y$, we are solely interested in the generic probes. Let us first
look for a lower bound on the maximum possible value of $K_{\a\,,\, \b}$
using equation (\ref{suN-dim-kostka}) in this case. Since it is the maximum, it is
bigger or equal to the average value of $K_{\y \,,\, \b}$ in this equation. To
calculate this average we need to know the total number of terms in
(\ref{suN-dim-kostka}). Using the estimate
(\ref{nconstrained-filling-fixed-length}), an over-estimate of the total number
of terms is given by:
$$ \caln_{tot} = \sum_{k =0}^N C_N^{k} \, C_{h-1}^{k-1} < 2^{N+h} \;. $$
The combination of this estimate together with equation (\ref{suN-dim-kostka}),
the fact that $\log \, \tdim_N \y \sim h \, \log N$ (see subsection
\ref{probes-section}), gives the following approximate value for $K_\a^{max}$
the maximum value for the Kostka numbers:
\begin{equation}
 \log \, K_\y^{max} \approx \log \, \tdim \, \y \sim  h \, \log N \;.
\label{approx-max-kostka-h-sim-N}
\end{equation}
This means that there are some SSYTx $\y^\ast$ whose Kostka number is big enough
to give rise to the leading term in $\tdim_N \, \y$. What are the fillings of
such SSYTx $\y^\ast$? According to our discussion in
appendix-\ref{Kostka-numbers-app}, the fillings $\b^\ast$ are such that all the
labels $1 \,,\, 2 \,,\, \ldots \,,\, N$ are present with almost equal
frequency.

Based on this results, one expects that there are few\footnote{Few in the sense
that the leading behavior of the log of their number is subleading with respect
to $h \, \log N$.} YD $\vf_k^\ast$ in the decomposition
(\ref{tensor-prod-decomp-schem}) whose degeneracy $d_k^\ast$ has the leading
behavior:
\begin{equation}
 \log \, d_k^\ast \approx \log \, \tdim_N \, \y \approx h \, \log \frac{N}{\y_0}
\;.
\label{tensor-prod-info-h-sim-gg-N}
\end{equation}
Once again, we need to check the effect of taking into account the YD and the
antisymmetry rules. We will discuss the inclusion of the YD rule here as it is
relatively easier to discuss than the inclusion of the other rule, the
antisymmetry rule, which will be dealt with in section
\ref{h-sim-gg-n-discussion} of appendix-\ref{app-inclusion-yd-antisym-rules}.

The fillings of these special SSYTx $\y^\ast$ include all the numbers from $1$
to $N$. So, the potential associated YD $\vf_k^\ast$ will be obtained by adding
boxes to all the rows of $\calo$. The YD rule then implies that the length of
the new rows should be decreasing from top to bottom. The invariance of the
Kostka numbers $K_{\y \,,\, \b}$ under the reshuffling of the entries $\b_i$ of
the filling $\b$ comes to our rescue, see appendix-\ref{Kostka-numbers-app}.
Using this invariance, we choose the ordered filling $\b^\ast$: $\b_1 \geq \b_2
\, \ldots \geq \b_N$ to correspond to one of theses possible $\vf_k^\ast$. At
this point one might be tempted to declare victory and conclude that taking into
account the YD rule does not change the results obtained through the Kostka
number means. However, one should remember that the YD rule is not limited to
the final $\vf_k$, but it is enforced after the addition of each box to the YD
$\calo$. Although we could not come up with a satisfactory argument to why
the final conclusion will not change even after fully taking into account the
YD rule, it does not seems to be that crazy to conjecture that this is the case.
A piece of evidence has to do with the entries $\b_i^\ast$ of the filling
$\b^\ast$. Remember that for maximum Kostka numbers in our case of interest, we
have (see appendix-\ref{Kostka-numbers-app} for details):
$$ \forall \; i \in \{1 \,,\, 2 \,,\, \ldots \,,\, N \} \;\;; \quad \b^\ast_i
\approx \frac{h}{N} \;.$$
Using that for our probes $h \ll N^2$, we conclude that $\b_i^\ast \ll N$. This
implies that it is more probable for the newly added box to end up in different
row than the previous one, rather than the same row. So, one can think of the
cases where the YD rule is violated as organized constructions in contrast to
the cases where it is satisfied, which can be thought of as random constructions.
As a result, one can conclude that the number of cases where the YD rule is
violated is subleading with respect to the total number of possible $\vf^\ast_k$.
This conclusion is in conformation with the assumption that the leading term of
$\log \,\lag \lag \y \; \y \rag \rag_\calo$ is continuous, see the end of the next
section and the conclusions for more details.

The inclusion of the antisymmetry rule is more involved and is left to
section-\ref{h-sim-gg-n-discussion} of
appendix-\ref{app-inclusion-yd-antisym-rules} as already advertised to. The
final conclusion is that, in the decomposition of the tensor product $\calo
\otimes \y$, for $\y$ a generic probe and $h \sim N$ or $h \gg N$, there are
special YDs $\vf_k^\ast$ whose degeneracy $d_k^\ast$ reproduces the leading
behavior of $\tdim_N \, \y$, which is given in
(\ref{tensor-prod-info-h-sim-gg-N}). Restricting ourselves to one of these YDs
will give us the leading order of the log of the  two point function
(\ref{two-pt-bc}) that we are after. For more details see subsection
\ref{two-pt-funct-full-gen-h-sim-gg-n}.
%
%
%
\section{The full two point function and its universality}
\label{two-pt-funct-full-section}
%
%
%
After we managed to gather the information we need on the decomposition of
the tensor product $\calo \otimes \y$, we are almost ready to solve the last
piece of our puzzle: evaluate the leading term of the two point function
(\ref{two-pt-bc}), whose explicit expression is given by
(\ref{two-point-funct-explicit}), in the case of generic probes. The only
remaining step is to evaluate the quantity $\lag \vf_k \; \vf_k \rag_\calo$. Our
aim in this section is to evaluate the latter quantity, then get the leading
term of the log of the full two point function (\ref{two-pt-bc}). Before
continuing let us, for completeness, remind ourselves of the leading term of
the two point function in the case of non-generic probes. This was derived in
subsection \ref{ngen-probes-two-pt-funct-section} and is given by equation
(\ref{two-pt-funct-ngen-univ}), which we rewrite here for convenience:
\begin{equation}
 \log \, \lag \lag \y \; \y \rag \rag_\calo \approx h \, \log \y_0 \;,
\label{two-pt-funct-ngen}
\end{equation}
where $h$ is the total number of boxes of the probe $\y$, $\y_0 = \tmax \, \{d
\,,\, n \}$, $d$ is the total number of columns of $\y$, and $n$ is the total
number of its rows. In the formula above, we should only keep the leading term
of $h$ and just the leading behavior of $\y_0$ i.e. its scaling with $N$. We
want to make a small observation before moving on to the case of the generic
probes. Notice that this formula does not have an explicit dependence on $N$,
except for linear probes where $\y_0 \sim N$. This suggests that it might be
better to see the long probes ($\y_0 = d \gg N$) as living in an SU$(d)$ field
theory rather than an SU$(N)$ one.  We will discuss this observation in some
detail in the conclusions.

Back to our generic probes and their two point function. First of all, we need
to deal with the quantity $\lag \vf_k \; \vf_k \rag_\calo$. To do so and in
order to have a unified discussion below, let us introduce some new notations as
well as remind ourselves of old ones. $\calo_i$ is the length of the row $i$ of
the background YD $\calo$. For almost all the values of $i$, $\calo_i$ is of
order $N$. Remember that $i$ runs from $1$ to $N$. $\d_i^{(k)}$ denotes the
number of added boxes to row $i$ of $\calo$ to form the YD $\vf_k$, and $n_\d$
stands for the total number of non-zero $\d_i^{(k)}$. The ordering rule implies
that:
\begin{equation}
 n_\d \geq n \;, \quad \forall\; k \;, \; \forall \; i \;; \quad \d_i^{(k)} \leq
d \;. \label{cond-delta-n-delta}
\end{equation}
These inequalities are also a straightforward implications of the properties of
the SSYT labelings in conformation with the equivalence between the ordering
rule and the SSYTx $\y$. These inequalities will play an important role in the
following. Our starting point is equation (\ref{two-point-funct-summ-1-extra})
adapted to the present case:
\begin{align*}
 \log \, \lag \vf_k \; \vf_k \rag_\calo \approx& - h + \sum_{i=1}^N (N +
\calo_i + \d_i^{(k)} -i) \, \log (N + \calo_i + \d_i^{(k)} -i) \nonumber \\
                                               & \quad \qquad - \sum_{i=1}^N (N
+ \calo_i -i) \, \log (N + \calo_i -i) \;.
\end{align*}
To proceed further, remember that for the generic probes we have $d \ll N$. This
implies, according to (\ref{cond-delta-n-delta}), that $\d_i^{(k)} \ll N$. Using
that for almost all $i$'s $\calo_i \sim N$, we can expand the $\log$ term in
the equation above and perform the sum to get:
\begin{align}
 \log \, \lag \vf_k \; \vf_k \rag_\calo \approx& - h + \sum_{i=1}^N
\d_i^{(k)} \, \log (N + \calo_i -i) \approx h \, \log N \;.
\label{two-point-function-summand-gen}
\end{align}
A legitimate objection to the use of this result in the expression
(\ref{two-point-funct-explicit}) is that the rows where $\calo_i \ll N$ might
spoil the final result. An in depth discussion of this point depends on
the type of dominant YDs $\vf_k$, which in turn depends on whether $h \ll N$,
or $(h \sim N \,,\, h \gg N)$. Since the leading term of the degeneracy $d_k$
depends also on this splitting in the regimes of $h$, we will specialize below
to these two distinct cases for the evaluation of the full two point function
(\ref{two-pt-bc}).
%
%
\subsection{Generic probes with $h \ll N$} \label{two-pt-funct-full-gen-h-ll-N}
%
%
We concluded at the end of subsection \ref{tensor-prod-h-ll-N} that, in this
case, a good set of representative YDs $\vf_k^0$ are the ones resulting from
adding at most one box to each row of $\calo$. The number of these YDs $\caln_0$
as well as their degeneracy $d_k^0$ have the leading terms that are given in
equation (\ref{tensor-prod-info-h-ll-N}), reproduced here for convenience:
\begin{equation}
 \log \, d_k^0 \approx h \, \log \frac{h}{\y_0} \;, \quad \log \caln_0 \approx
h \, \log \frac{N}{h} \;. \label{tensor-prod-info-gen-h-ll-n}
\end{equation}
We need to settle the question of the usage of the expression
(\ref{two-point-function-summand-gen}) for all $\lag \vf_k \; \vf_k \rag_\calo$
in the expression (\ref{two-point-funct-explicit}). The issue as pointed out
earlier is that the behavior $\calo_i \sim N$ used in deriving this expression
is not valid always. Although the ratio of the number of such rows in $\calo$
to the total number of rows $N$ tends to zero in the large $N$ limit, which is
the limit we are interested in, one still should check that things do not go
astray. First of all, notice that we only need to worry about these rows in the
case where the number of boxes added to these rows is a finite fraction of the
total number of boxes $h$. This is because we are interested in leading order
terms. From the properties of the background YDs $\calo$, see section
\ref{superstar-ensemble}, one easily infers that these problematic rows sit at
the tail of the YD $\calo$. Let us denote their number by $m$, $m \ll N$ but can
be as big as we want. The probability that a box from $\y$ ends up in one of
these rows is $p = (m/N)$. Let us suppose that $\vf$ is a YD among the dominant
$\vf_k$ YDs where $(\k \, h)$ boxes from $\y$ are added to some rows among these
$m$ rows. Then the probability of finding such a YD among all the available YDs
$\vf_k$ is bounded from above by:
$$ \calp \leq C_h^{\k \, h} \; p^{\k \, h} \; (1-p)^{(1-\k) \, h}
\longrightarrow 0 \;, \quad \text{for} \quad N \longrightarrow \infty \;, $$
as one expects. This is an upper limit since we did not take into account that
we still have a further contribution coming from the ratio between the total
number of possibilities to distribute $h$ boxes on $N$ rows and the number of
possibilities to distribute $(\k \, h)$ boxes on $m$ rows and the remaining
boxes on $(N-m)$ rows. This contribution reduces the probability further, but
its effect is small, hence can be neglected.

As a result, when evaluating the contribution of the YDs $\vf_k^0$ to the
leading term of the two point function (\ref{two-pt-bc}) using the explicit
expression (\ref{two-point-funct-explicit}), we can safely use
(\ref{two-point-function-summand-gen}) for $\lag \vf_k^0 \; \vf_k^0 \rag_\calo$,
the degeneracy $d_k^0$ as well as the number $\caln_0$ given by equation
(\ref{tensor-prod-info-gen-h-ll-n}). At the end, we get the following leading
behavior of the contribution of the YDs $\vf_k^0$.
\begin{equation*}
 \log \lag \lag \y \; \y \rag \rag_\calo^0 \approx h \, \log N + 2\, h \, \log
\frac{h}{\y_0} + h\, \log \frac{N}{h} \approx h \,  \log \left[ h \; \left(
\frac{ N}{\y_0} \right)^2 \right] \;.
\end{equation*}
We need to worry about possible corrections to this leading behavior due to the
other YDs. For this, we need to remember that the degeneracies $d_k^0$ take the
maximum possible value of $d_k$. We need also to remember that the leading term
of the maximum of $\lag \vf_k \; \vf_k \rag_\calo$ is the value used for $\lag
\vf_k^0 \; \vf_k^0 \rag_\calo$. The final piece of information that we need is
that the number of all possible YDs $\vf_k$ is bounded from above by:
$$ \overline{\caln} = C_N^h \, 2^h \approx 2^h \, \caln_0 \;, $$
see subsection \ref{tensor-prod-h-ll-N} for more details. Using all these
information, it is easy to conclude that the correction to the log of the full
two point function $\lag \lag \y \; \y \rag \rag_\calo$ when we include all the
other YDs $\vf_k$ is subleading. Hence, the log of the two point function for
a generic probe with $h \ll N$ in the background of a typical state $\calo$ of
the superstar ensemble has the following leading term:
\begin{equation}
 \log \lag \lag \y \; \y \rag \rag_\calo \approx h \,  \log \left[ h \; \left(
\frac{ N}{\y_0} \right)^2 \right] \;. \label{two-point-funct-h-ll-N}
\end{equation}
This expression should be understood as follows. We should only take into
account the leading behavior of the terms inside the log, whereas for $h$
outside the log, we keep only its leading term. Notice the huge difference
between this leading term and the corresponding one in the case of the
backgrounds discussed in subsection \ref{toy-model-section}, given by equation
(\ref{two-pt-funct-ntyp-gen}). Due to this difference, one concludes that this
class of probes $\y$ probe deep into the geometry, which allows us to
distinguish between typical states and a random one, even if it is very close
to a typical state. 
%
%
\subsection{Generic probes with either $h\sim N$ or $h \gg N$}
\label{two-pt-funct-full-gen-h-sim-gg-n}
%
%
In these cases there are few dominant YD $\vf^\ast_k$ that are the result of
adding $\d_i^{(k)} \approx (h/N)$ boxes to each row of $\calo$. Their
degeneracy is such that its leading term is:
\begin{equation}
 \log \, d^\ast_k \approx \log \, \tdim_N \, \y \approx h \, \log
\frac{N}{\y_0} \;. \label{tensor-prod-info-gen-h-sim-gg-n}
\end{equation}
See section \ref{tensor-prod-h-sim-gg-N} and
appendix-\ref{app-inclusion-yd-antisym-rules} section
\ref{h-sim-gg-n-discussion} for more details. The reason we can use
expression (\ref{two-point-function-summand-gen}) for $\lag \vf_k^\ast \;
\vf_k^\ast \rag_\calo$ is much easier to understand in the present situation.
Due to the way we build the dominant YDs $\vf^\ast_k$, one easily concludes
that the number of boxes added to the $m$ rows whose length $\calo_i
\ll N$ is less than:
$$ \frac{h}{N} \, m \, \log m \sim \frac{m}{N} \, h \, \log N \ll h \, \log N
\;, $$
which implies that the leading order of $\lag \vf_k^\ast \; \vf_k^\ast
\rag_\calo$ is the same as in the equation
(\ref{two-point-function-summand-gen}).

Now using the fact that $\tdim_N \, \y$ is an upper bound on the total number
of YDs $\vf_k$, the leading term of the degeneracy $d_k^\ast$ of the dominant
YDs $\vf_k^\ast$, as well as the leading behavior of $\lag \vf_k^\ast \;
\vf_k^\ast \rag_\calo$, we find the following leading term of the two point
function (\ref{two-pt-bc}):
\begin{equation}
 \log \, \lag \lag \y \; \y \rag \rag_\calo \approx h \, \log N + 2 \, h \log
\frac{N}{\y_0} \approx h \, \log \left[ N \; \left(\frac{ N}{\y_0} \right)^2
\right] \;. \label{two-point-funct-h-sim-gg-N-generic}
\end{equation}
Once again, we only care about scaling with $N$ for what is inside the log, and
for $h$ outside the log we only keep its leading term in this expression. Using
the same ideas as in the previous case, we can show that the inclusion of the
other YDs $\vf_k$ adds only subleading corrections to this formula. Notice
that in this case as well, the leading term of the two point function is different
from the corresponding one discussed in subsection \ref{toy-model-section} and given
by equation (\ref{two-pt-funct-ntyp-gen}). Hence, we conclude that the generic probes
are very efficient in distinguishing typical states from other backgrounds as well
as other states in the superstar ensemble.

What about the analytic properties of the leading term of the two point
function in the background of typical states of the superstar ensemble? By
looking at equations (\ref{two-pt-funct-ngen}), (\ref{two-point-funct-h-ll-N}),
and (\ref{two-point-funct-h-sim-gg-N-generic}) it clear that the leading term
of the two point function is continuous in all the regime of parameters we are
interested in. As a by-product, this continuity in the case of generic probes
gives a further support of our conjecture that the leading term of the
degeneracies $d^\ast_k$ is given by the leading term of the associated Kostka
number\footnote{One should take these claims with a grain of salt. According
to \cite{Balasubramanian:2005mg}, the dual effective description of the
superstar ensemble is a singular geometry. So, a priori, it is not clear that
the two point function should be continuous.}. See section
\ref{tensor-prod-h-sim-gg-N} as well as
appendix-\ref{app-inclusion-yd-antisym-rules} section
\ref{h-sim-gg-n-discussion} for more details. On top of being continuous, the
leading term is differentiable except at two ``distinct'' points. The first
happens at $\y_0 \sim N$ and is the same as the one we stumbled onto in
subsection \ref{toy-model-section}. As already discussed in the latter
subsection, this non-differentiability can be attributed to the phase transition
gravitons/(dual-)giant gravitons in the bulk. The second non-differentiability
is new and happens at the point $h \sim N$. This non-differentiability does
not have a counterpart in the case of the background $\calq$ of subsection
\ref{toy-model-section}. It seems also that the point $h \sim N$ is intimately
connected to the fact that the typical states of the superstar ensemble have
order $N$ corners, and we expect that this point will shift for other
backgrounds. So, this non-differentiability tells us something about the
background we are probing rather than the probes. Although we lack at the moment
a complete understanding of this non-differentiability, we will advance some
proposals in the conclusions section below.
%
%
%
\section{Discussion and conclusions}
%
%
%
The fuzzball proposal for black holes \cite{Lunin:2001jy, Mathur:2002ie,
Mathur:2005zp} has a lot of potential to solve the black hole puzzles. Even
though there are several tests of this proposal (see the reviews
\cite{Mathur:2005zp, Mathur:2005ai, Bena:2007kg, Skenderis:2008qn,
Mathur:2008nj, Balasubramanian:2008da, Simon:2011zza} and the
references therein.), it is not fully clear yet whether it is
right or wrong, or needs just a better phrasing. In this paper, we started a
further test of this proposal in the case of the superstar of
\cite{Myers:2001aq} and its conjectured ensemble, the superstar ensemble, that
was introduced in \cite{Balasubramanian:2005mg}. We probed the different typical
states of this ensemble using light half-BPS probes. Of course, to complete the
test one needs to do the dual calculation in the superstar geometry. We hope to
come back to this issue in the future.

Although what we did in this paper was half the needed job, the results we obtained
puts a question mark on the validity of the conjecture of \cite{Balasubramanian:2005mg}. Performing the dual gravity calculation is bound to bring more clarification on this question mark.
We find the following universal leading term of the two point
function (\ref{two-pt-bc}):
\begin{equation}
 \log \lag \lag \y \; \y \rag \rag_\calo \approx \a \; h \, \log N \;,
\label{main-eqn}
\end{equation}
where $h$ is the energy/conformal weight of the probe $\y$, $N$ the flux of the
background we are probing, and $\a$ is a constant that depends heavily on the
shape of the probe $\y$ and, on a lesser level, on the scaling of $h$ with $N$,
see equations (\ref{two-pt-funct-ngen}), (\ref{two-point-funct-h-ll-N}), and
(\ref{two-point-funct-h-sim-gg-N-generic}). By the shape of the YD $\y$, we mean
which of the two numbers: the number of columns $d$ or the number of rows $n$
dominates in the large $N$ limit. Notice that the value of $\a$ is invariant
under the exchange (rows $\leftrightarrow$ columns). This invariance can be seen
as a manifestation of the symmetry (particle $\leftrightarrow$ hole) in the
quantum hall description of the half-BPS sector of type-IIB string theory on
asymptotically AdS$_5 \times$ S$^5$ \cite{Berenstein:2004hw, Ghodsi:2005ks}.
Two questions beg for an answer at this stage: Is there an intuitive explanation
for the form of the leading term of $\log \, \lag\lag \y \; \y \rag\rag_\calo$
given above? What can we say about the different values of $\a$?

Before attempting to answer these two questions, let us first discuss a subtle
issue about the value of $\a$. To be fair, we cheated a little bit when we declared
that $\a$ did not depend on the background. Naively, one would expect it to depend
on the shape of the background. This is the case and this dependence is encoded in
the tensor product decomposition of $\calo \otimes \y$, see section
\ref{tensor-prod-section} and appendix-\ref{app-inclusion-yd-antisym-rules}.
However, since we are dealing with typical states of an ensemble which we claim
that it has a well defined effective gravity description
\cite{Balasubramanian:2005mg}, the leading order of $\a$ should have a definite
value irrespective of the typical state we pick from the ensemble. This is
precisely what we find. Actually we find more. If we try to move a little bit
out of typicality, the value of $\a$ changes in the case of generic probes. This
can be observed in the discussion of appendix-\ref{app-inclusion-yd-antisym-rules}.
There, when we discussed the possible modifications to $\a$ when we change the
background YD $\calo$, we concluded that as far as the number of corners of the
background YD is of order $N$, $\a$ remains intact. This was only satisfied by
typical states. We can actually construct YDs $\calo$ that are very close to
typicality but with a different value of $\a$. These are YDs $\overline{\calo}$
where the rows are grouped into $N^{1-a}$ sets of rows of equal length, where each
set has $N^a$ rows, or YDs that are close enough to these. It is clear from the discussion in
appendix-\ref{app-inclusion-yd-antisym-rules} that even for very small $a \ll
1$, we still get a different value of $\a$. The deviation is proportional to
$a$, and hence small for small $a$. So, we can talk about near typical states.

By assuming that each box in the probe YD can be thought of as a free graviton, one can intuitively understand the dependence of the two point function $\lag\lag \y\; \y \rag\rag_\calo$ on $N^h$ as follows. The only thing
that the graviton will see is the mass $N^2$ of the background. So, one expects that the
two point function of a single free graviton to be proportional to $N^2$. As a result,
once we are dealing with $h$ free gravitons, we should get something proportional to $N^{2\,h}$.
However, this result should be modified since these gravitons are not completely free as they are in a complicated bound state. Intuitively, one can think of the interaction between these gravitons as effectively reducing the number of free gravitons. Hence, one expects the two point function to take the form $N^{\a\, h}$. What about the value of the constant $\a$?

Roughly speaking, one can relate the dependence of $\a$ on the shape of the probe
$\y$ to the effects of angular momentum. 
The presence of the latter gives rise to a centrifugal force that pushes the particles away from
the center. When we are dealing with a gas of bound gravitons, there are two
competing effects: Attraction due to gravity which is proportional to its mass
$h$, and ``repulsion'' due to the centrifugal force which is proportional to the
angular momentum $\y_0$. Hence, one expects that $\a$ will depend on the ratio
between $h$ and $\y_0$. This intuition works perfectly for generic probes when
$h \ll N$ or $h \sim N$ in the superstar ensemble, equation
(\ref{two-point-funct-h-ll-N}). However, things seem to go wrong once we cross
the $h \sim N$ point, equation (\ref{two-point-funct-h-sim-gg-N-generic}). Notice that
the leading term of the two point function is not differentiable at this point.
Unfortunately, we do not have a complete understanding of this behavior. However,
we will discuss some suggestions to explain the origin of such behavior later on
in this section.

To complete the previous discussion, let us deal with the non-generic probes case.
Remember that these probes are dual to (dual-)giant gravitons. For these type of
probes, things fall nicely into place. Remember that the radius of a (dual-)giant
graviton is proportional to its angular momentum only \cite{McGreevy:2000cw,
Hashimoto:2000zp}. As a result, one expects that the leading term of the two
point function will only depend on the angular momentum $ \y_0$. This is
precisely what we find in equation (\ref{two-pt-funct-ngen}). Actually, in this
regime of $\y_0$, the leading term of the two point function is universal in
the sense that it is insensitive to the background. This universality can be
attributed to the position of the (dual-)giant which is at the outer region of
the background geometry. The latter is fixed to be AdS$_5 \times$ S$^5$ which
depends essentially on the flux of the background and not its details.

An observation that was made after equation (\ref{two-pt-funct-ngen}), and
deserves a further discussion, has to do with YDs $\y$ whose number of columns
$d$ is such that $d \gg N$ (long probes). It was observed that the leading term of the log of
two point function $\lag \lag \y \; \y \rag\rag_\calo$, given by equation
(\ref{two-pt-funct-ngen}), did not depend on $N$ explicitly. Taking into account
that this formula is valid for both regimes $\y_0 \sim N$ and $\y_0 \gg N$, the
two point function for the long probes ($\y_0 = d \gg N$) will have the same
leading order if we were dealing with backgrounds with $d$ units of flux
instead of $N$. This observation suggests that one can pretty much conjecture that
these probes have a non-trivial backreaction on the background. This could be
the case when interpreting these probes as $d$ giants in S$^5$. However, the
picture is more obscure on the dual-giants side. It could very well be the case
that since the dual-giants are close to the boundary, one should be careful
about interpreting calculations on the dual field theory, see for example
\cite{Seiberg:1999xz}. 
On the other hand, if we would have naively used very
heavy probes $h \gg N^2$, we will end up always with these kind of YDs, since
the number of rows is bounded by $N$. In this case, we will safely conclude that
one should be dealing with a different dual field theory since these probes are
heavier that the background and their backreaction should be taken into account.
This gives a nice explanation for the ``absence'' of states of conformal weight
$h \gg N^2$ in $\caln = 4$ SU$(N)$ super Yang-Mills theory, even though there
is no good reason for that. It is just that these states belong to a different
theory when using AdS/CFT duality.

Finally, and before discussing possible extensions, ameliorations of this work, let us
go back to our open question regarding the non-differentiability of the leading term of $\log
\, \lag \lag \y \; \y \rag\rag_\calo $ at the point $h \sim N$ in the case of
generic probes. As already alluded to at the end of the section \ref{two-pt-funct-full-section},
this point of non-differentiability depends highly on the background we are
probing. Let us try to look for possible origins of such non-differentiability and
discuss them. To do so, we need to remember two important points. First of all,
in principle, we can do a dual gravity calculation. Secondly, as already
mentioned in the beginning of subsection \ref{two-point-function-section},
the spacetime dependence of the two point function was factored out. The latter
is trivial and is completely fixed by demanding the conformal invariance
\cite{Corley:2001zk}. Hence, the expression (\ref{main-eqn}) of the leading
term of the two point function includes only information about the type of
probes and backgrounds. In light of these two remarks, let us examine possible
origins of non-differentiability of the two point function. We have the following
possible scenarios:
\begin{itemize}
  \item[{\emph{1.}}] The two point function is continuous but not differentiable in general.
  On the gravity side, this might be the result of adding $\d$-like sources when
doing the probe analysis. However, such an effect would manifest itself in the
spacetime dependence of the two point function which we have already factored out.
Hence, this possibility is ruled out. Another reason in support of ruling out this proposal is that, the leading term of $\log \,\lag\lag \y \; \y \rag\rag_\calo$ is continuous and differentiable except at certain points. Hence, it is more logical that there is a physical explanation of such non-differentiability rather than declaring that it is a generic property of the two point function.
  \item[{\emph{2.}}] Subleading terms of the two point function contribute in such a way to kill
  this non-differentiability. This is very plausible, since when taking the
derivative we end up reducing the order of some terms. However, trying to remedy
the non-differentiability using subleading terms in our case spoils either
continuity or the leading term expression. Hence, this option is ruled out.
  \item[{\emph{3.}}] There is a further phase transition from a gas of gravitons to either
  (dual-)giants or a new unknown state. This is a very plausible possibility.
  As we mentioned when arguing for the shape of the leading term (\ref{main-eqn})
  of the two point function, we can think of the YD as a bound state of graviton with
  certain quantium numbers (mass and angular momentum). When moving in the space of
  parameters, we are either increasing the mass or the angular momentum of the probe.
  So one expects that when the probes are heavy enough, or the angular momentum is large
  enough, that there will be a phase transition. We have already encountered a similar
  phenomenon at the point $\y_0 \sim N$, where the leading term of the two point function was
  non-differentiable. The latter was associated to the well known phase
  transition gravitons/(dual-)giant gravitons \cite{McGreevy:2000cw, Hashimoto:2000zp}.
  This is because this phase transition occurs when the angular momentum $\y_0$ is large
  enough. Notice that this phase transition occurs irrespective of the details of
  the background, in contrast to our mysterious phase transition (if this is the right explanation) at $h \sim N$. As already mentioned, the point of non-differentiability
  $h \sim N$ depends on more details than just the mass of the background.
  Tracing back our derivations, it seems that this point is tightly related to the two
  facts that the typical YDs have order $N$ corners, and the length of almost all of its
  rows is of order $N$. Due to such dependence, one can safely rule out this explanation.
  \item[{\emph{4.}}] A manifestation of the stringy exclusion principle \cite{Maldacena:1998bw, Jevicki:1998rr, Ho:1999bn}. The reason behind this proposal is as follows. Imagine that our two point function is the result of scattering some bound state of graviton off the superstar. In this process, 
  a ``bound'' state of our probe and the superstar will form in the intermediate stages. Now, due to the stringy exclusion principle, some of these intermediate bound states will not be allowed reducing the value of the two point function. This reduction can lead to a non-differentiability of the two point function when the effects of the stringy exclusion principle are not negligible. This explanation seems very plausible, however, the fact that this point of non-differentiability depends on more than just the conserved charges of the background puts this proposal on shaky grounds.
  \item[{\emph{5.}}] A tempting explanation would be to declare that this non-differentiability is, somehow,
  connected to the singularity of the background geometry. An argument in favor of such claim
  is that, in absence of a phase transition, we seem to have lost some information about our
  probes. Such a proposal, if true, gives a strong support for the fuzzball proposal. It also gives a hope that we can actually see the imprint of the singularity of the bulk geometry in the dual field theory. One can object to this explanation by advocating the different nature of the bulk singularity (spacetime) and the non-differentiability of the two point function (parameter space of the probes).   
  \item[{\emph{6.}}] The generic probes can actually see some inner-structure of the superstar geometry. Remember that the value of the leading term of the two point function differs depending on whether we probe typical or non-typical states of the superstar ensemble. If this explanation turns to be true, then we expect that by varying our probes, we can see behind the horizon of black holes. Notice that, at best, we can only distinguish between typical and non-typical states by doing so. This is very encouraging from the point of view of the fuzzball proposal, as we can have an idea about the ensemble to associate to black holes using gravity probes. However, it is kind of hard to imagine that by just probing a single geometry, we can get information about its typical states.
%
  \item[\emph{7.}] The gravity calculation in the superstar background does not reproduce this behavior of the two point function. This is a bold proposal and, if it is true, invalidates the proposal of \cite{Balasubramanian:2005mg}. Notice that this does not prove that the fuzzball proposal is completely wrong, but puts it still in a bad position. 
\end{itemize}

It is not very clear to us which of the proposals above is the right one. Our confusion stems from the fact that in the bulk we are, in principle, evaluating supergravity quantities whereas the point of non-differentiability seems to be intimately connected to some stringy phenomena in general. It is clear that a bulk calculation is required to solve this puzzle. The only issue there is that the type of probes we are dealing with, the generic probes, are a beast on the gravity side: they are a complicated bound states of a gas of gravitons. Hence, the needed dual calculations are very challenging and we seem to not have the right technology to tackle them at the moment.


We close this section by discussion some further open questions and possible
future directions of research. The first one, which was mentioned already, is to
do the dual calculation in the superstar background. Another direction of
research has to do with the simple final result (\ref{main-eqn}) that we got.
Its simplicity begs for a better approach to derive it instead of our
``tour-de-force'' approach. A third open question has to do with beyond the
leading term calculation. Unfortunately, the techniques developed here are
limited to the leading order, and things get wild when going beyond except for
long probes where one can easily calculate the first subleading term. Given that
the long probes are the worst to use, this does not seem to be such an
interesting progress. Maybe a different approach to the problem can help here.
On a slightly related note, one of the original motivations to this work was to
try to distinguish the typical states of the superstar ensemble. We found, as
expected, that we need to go beyond the leading term for that. The question is then,
at which subleading term can we start to distinguish some typical states from the
others. According to the calculations in this paper, it seems that we might have
some chance in the first subleading correction in the case of generic probes.
However, our technology that we used here comes short to this task. In the same
line of thoughts, one wonders if there is a nice expansion parameter for $\log \,
\lag\lag \y \; \y \rag\rag_\calo$? It seems that the general structure will be an
alternation of terms that contain $\log N$, and terms without this log. It does
seem also that for generic probes, $(1/\y_0)$ might be the parameter of expansion,
however we could not come up with a nice reason to why this is the case. It is
possible that this parameter will work only for the first few subleading terms since
it reflects the separation of the (dual-)giants in the background geometry, and one
would expect that at certain stage the flux of the background will kick in. This
happens for the second subleading term in the case of the long probes as the parameter
of expansion seems to be $(N/d)$. Actually, it is not even clear that there will
be a nice expansion in the first place. Hopefully, further works tackling the same
problem discussed in this paper using a better technology will help shed light on
these open questions and others.
%
%
%
\section*{Acknowledgements}
%
%
%
The author wishes to thank Vishnu Jejjala and Robert de Mello Koch for their collaboration
in the early stages of this project, and for the subsequent useful discussions.
The author is also very grateful to Robert de Mello Koch for carefully reading
the first version of this paper, as well as to Dimitrios Giataganas
for useful discussions. This project is financially supported by NITheP fellowship.
%
%
%
%
%
%

%
%
%
\appendix
%
%
%
\section{Conventions and notation}\label{conventions}
%
%
%
In this appendix we will summarize our conventions and notation that are used
throughout this paper.
\paragraph{Young diagrams (YD)} Throughout this paper we will use YD as a short
notation for a Young diagram and YDs for more than one. A YD is a collection of
boxes, arranged in left-justified rows, with the row lengths weakly decreasing
from top to bottom. For example:

\begin{align*}
 \y = \quad \yng (4,3,2,2) \quad ,
\end{align*}
\\
is a YD. We start enumerating rows from top to bottom and columns from left to
right. Our main use of YDs in this paper is through their connection with
irreducible representations of the unitary and the permutation groups.
\paragraph{The shape of a YD} is the collection of the lengths of the rows of
the YD into an $n$-tuple of integers. For example the shape of the previous YD
is $\y = (4\,,\,3\,,\,2\,,\,2)$. We will be using the same symbol to denote the
YD as well as its shape.
\paragraph{The hook length} A box in a YD is denoted by the pair $(i
\,,\, j)$ where $i$ is the number of the row and $j$ is the number of the column
whose intersection is this box. $h_{(i \,,\, j)}$  stands for the hook length
associated to the box $(i \,,\, j)$. The hook length associated to a box is the
number of boxes below and to the left of it plus one. As an example, we depict
below a YD whose boxes are filled with their associated hook lengths $h_{(i
\,,\, j)}$:

\begin{align*}
 \yng(4,3,2,2) \quad \longrightarrow \quad \young(7621,541,32,21) \quad .
\end{align*}
\\
The product of all the hook lengths plays an important role in the formulation
of the dimension of the SU$(N)$ group or the permutation group representation
given by this YD, see equations (\ref{dim-hooks}) and (\ref{dim-perm}).
\paragraph{Young tableau (YT)} Throughout this paper we will use YT as a short
notation for a Young tableau and YTx for more than one. A YT is a YD where each
box is filled with a number\footnote{This is not the only possibility as we can
use letters for example to label the boxes. However, we will be mainly
interested in labels that are strictly positive integers}. The collection of
these number is called a {\bf filling}, and will be denoted by a Greek letter
e.g. $\b$. We will use the word {\bf labeling} of the YD to refer to all
possible fillings $\b$. It is more convenient to denote a filling $\b$ by $ \b =
(\b_1 \,,\, \b_2 \,,\, \ldots \, \b_m)$, where $\b_i$ denotes the number of
times the integer $i$ appears. We denote by $|\b|$ the {\bf length} of the
filling $\b$ which is the number of its non-zero entries $\b_i \neq 0$. As an
example, the following YTx:

\begin{align*}
 \young(1122,244,45,57) \quad, \quad \young(5722,444,25,11) \quad, \quad
\young(5725,142,22,44) \quad , \quad \dots
\end{align*}
\\
correspond to the YD of the previous example with the filling $ \b = (2 \,,\, 3
\,,\, 0 \;,\, 3 \,,\, 2 \,,\, 0 \,,\, 1)$. The length of this filling is $|\b|
= 5$.
\paragraph{Standard Young tableau (SYT)} Throughout this paper we will use SYT
as a short notation for a standard Young tableau, and SYTx for more than one.
Let $h$ be the number of boxes of a YD under consideration. A STY is a YD whose
boxes are filled with numbers $1 \,,\, 2\,,\, \ldots h$, each one occurring once
such that, the numbers on the same row are strictly increasing from left to
right, and the numbers on the same column are strictly increasing from top to
bottom. The following YTx depict some examples of SYT.

\begin{align*}
\young(1234,567,89) \quad , \quad \young(1345,269,78)
\quad , \quad \young(1497,258,36) \quad , \quad \ldots
\end{align*}
\\
The number of all SYTx associated to a YD $\y$ gives the dimension of the
representation of the permutation group $S_h$ given by this YD $\y$, see for
example \cite{Fulton}.
\paragraph{Semi-standard Young tableau (SSYT)} Throughout this paper we will use
SSYT as a short notation for a semi-standard Young tableau and SSYTx for more
than one. A SSYT is a YD whose
boxes are filled with positive integers such that the numbers on the same row
are weakly increasing from left to right and the numbers on the same column are
strictly increasing from top to bottom. The following YTx are examples of SSYT.

\begin{align*}
\young(1122,244,45,57) \quad , \quad \young(1127,224,44,55) \quad , \quad
\young(1125,224,44,57)  \quad \ldots
\end{align*}
\\
The number of all possible SSYTx associated to a YD $\y$, whose labels are
integers running from $1$ to $N$ is the dimension of the representation of the
group SU$(N)$ given by this YD $\y$, see for example \cite{Fulton}.
\paragraph{Kostka numbers} The Kostka number $K_{\y \,,\, \b}$ is the number of
SSYTx associated to the YD of shape $\y$ and the filling $\b$.
\paragraph{Background YD conventions} In this paper we will use the letter
$\calo$ to denote the background under consideration, be it a state, a
geometry, or a dual YD. We will denote by $\calo_i$ the length of the
$i^{\text{th}}$ row of the YD $\calo$. We will reserve the notation $\calo_0$ to
the limit shape YD of the ensemble under consideration. The energy/conformal
dimension/number of boxes of the background will be denoted by $\D$, whereas the
number of its columns will be denoted by $D$. For characteristics of these
backgrounds see section \ref{superstar-ensemble}.
\paragraph{Probe YD conventions} In this paper we will use the letter $\y$ to
denote the probe, be it a state, or a dual YD. Its energy/conformal
dimension/number of boxes will be denoted by $h$. $\y_i$ will denote the length
of its $i^{\text{th}}$ row. We will also reserve the letter $d$ to denote
the number of its columns ($d=\y_1$), the letter $n$ to denote the number of its
rows, $\y_0 = \tmax \, \{n \,,\, d\}$, and the letter $\calh$ to denote the
product over all its hook lengths. See section \ref{probes-section} for
characteristics of the probe YDs that will be used in this paper.
\paragraph{The dimension of a YD} Two dimensions associated to the probe YD $\y$
will play an important role in this paper. $\tdim_N \, \y$ will denote the
dimension of $\y$ seen as an irreducible representation of the group SU$(N)$,
whereas $\tdim_h \, \y$ stands for its dimension seen as an irreducible
representation of the permutation group $S_h$.
\paragraph{Leading behavior vs leading term} Two notions that will be used
extensively in the paper: leading behavior and leading term. By the leading
behavior of a quantity $A$ we mean its scaling behavior with $N$. We use the
symbol ``$\sim$'' to express the leading behavior. On the other hand, the
leading term of a quantity $A$ means the dominant term in an expansion of $A$
in the large $N$ limit. We will use the symbol ``$\approx$'' to express the
leading term. For example:
$$ A = \sum_{k=1}^N k^2 = \frac{1}{6} \, N\, (N+1) \, (2\, N+1) \;, \quad A \sim
N^3 \;, \quad A \approx \frac{1}{3} \, N^3 \;. $$
%
%
%
%
\section{A collection of some useful formulas} \label{useful-form}
%
%
%
Most of the results derived in this paper rely on formulas that will be
discussed in this appendix. The first kind of formulas has to do with
approximating sums of the form:
$$ \cals_\ell (h\,,\,a\,;\,n) = \sum_{k=0}^n (h\,k + a)^\ell \, \log (h\,k + a)
\;, $$
for $n \gg 1$, and $\ell \leq 2$, which will be the topic of the first section
below. The second formula has to do with counting the possible ways to partition
a positive integer $n$ into $m$ positive integers $n_i$; $i=1\,,\, 2\,,\, \ldots
m$, where the order of the index $i$ is important. We will be more precise about
what we are really counting in the second section below. In the last section we
will discuss some useful properties of polylogarithm function $\Li_n (x)$ that
will be used in appendix-\ref{app-thermo-dyn}.
%
%
\subsection{Sum approximation} \label{sums-approx}
%
%
In the following we will discuss an approximation to sums of the form:
\begin{equation}
 \cals_\ell (h\,,\,a\,;\,n) = \sum_{k=0}^n (h\,k + a)^\ell \, \log (h\,k + a)
\;,
\label{the-master-sum-def}
\end{equation}
for $n \gg 1$. There are different approaches to deal with this kind of sums.
We will use the Euler-Maclaurin approximation which is valid for all smooth real
functions $f (x)$. We have (see \cite{math-funct} for example):
\begin{equation}
 \sum_{k=0}^n f(h\,k + a) = \frac{1}{h} \, \int_a^b f(x)\, dx +
\frac{f(a)+f(b)}{2} + \sum_{k=1}^p \frac{h^{2\,k-1}}{(2\,k)!} \,B_{2\,k}\,
\left( f^{(2\, k-1)} (b) - f^{(2\, k -1)} (a) \right) + \calr_h \;,
\label{EM-approx-gen}
\end{equation}
where $b=a+ h\, n$, and $\calr_h$ is the remainder and is given by:
\begin{equation*}
 \calr_h = \frac{h^{2\,p+2}}{(2\, p+2)!} \, B_{2\, p+2} \,\sum_{k=0}^{n-1}
f^{(2p+2)} (a+ h\,k + \q)
\;,
\end{equation*}
where $\q$ is some constant between $0$ and $1$ i.e. $0<\q<1$. It can also be
shown that $\calr$ is bounded by:
\begin{equation*}
 |\calr| \leq \frac{2\, \z (2\, p)}{(2\, \p)^{2\, p}} \, \int_1^n |f^{(2\, p)}
(x)| \, dx \;,
\end{equation*}
where $\z (t)$ is the Riemann zeta function. The constants $B_n$ that appear in
the relation (\ref{EM-approx-gen}) are the Bernoulli numbers with $B_1 =1/2$.
Their generating function is given by:
\begin{equation*}
 \sum_{m=0}^\infty B_m \, \frac{t^m}{m!} = \frac{t}{1-e^{-t}} := \calb (t) \;.
\end{equation*}
It is clear from the generating function above that if $n \geq 1$ then $B_{2\,
n+1} = 0$. The easiest way to see this is by using the observation:
$$ \calb (t) - \calb (-t) = t \;. $$

To apply the Euler-Maclaurin formula to our sum (\ref{the-master-sum-def}), we
need to calculate the integral as well as the derivatives of the function $ f(x)
= x^\ell \, \log x$. We have:
\begin{align*}
 \int f(x)\, dx &= \frac{1}{\ell+1} \, x^{\ell+1} \, \left(\log x -
\frac{1}{\ell+1} \right)\;, \\
 f^{(i)} (x) &= \frac{\ell!}{(\ell-i)!} \, x^{\ell-i} \, \left(\log x +
\sum_{j=1}^i \left[ \frac{1}{\ell+1-j}  \right]\right) \;; \qquad \text{for}
\quad i < \ell \;,\\
 f^{(m)} (x) &= \ell! \, \log x + \ldots
\end{align*}
The last ingredient we need is the values of the first few Bernoulli numbers.
We have:
$$ B_0 = 1 \;, \quad B_1 = \frac{1}{2} \;, \quad B_2 = \frac{1}{6} \;, \quad
B_4 = - \frac{1}{30} \;.$$
Putting everything together we get the following approximations to the first
few sums:
\begin{align}
 \cals_0 (h\,,\,a\,;\,n) &\simeq  \frac{1}{h} \, \left[\left( b +\frac{h}{2}
\right) \, \log b - \left( a -\frac{h}{2} \right) \, \log a \right] - n \;,
\label{gen-sum-m-0}\\
 \cals_1 (h\,,\,a\,;\, n) &\simeq \frac{1}{2\, h} \, \left[ \left( b\, (b+h)
+ \frac{h^2}{6} \right) \, \log b - \left( a\, (a-h) + \frac{h^2}{6} \right) \,
\log a \right] - \frac{1}{4} \, n \, (b+a) \;, \label{gen-sum-m-1} \\
 \cals_2 (h\,,\,a\,;\,n) &\simeq \frac{1}{3 \, h} \, \left[ b \, (b+h) \,
\left( b+ \frac{h}{2} \right) \, \log b - a\, (a-h) \, \left( a- \frac{h}{2}
\right) \, \log a \right] - \nonumber \\
             & \qquad  \qquad - \frac{1}{9} \, n \, (a^2 + b^2 + a\,b) +
\frac{h^2}{12} \, n \;, \label{gen-sum-m-2}
\end{align}
where we truncated the sums to the log order which is enough for our purposes.

In the remaining of this subsection, we will use the approximations above to
evaluate a double sum that will appear frequently in appendix-\ref{sample-YD}.
This double sum is given by:
\begin{equation}
 \log \, \calh_0 = \sum_{i=0}^{n-1} \sum_{j=0}^{d-1} \log (\caln +1 + i + j)
\;, \label{sym-hook}
\end{equation}
where $\caln$ is some positive integer. The idea is to use the special
structure of the summand to change the double sum over $i$ and $j$ to a single
one over $k = i+j$. For that we need to introduce the following quantities:
  $$ a = \tmin\, \{n \,,\, d\} \quad , \qquad b = \tmax \, \{n \,,\, d\} \;. $$
Notice that $a+b = n+d$. It is easy to see that one gets:
\begin{align}
 \log \, \calh_0 &= \sum_{k=1}^{a-1} k \, \log (\caln + k) + a \, \sum_{k=a}^b
\log (\caln+ k) + \sum_{k=b+1}^{a+b -1} (a+b - k) \, \log (\caln + k) \nonumber
\\
                 &= H_\caln (n+d) - H_\caln (n) - H_\caln (d) \;,
\label{h-h-funct}
\end{align}
where $H_\b (\a)$ is a short notation for the sum:
\begin{equation}
 H_\b (\a) = \sum_{k=1}^{\a-1} (\a-k) \, \log (\b + k) \;, \label{H-funct}
\end{equation}
It is a straightforward exercise to get an approximate expression for $H_\b
(\a)$ using equations (\ref{gen-sum-m-0}) and (\ref{gen-sum-m-1}). We get:
\begin{align*}
 H_\b (\a) &= (\a + \b) \, \sum_{k=0}^{\a-2} \, \log (\b+1 +k) -
\sum_{k=0}^{\a-2} (\b+1+k) \, \log (\b+1+k) \nonumber \\
           &\simeq \frac{1}{2} \, \left[ (\a+\b)^2 - \frac{1}{6} \right] \,
\log (\a+\b) - \frac{1}{2} \, \left[\b^2 + \a\, (2\, \b + 1) - \frac{1}{6}
\right] \, \log \b - \frac{1}{4} \, \a \, (2\, \b+ 3\, \a) \;.
\end{align*}
Plugging in this result into the expression (\ref{h-h-funct}) of $\log \calh_0$
above, we find:
\begin{align}
 \log \, \calh_0 &\simeq \frac{1}{2} \, \left(\caln + n + d - \frac{1}{6}
\right) \, \log (\caln + n + d) + \frac{1}{2} \, \left(\caln - \frac{1}{6}
\right) \, \log \caln - \nonumber \\
                 &\quad - \frac{1}{2} \, \left(\caln + n - \frac{1}{6} \right)
\, \log (\caln + n) - \frac{1}{2} \, \left(\caln + d - \frac{1}{6} \right) \,
\log (\caln + d) - \frac{3}{2} \, n \, d \;. \label{sym-hook-approx}
\end{align}
%
%
\subsection{A counting formula} \label{counting-formula}
%
%
We will need at different places in this paper to count the possible ways to
partition a positive integer $n$ into $m$ positive integers $n_i$; $i=1\,,\,
2\,,\, \ldots m$. More precisely we are interested in counting the $m$-tuples
$(n_1 \,,\, n_2 \,,\, \ldots \,,\, n_m)$ such that:
$$ \forall \; i \;; \quad n_i \geq 0 \;, \qquad \sum_{i=1}^m n_i = n \;. $$
Notice that this is different, and much easier, than the usual notion of
partition of integers. In the latter case the order of the index $i$ is not
important i.e. we are counting sets of integers instead of $m$-tuples. On
top of that, in the usual counting of partitions, we require the integers $n_i$
to be strictly positive.

The answer to our counting question is easy. It is the same as counting the
number of ways to distribute $n$ balls on $m$ boxes. The answer to the
latter question is given by:
\begin{equation}
 \calp_m (n) = C_{n+m-1}^n \;, \label{counting-form}
\end{equation}
where $C_a^b$ is the binomial coefficient:
$$ C^a_b = \frac{b!}{a! \, (b-a)!} \;.$$
%
%
\subsection{The polylogarithm function} \label{app-polylog}
%
%
In this section, we will summarize some of the important properties of the
polylogarithm functions $\Li_n (z)$, that will be of importance to us in
appendix-\ref{app-thermo-dyn}. Since, in appendix-\ref{app-thermo-dyn}, we will
be dealing with quantities that are of the form $e^{-\b \, x - \l \, y}$, where
$\b$, $x$, $\l$ and $y$ are all positive, we restrict the argument $z$ of the
function $\Li_n (z)$ to lay in the interval $0 < z < 1$. We have by
definition:
\begin{equation}
 \Li_n (z) = \sum_{k=1}^\infty \frac{z^k}{k^n} \; ; \quad \forall \; n>0\;.
\label{polylog-def-n-positive}
\end{equation}
It is easy to see, using the definition above that:
\begin{align}
 \int_0^z \frac{\Li_n (y)}{y} \; dy &= \Li_{n+1} (z) \;; \quad \text{for} \quad
0 < z <1 \;,
\label{polylog-integ-property} \\
 z \, \pt_z \, \Li_n (z) &= \Li_{n-1} (z) \;. \label{polylog-diff-property}
\end{align}
One can use the second property to define polylog for negative $n$. We find:
\begin{equation}
 \Li_{(-n)} (z) = \left( z\, \pt_z \right)^n \, \frac{z}{1-z} \;; \quad \forall
\; n \geq 0\;, \label{polylog-neg-def}
\end{equation}
where the case $n=0$ corresponds to the function that the operator$(z\,
\pt_z)^n$ acts on.

The last property we would like to discuss is the Taylor expansion of the
polylogarithm function whose argument takes the special form $z = z_0 \, e^\a$,
where $\a \ll 1$. It can be easily proved, using the defining equation of the
polylogarithm (\ref{polylog-def-n-positive}) that:
\begin{align}
 \Li_n \left(z_0 \, e^\a \right) - \Li_n (z_0) \simeq \sum_{k=1}^{n}
\frac{\a^k}{k!} \, \Li_{n-k} (z_0) + \ldots \;. \label{polylog-expan}
\end{align}
This expansion is more than enough for our purposes even though we truncated
the sum to $n$.
%
%
%
\section{Thermodynamics of the superstar ensemble} \label{app-thermo-dyn}
%
%
In this appendix, we will evaluate some thermodynamical quantities of our
superstar ensemble of background YDs defined in subsection
\ref{superstar-ensemble}. Our first step in our study, is to evaluate the log of the partition
function $\calz$ defined in equation (\ref{part-funct}), using the approximation
(\ref{EM-approx-gen}). We will check the validity of our approximation at the
end. After that, we will move on to fix the values of the parameters $\b$ and
$\l$. All the expressions derived below can be found in
\cite{Balasubramanian:2005mg} with appropriate modifications.

Our starting point is the defining equation of $\calz$, equation
(\ref{part-funct}), that we rewrite here for convenience.
\begin{equation}
 \log \calz = - \sum_{j=1}^N \log (1 - p\, q^j) = \sum_{j=1}^N  \Li_1 \left(p \,
q^j \right) \;, \label{app-part-funct}
\end{equation}
where $\Li_n (x)$ stands for the polylog function defined in
(\ref{polylog-def-n-positive}). To evaluate the sum that appears in the
expression of $\log \calz$ above using the approximation
(\ref{EM-approx-gen}), we need
to take $h=1$, $a = 1$, $b=N$, and $f(x) = \Li_1 (p\, q^x)$. We have:
\begin{align*}
 \int_1^N \Li_1 \left(p \, q^x \right) \, dx= \frac{1}{\log q} \,
\int_{p\,q}^{p\, q^N} \frac{Li_1 (y)}{y} \; dy = \frac{1}{\log q} \;
\left[\Li_2 \left( p\, q^N \right) - \Li_2 (p \, q) \right]\;, 
\end{align*}
where we used the change of variables $y = p\, q^x$, then the property
(\ref{polylog-integ-property}) of the polylog functions since both $q$ and $p$
are positive and less than one, and $N$ is a positive integer. Next, we need to
evaluate the derivatives of $f(x)$. One can prove using the property
(\ref{polylog-diff-property}) and the definition (\ref{polylog-neg-def}) that:
\begin{align*}
 f^{(n)} (x) = (\log q)^n \, \Li_{(1-n)} \, \left( p \, q^x \right) \;.
\end{align*}
The proof proceeds by using once again the variable $y = p\, q^x$, then the
chain rule to change the derivative with respect to $x$ to:
$$ \pt_x = (\log q) \, (y\, \pt_y) \;. $$
Collecting everything, we find the following approximate expression for $\log
\calz$:
\begin{align*}
 \log \calz \simeq & \frac{1}{\log q} \, \left[ \Li_2 \left( p\, q^N \right) -
\Li_2 (p \, q) \right] + \frac{1}{2} \, \left[ \Li_1 \left( p\, q^N \right) +
\Li_1 (p \, q) \right] \nonumber \\
                    & \quad + \sum_{k=1}^\infty \frac{B_{2\,k}}{(2\, k)!}
\, (\log q)^{(2\, k -1)} \, \left[ \Li_{(2-2\,k)} \left(p \,q^N \right) -
\Li_{(2-2\,k)} \left(p \,q \right) \right] \;, 
\end{align*}
Although we managed to evaluate $\log \, \calz$, we need to check the validity
of our approximation. Luckily enough, our $\b = - \log q$ is very small
\cite[section 3.3]{Balasubramanian:2005mg}. In this regime, we can safely neglect
the sum over $k$ in equation above, to get the more manageable approximation:
\begin{align}
 \log \calz \simeq \frac{1}{\log q} \, \left[ \Li_2 \left( p\, q^N \right)
- \Li_2 (p\, q) \right] + \frac{1}{2} \, \left[ \Li_1 \left( p\, q^N \right) +
\Li_1 (p \, q) \right] \;, \label{part-funct-approx-2}
\end{align}
To evaluate $\D$, we plug this expression in the relation
(\ref{beta-lambda-const-1}) to get:
\begin{align}
 \D \simeq& \frac{1}{(\log q)^2} \, \left[ \Li_2 (p) - \Li_2 \left(p\, q^N
\right)\right] - \frac{1}{(\log q)} \, \left[\Li_1 (p) - N \,\Li_1 \left(p\,
q^N \right)\right] + \frac{N}{2} \, \Li_0 \left( p\, q^N \right) \;,
\label{app-boxes-expression}
\end{align}
where we used the property (\ref{polylog-diff-property}). To make connection
with a similar expression that appears in \cite{Balasubramanian:2005mg}, we
need to use the fact that $\Li_1 (x) = - \log (1-x)$ and the identity:
$$ \Li_2 (y) + \Li_2 (1-y) = \frac{\p^2}{6} - \log y \, \log (1-y) \;. $$

The expression (\ref{part-funct-approx-2}) above together with the relation
(\ref{beta-lambda-const-2}), gives for $D$ the following
expression\footnote{It turns out that keeping only the leading term is
enough for our purposes.}:
\begin{equation}
 D \simeq \frac{1}{\log q} \, \left[\log (1-p) - \log \left( 1- p\, q^N
\right)\right] \;. \label{app-columns-expression-1}
\end{equation}
Observe that we can arrive at the same expressions as
(\ref{app-boxes-expression}) and (\ref{app-columns-expression-1}) by starting
from (\ref{beta-lambda-const-1}) and (\ref{beta-lambda-const-2}), then
performing the sum following the same steps we did for $\log \calz$. The
value of $p$ can be already fixed using the constraint
(\ref{app-columns-expression-1}). We
find:
\begin{equation}
  p = \frac{1-q^D}{1-q^{D+N}} \;. \label{app-p-value}
\end{equation}

We need also a way to fix $\b$, however the expression
(\ref{app-boxes-expression}) is complicated,even after substituting $p$ by its
value, and the solution depends on the regime of $\b$. We will take a
slight detour here to have an idea about the regime of $\b$, then we will come
back to its value later on. The idea is to use the constraint on the limit
shape YD of this ensemble \cite{Balasubramanian:2005mg}. Remember that we want
the limit shape curve to be a line. This is because the superstar maximized the
entropy. The latter is associated to moving around the outer boxes of the YDs.
So, naively one would expect that the typical YDs of these ensemble should be
closer to a triangular YD \cite{Balasubramanian:2005mg}.

Following \cite{Balasubramanian:2005mg}, we start by the expression\footnote{We
have a different expression for $y(x)$ than the one in
\cite{Balasubramanian:2005mg} since we have different conventions.}:
$$ y(x) = \sum_{i=x}^N \lag c_i \rag \;, $$
where $x$ is the coordinate along the rows that increases from the top to the
bottom of YDs, and $y(x)$ is the length of the row $x$. Using the expression
for $\lag c_i \rag$ that appears in (\ref{beta-lambda-const-2}), we find the
following equation for the limit shape:
\begin{equation*}
 \left(1- p \, q^N\right) \, q^y + p \, q^x = 1 \;.
\end{equation*}
One can easily check that the box $(N,0)$ is part of this limit shape curve.
Requiring, on the other hand, that $(0,D)$ is also part of this limit shape curve
leads to the same value of $p$ as before, equation (\ref{app-p-value}).
Plugging this value into the expression above, leads to the following limit
shape curve equation:
\begin{equation}
 \left(1 - q^N \right) \, q^y + \left( 1 - q^D\right) \, q^x = 1 - q^{D+N} \;.
\label{app-limit-shape-eqn-1}
\end{equation}
To get a straight line out of this equation, we need to bring down $x$ and $y$
from the exponent. Since both $x$ and $y$ are of order $N$, we need $(\b \, N)
\ll 1$ i.e. $\b \ll (1/N)$. If this is satisfied, one gets for $p$ the
following approximate value:
\begin{equation}
  p \approx \frac{D}{D+M} \;. \label{app-p-value-app}
\end{equation}
The limit shape curve becomes in this case:
\begin{equation}
  y \approx D \, \left( 1 - \frac{x}{N} \right) \;, \label{app-limit-shape-2}
\end{equation}
which is a straight line.

To fix completely $\b$, one can try to use the fact that $(\b \, N \ll 1)$,
expand the expression (\ref{app-boxes-expression}) appropriately using equation
(\ref{polylog-expan}) and equate $\D$ to $(N \, D)/2$. However, one gets exactly
the latter equality independently of $\b$. One could have anticipated this
since the limit shape YD is a triangle and $\lag \D \rag$ should give to
leading order the same number of boxes as in this YD, which is precisely $(N \,
D)/2$, a value that is independent of $\b$.

Our last card is to use the entropy $S$ of this ensemble
\cite{Balasubramanian:2005mg}. We have:
\begin{align*}
 S = \b \, \D + \l \, D + \log \calz \approx - N \, \log (1-p) - D \, \log p +
\calo (\log^2 q) \sim N \;.
\end{align*}
Notice that there is no linear term in $\b = \log q$, so the superstar ensemble
extremizes the entropy. It can be shown easily by calculating the coefficient
in front of $\log^2 q$, that the superstar ensemble maximizes the entropy. Using
the relation between the inverse
temperature $\b$, the energy $\D$ and the entropy $S$, we find:
\begin{equation}
 \b = \frac{\pt S}{\pt \D} \sim \frac{1}{N} \;, \label{beta-scaling-N}
\end{equation}
where we used that $\D \sim N^2$. Finally, we should remember that in the
expressions above we set the Plank constant to one $\hbar =1$
\cite{Balasubramanian:2005mg}. Once we restore it, we find $N \, \log q \ll 1$
even though the scaling of $\log q$ with $N$ is $\log q \sim 1/N$.
%
%
%
\section{A non-trivial simple family of Young diagrams} \label{sample-YD}
%
%
%
In the following, we are going to evaluate the dimension of a family of YDs that
we call from now on the optimum family. This family is generic enough for our
purposes and, at the same time, simple enough to allow for exact evaluation of
the leading term of the log of the dimensions of its YDs. In contrast to the
general discussion about the dimension of different probe YDs in section
\ref{probes-section}, where equation (\ref{dim-step}) was mainly used, we will
instead be using equation (\ref{dim-hooks}) in this appendix. We summarized the
results at the end of this appendix for the convenience of the reader. This
allows for an easier comparison with the general discussion in section
\ref{probes-section}.

Let us first start by introducing this family of YDs. The rows of a YD in this
family are grouped into groups of $n_0$ rows of equal length. Two consecutive
groups will have a constant shift in length given by $d_0$. The number of sets
of equal length rows will be $m \leq [N/n_0]$ and the last rows will be of
length $d_0$. The total number of boxes is:
\begin{equation}
 h = \frac{1}{2} \, n_0 \, d_0\, m\, (m+1) \;. \label{numb-boxes-opt-family}
\end{equation}
Let us denote by $0 \leq k \leq (m-1)$, the ``group'' of rows number, and by $ 1 \leq a
\leq n_0$ the row number in each group, then the row length takes the form:
  $$ \y_{k\, n_0 + a} = d_0 \, (m - k) \;; \qquad \text{if} \quad 1 \leq k \leq
m -1 \;, $$
and vanishes for $k \geq m$. To make connection with our conventions in
appendix-\ref{conventions}, notice that the number of rows of a YD in this
family is $n = n_0 \, m$, and the number of columns is $d=d_0 \, m$. Notice also
that the YDs in this family are homogeneous as can be checked easily, even though
not all the lengths of columns/rows scale in the same way with $N$.

We will start our discussion by looking for an approximate expression of the
log of the numerator of equation (\ref{dim-hooks}) for this family, in the case
where $N$, $n_0$, $d_0$, are very large. From now on, we will call this numerator
$ \tnum \, \y_{opt} $. Before that, let us  first fix the parametrization of the
position of a box in a YD of this family. Since the YD is divided into $m$
blocks of $n_0$ rows and $m$ blocks of $d_0$ columns, one will need four
integers $i$, $j$, $s$, and $r$ where $0 \leq i \leq m-1$, $0\leq j \leq m-1-i$,
$ 1 \leq s \leq n_0$, and $1 \leq r \leq d_0$ to parametrize a position of a
box. The latter will be characterized by its row number $( i\, n_0 + s)$, and its
column number $(j\, d_0 + r)$. We will denote such information by $(i\,,\,j \,;\,s
\,,\,r )$. Before continuing, let us also introduce a set of short notations:
\begin{align}
 \tm = n_0 + d_0 \;, \qquad \ty =& m \, \tm = d + n \;, \qquad d = d_0\, m\;,
\qquad n = n_0\, m \;, \nonumber \\
\tN = N- n & \;, \qquad\qquad \on = N + d = \tN + m \, \tm \;.
\end{align}

We will subdivide this family of YDs into two subfamilies corresponding to the
case $m \sim N^0$ and the case $m \sim N^\m$, $\m >0$ and we will discuss them
independently. We will call the first subfamily $opt_0$ and the second subfamily
$opt_\m$. Before specializing to the two subfamilies, we will approximate
$\log \, \tnum \, \y_{opt}$ and $\log \, \calh_{opt}$ by carrying out all the
sums involved except for the sum that depends on $m$.

As declared above, we start by evaluating $\log \, \tnum\, \y_{opt}$. Using the
parametrization of the position of boxes in $\y_{opt}$ we get:
\begin{align}
 \log \, \tnum \, \y_{opt} &= \sum_{i=0}^{m-1} \sum_{j=0}^{m-1-i}
\sum_{s=1}^{n_0} \sum_{r'=1}^{d_0} \log (N + j\, d_0 + r' - i \, m - s )
\nonumber\\
                           &= \sum_{k=0}^{m-1} \left[ \sum_{t=0}^{n_0-1}
\;\sum_{u=0}^{(k+1)\,d_0-1} \log (\tN + k\,n_0 +t+u+1) \right] \nonumber
\end{align}
where we introduced the new variables $k = m-1-i$, $t=n_0-s$, $r = r'-1$, and $u
= d_0\, j + r $. Notice that the sum inside the two brackets is of the form of
the sum (\ref{sym-hook}) whose approximate expression is given by
(\ref{sym-hook-approx}). Using this observation, we get:
\begin{align}
 \log \, \tnum \, \y_{opt} &\approx - \frac{3}{2} \, h + \frac{1}{2}
\,\sum_{k=0}^{m-1} \left(\left[ (\tN + (k+1)\, \tm )^2 - \frac{1}{6} \right]
\, \log (\tN + (k+1)\, \tm ) \right. \nonumber\\
                           & \qquad \qquad \quad \left. - \left[ (\tN + (k+1)\,
\tm - n_0 )^2 - \frac{1}{6} \right] \, \log (\tN + (k+1)\, \tm - n_0) \right)
\nonumber\\
                           & \quad + \frac{1}{2} \, \left( \left[ \tN^2 -
\frac{1}{6} \right] \, \log \tN  -  \left[ N^2 - \frac{1}{6} \right] \, \log N
\right) \;. \label{opt-family-num-nosum-n}
\end{align}

Next, we need to evaluate $\log \, \calh_{opt}$. Once gain our starting point is
the parametrization of the position of boxes in a YD of this family. It is easy
to see that the hook length associated to a box $(i\,,\,j\,;\,s\,,r)$ is given
by:
$$ h_{(i\,,\,j\,;\,s\,,\,r)} = [m -(i+j)] \, \tm + 1 - (r+s) \;,$$
where $1 \leq s \leq n_0$, $1 \leq r \leq d_0$, and $i+j \leq m-1$. So we need
to evaluate:
\begin{align}
 \log \, \calh_{opt} &= \sum_{i=0}^{m-1} \sum_{j=0}^{m-1-i} \sum_{s=1}^{n_0}
\sum_{r=1}^{d_0} \log ([m-(i+j)] \, \tm + 1 - (r+s)) \nonumber\\
                     &= \sum_{k=0}^{m-1} \sum_{u=0}^{n_0-1} \sum_{v=0}^{d_0-1}
(n_0-k) \, \log (k\, \tm + 1 + u+v) \nonumber\\
                     &\approx - \frac{3}{2} \, h + \frac{1}{2} \,
\sum_{k=0}^{m-1} (m-k) \left\{ \left[(k+1)^2 \, \tm^2 - \frac{1}{6}
\right] \, \log [(k+1) \, \tm] \right. \nonumber\\
                     & \left. - \left[ (k\, \tm + n_0)^2 - \frac{1}{6} \right]
\, \log (k\, \tm + n_0) - \left[ (k\, \tm + d_0)^2 - \frac{1}{6} \right] \, \log
(k\, \tm + d_0) \right\} \nonumber\\
& \; + \frac{1}{2} \, \sum_{k=1}^{m-1} (m-k) \, \left[k^2 \, \tm^2
- \frac{1}{6} \right] \, \log (k \, \tm)\;, \label{opt-family-hook-nosum-n}
\end{align}
where we introduced the new variables $k = m-1-(i+j)$, $u=n_0-s$, and $v=d_0-r$
to move from the first to the second line. The evaluation of the sum over $u$
and $v$ follows the same steps as in evaluating $\calh_0$ defined in equation
(\ref{sym-hook}) whose approximate expression is given by equation
(\ref{sym-hook-approx}).

We have two options in evaluating the leftover sum over $k$ in the equations
(\ref{opt-family-num-nosum-n}) and (\ref{opt-family-hook-nosum-n}). We can
either look for an approximate expression of the sum (\ref{the-master-sum-def})
for $\ell \leq 3$, perform the sum over $k$, then expand the final result
depending on the regime of the parameters $n_0$, $d_0$, and $m$. Or, we can
first expand the summands in equations (\ref{opt-family-num-nosum-n}) and
(\ref{opt-family-hook-nosum-n}), then perform the sum over $k$ using, at worst,
the approximations (\ref{gen-sum-m-0}), (\ref{gen-sum-m-1}) and
(\ref{gen-sum-m-2}). We find it easier to follow the second option and that is
what we are going to do in the following for each subfamily. Another reason to
follow such route is that we want to make our $opt$ family of YDs as large as
possible. In doing so, the number $m$ can be finite and small and hence the
techniques of section \ref{sums-approx} of appendix-\ref{useful-form} will not
be applicable. Before going ahead with our task, let us introduce some
notations for the sake of unifying the discussion below. We will assume the
following leading behavior of $n_0$, $m$ and $d_0$:
\begin{align}
 n_0 \approx \bn\, N^{\n} \;, \quad & m \approx \bm \, N^{\m} \;, \quad d_0
\approx \bd \, N^{\d}\;, \nonumber \\
 n \approx \bn\, \bm \, N^{\m+\n} \;, \quad & d \approx \bm \, \bd \, N^{\m+\d}
\;, \quad h \approx \bh \, N^{2\, \m + \n +\d} \;,
\end{align}
where $\bn$, $\bm$, $\bd$ are $N$-independent and $\bh = (1/2) \, \bn\, \bd\,
\bm \, (\bm+1)$. Remember that we are interested in the regime of parameters:
\begin{equation}
 2\, \m + \n +\d < 2 \;, \quad \qquad \n+\m \leq 1 \;. \label{param-regime}
\end{equation}
%
\subsection{The $opt_0$ family}
%
This case corresponds to $\m=0$. In the present case our regime of
parameters (\ref{param-regime}) simplifies to:
$$ \n+\d<2 \;, \qquad \n \leq 1 \;,$$
and since $N$ enters in the expression (\ref{opt-family-num-nosum-n}), we
distinguish the following cases:
\begin{itemize}
   \item \underline{$1<\d<2$:} This implies that $\n<1$, and hence $\tm \approx
d_0$. We get:
\begin{align}
  \log \, \tnum \, \y_{opt} &\approx \d\, h\, \log N - h + \frac{2\,
h}{m\,(m+1)} \, \left( \sum_{k=1}^m k\, \log (k\, \bd) \right) + m\,\bn\,
N^{1+\n} \, \log \frac{d_0}{N}\;, \nonumber \\
  \log \, \calh_{opt} \approx& \d\, h\, \log N - h + \frac{2 \,h}{m\,(m+1)} \,
\left( \sum_{k=1}^m k\, \log (k\, \bd) \right)+ \frac{1}{2} \, (\d-\n) \,
N^{2\, \n} \, \log N \;, \nonumber \\
  \log \, \tdim_N \, \y_{opt} &\approx (\d-1)\, m \,\bn \, N^{1+\n} \, \log
N \approx n \, N \, \log \frac{d}{N}\;.
\end{align}
   \item \underline{$\d=1$:} Once again we have $\n <1$ and hence $\tm \approx
d_0$. We find:
   \begin{align}
     \log\, \tnum\, \y_{opt} &\approx h\, \log N - h + \frac{2\,h}{\bd\, m\,
(m+1)} \, \left[\sum_{k=1}^{m} (1+ k\, \bd) \, \log (1+k\, \bd) \right]
\;,\nonumber\\
     \log\, \calh_{opt} &\approx h \log N  - h  + \frac{2\, h}{m\, (m+1)} \,
\left( \sum_{k=1}^m k\, \log (k\, \bd) \right)\;,\nonumber \\
     \log\, \tdim_N \, \y_{opt} &\approx \frac{2\, h}{\bd \, m\, (m+1)} \,
\sum_{k=1}^m \left[(1+ k\, \bd) \, \log (1+k\, \bd) - (k\,\bd) \, \log (k\, \bd)
\right]\;.
   \end{align}
   \item \underline{$\d<1$:} We need to distinguish between four subcases:
       \begin{itemize}
          \item[$\ast$] \underline{$\n < \d$:} This is the simplest case as $\tm
\approx d_0$ as before. We find:
              \begin{align}
               \log\, \tnum\, \y_{opt} &\approx h \, \log N + 0 \,h \;,
\nonumber \\
               \log\, \calh_{opt} &\approx \d \, h \, \log N - h + \frac{2\,
h}{m\,(m+1)} \, \left( \sum_{k=1}^m k\, \log (k\, \bd) \right) \;, \nonumber\\
               \log\, \tdim_N \, \y_{opt} &\approx (1-\d) \, h\, \log N  \approx
h \, \log \frac{N}{d}\;, \\
               \log\, \tdim_h \, \y_{opt} \approx& \, \n \,h\, \log N + \left(
\log \left[ \frac{1}{2} \, \bn \, m \,(m+1) \right] - \frac{2}{m\, (m+1)} \,
\sum_{k=1}^m k\, \log \, k \right)\, h \nonumber \\
                                          \approx& h \, \log \frac{h}{d} + a
\, h\;,
              \end{align}
where $a$ is some constant. The reason we included the next subleading term in
the last expression is that $\n=0$ is a valid point in our space of parameters.
It is easy to check that this subleading term is positive when $\n=0$ as it
should be. This is because in this case $\bn \geq 1$ and:
$$ \sum_{k=1}^m k \, \log k \leq \sum_{k=1}^m k \, \log m = \frac{1}{2} \, m \,
(m+1) \, \log m \;. $$
          \item[$\ast$] \underline{$\n = \d$:} This case is slightly more
complicated than the previous one. We find:
              \begin{align}
                \log\, \tnum\, \y_{opt} &\approx h\, \log N + 0\, h
\;,\nonumber\\
                \log\, \calh_{opt} &\approx \d\, h \, \log N + \calo \, (h)
\;, \nonumber \\
                \log\, \tdim_N \, \y_{opt} &\approx (1-\d)\, h \, \log N \approx
h \, \log \frac{N}{d} \approx h \, \log \frac{N}{n} \;,\\
                \log\, \tdim_h \, \y_{opt} &\approx \d \, h \, \log N \approx h
\, \log \frac{h}{d} \approx h \, \log \frac{h}{n}\;,
              \end{align}
where we used that $\n = \d \neq 0$.
          \item[$\ast$] \underline{$1 > \n >\d$:} In this case we have $\tm
\approx
n_0$. We find:
              \begin{align}
                \log\, \tnum\, \y_{opt} &\approx h\, \log N + 0\, h \;,
\nonumber \\
                \log \, \calh_{opt} &\approx \n\, h\, \log N - h +
\frac{2\, h}{m \, (m+1)} \left( \sum_{k=1}^m k\, \log (k\, \bn) \right) \;,
\nonumber \\
                \log\, \tdim_N\, \y_{opt} &\approx (1-\n) \,h\, \log N \approx h
\, \log \frac{N}{n} \;,\\
                \log\, \tdim_h\, \y_{opt} &\approx \d \,h\, \log N + \left( \log
\left[ \frac{1}{2} \, \bd \, m \,(m+1) \right] - \frac{2}{m\, (m+1)} \,
\sum_{k=1}^m k\, \log \, k \right)\, h \nonumber \\
                                          &\approx h \, \log \frac{h}{n} + a
\, h \;,
              \end{align}
where $a$ is a constant. Notice that the roles of $n_0$ and $d_0$ here are
switched with respect to the case $\d>\n$, even though the approximate
expression (\ref{opt-family-num-nosum-n}) is not symmetric under the exchange $
n_0 \leftrightarrow d_0$. This is a manifestation the duality discussed in
subsection-\ref{approx-dual}.
          \item[$\ast$] \underline{$1=\n >\d$:} In this case $\tm \approx n_0
\sim N$.
We find:
              \begin{align}
                \log\, \tnum\, \y_{opt} &\approx h\, \log N - h - \frac{2\,
h}{\bm \, m\, (m+1)} \, \left[ \sum_{k=1}^m (1-k\, \bn) \, \log (1-k\, \bn)
\right] \;,\nonumber \\
                \log \, \calh_{opt} &\approx h\, \log N - h +
\frac{2\, h}{m \, (m+1)} \left( \sum_{k=1}^m k\, \log (k\, \bn) \right)
\;, \nonumber \\
                \log\, \tdim_N\, \y_{opt} &\approx - \frac{2\, h}{\bn\,
m\,(m+1)} \, \sum_{k=1}^m \left[ (1-k\, \bn) \, \log (1-k\, \bn) + (k\,
\bn) \, \log (k\, \bn) \right] \;.
              \end{align}
Notice that $\log \, \tdim_N \, \y_{opt}$ is well defined and positive
since $m\,\bn <1$
       \end{itemize}
\end{itemize}
%
\subsection{The $opt_\m$ family}
%
In this case, we have the following range of parameters:
$$ 0 < \m \leq 1 \;,\quad 0<\n+\m \leq 1 \;, \quad 0 < 2\, \m+ \n+ \d <2 \;. $$
The origin of the complication in this case is that we need to worry about $d =
m\, d_0$ and $n = n_0\, m$ on top of $n_0$ and $d_0$. As in the previous
subfamily we will use the different ranges of $\d$ as the main
classifying tool. We have the following cases:
\begin{itemize}
   \item \underline{$1 < \d < 2$:} This is the easiest case. We
find:
       \begin{align}
         \log \, \tnum \, \y_{opt} & \approx  \left( h+ n \, N -
\frac{1}{2} \, n^2 \right) \, \log \ty + \frac{h}{6\, m^2} \, \log m
\nonumber\\
                                   & \quad - n \, \left( N - \frac{1}{2} \,
n \right) \, \log N - \frac{3}{2} \, h + \frac{h}{2\, m} + \ldots \;,
\nonumber\\
               \log \, \calh_{opt} & \approx \left( h - \frac{1}{2} \, n_0\, n
\right) \, \log \ty + \frac{h}{6\, m^2} \, \log m - \frac{1}{2} \,n_0 \, n \,
\log n_0  - \frac{3}{2} \, h + \frac{h}{2\, m} + \ldots \;, \nonumber \\
       \log \, \tdim_N \, \y_{opt} & \approx (\m+\d-1)\, \bn\, \bm \,
N^{1+\n+\m} \, \log N \approx n \, N \, \log \frac{d}{N} \;.
       \end{align}
   \item \underline{$\d=1$:} Following the same steps as before we find:
       \begin{align}
         \log \, \tnum \, \y_{opt} & \approx \left( h+ \frac{h}{6\, m^2} + n
\, N - \frac{1}{2} \, n^2 \right) \, \log \ty \nonumber\\
                                   & \quad - \left(\frac{1}{6} + \frac{1+
\bd}{(\bd)^2} \right) \, \left( \frac{h}{m^2} \right) \, \log (N+d_0) \nonumber
\\
                                   & \quad - n \, \left( N - \frac{1}{2} \,
n \right) \, \log N - \frac{3}{2} \, h + \frac{h}{2\, m} + \ldots\;, \nonumber\\
       \log \, \calh_{opt}  \approx&  \left( h - \frac{1}{2} \, n_0\, n
\right) \, \log \ty + \frac{h}{6\, m^2} \, \log m - \frac{1}{2} \,n_0 \, n \,
\log n_0  - \frac{3}{2} \, h + \frac{h}{2\, m} + \ldots \;, \nonumber \\
       \log \, \tdim_N \, \y_{opt} & \approx \m\, \bm\, \bn \,
N^{1+\m+\n} \, \log N \approx n \, N \, \log \frac{d}{N} \;.
       \end{align}
   \item \underline{$\d < 1$:} As before we distinguish three subcases:
        \begin{itemize}
             \item[$\star$] \underline{$ \n < \d$:} We distinguish three
possibilities:
                 \begin{itemize}
                     \item \underline{$\d+ \m > 1$:} In this case we
need to keep terms that involve $d_0$ in the log expression. We find:
       \begin{align}
         \log \, \tnum \, \y_{opt} & \approx \left( h + n
\, N \right) \, \log \ty - n \, N \, \log N - \frac{3}{2} \, h + \ldots \;,
\nonumber\\
               \log \, \calh_{opt} & \approx  \left( h - \frac{1}{2} \, n_0\,
n \right) \, \log \ty - \frac{3}{2} \, h + \frac{h}{2\, m} + \ldots \;,\nonumber
\\
       \log \, \tdim_N \, \y_{opt} & \approx (\d+\m-1) \, \bm\, \bn \,
N^{1+\m+\n} \, \log N  \approx n \, N \, \log \frac{d}{N} \;.
       \end{align}
                     \item \underline{$\d+ \m = 1$:} Following the same steps
as in the previous case, we find:
       \begin{align}
         \log \, \tnum \, \y_{opt} & \approx \left(h - \frac{1}{2} \,
n^2 \right) \, \log N + \left( \frac{1+\oy}{\oy} \right)^2 \, h \, \log (1+
\oy) \nonumber \\
                                   & \qquad \qquad - \left(\frac{3}{2} +
\frac{1}{(\oy)} \right) \, h + \ldots\;, \nonumber\\
               \log \, \calh_{opt} & \approx \left( h - \frac{1}{2} \, n_0\, n
\right) \, \log \ty - \frac{3}{2} \, h + \ldots \;, \nonumber \\
       \log \, \tdim_N \, \y_{opt} & \approx \frac{1}{(\oy)^2}\, \left[
(1+\oy)^2 \, \log (1+ \oy) - (\oy)^2 \, \log \oy - \oy \right] \, h \;,
       \end{align}
where $\oy = \bm \, \bd$.
                     \item \underline{$\d+ \m < 1$:} We find in this case:
       \begin{align}
         \log \, \tnum \, \y_{opt} & \approx h \, \log N + 0\, h + ... \;,
\nonumber\\
               \log \, \calh_{opt} & \approx \left( h - \frac{1}{2} \, n_0\, n
\right) \, \log \ty - \frac{3}{2} \, h + \ldots  \;, \nonumber \\
       \log \, \tdim_N \, \y_{opt} & \approx (1- \m-\d) \, h\, \log N \approx h
\, \log \frac{N}{d}\;, \\
       \log \, \tdim_h \, \y_{opt} & \approx (\n+\m) \,h\, \log N \approx h \,
\log \frac{h}{d} \;.
       \end{align}
                 \end{itemize}
             \item[$\star$] \underline{$\d=  \n$:} We find:
       \begin{align}
         \log \, \tnum \, \y_{opt} & \approx h \, \log N  - \frac{3}{2} \,
\left(\frac{\bn}{\bd} \right) \, h + \ldots\;, \nonumber \\
               \log \, \calh_{opt} & \approx h \, \log \ty + \ldots \;,
\nonumber \\
       \log \, \tdim_N \, \y_{opt} & \approx (1- \m -\d) \, h \, \log N \approx
h \, \log \frac{N}{d} \approx h \, \log \frac{N}{n}\;,  \\
       \log \, \tdim_h \, \y_{opt} & \approx (\m+\d) \, h \, \log N \approx h
\, \log \frac{h}{d} \approx h \, \log \frac{N}{n}\;.
       \end{align}
             \item[$\star$] \underline{$\d < \n$:} We distinguish two
possibilities:
                 \begin{itemize}
                     \item \underline{$\m+ \n =1$:} In this case we need to
keep terms involving $n_0$ untouched when expanding the log terms. Then, we
evaluate the sum over $k$. We find:
       \begin{align}
         \log \, \tnum \, \y_{opt} & \approx  h\, \log N + \left(
\frac{1- \bm\, \bn}{\bm\, \bn} \right)^2 \, h\, \log (1-\bm\, \bn) + \left(
\frac{1}{\bm\,\bn} - \frac{3}{2} \right) \, h\;, \nonumber \\
               \log \, \calh_{opt} & \approx h\, \log \ty - \frac{3}{2} \, h +
\ldots \;, \nonumber \\
       \log \, \tdim_N \, \y_{opt} & \approx \frac{h}{(\bm\, \bn)^2} \, \left[
(1- \bm \, \bn)^2 \, \log (1- \bm \, \bn) - (\bm \, \bn)^2 \, \log (\bm \, \bn)
+ (\bm \, \bn) \right] \;.
       \end{align}
Notice that $\tdim_N \, \y_{opt}$ is positive as it should be. To see that
remember that\footnote{It is easy to prove that the function:
$ f(x)  = (1-x)^2 \, \log (1-x) - x \, \log x + x$, is an increasing function in
the interval $\left(0\,,\,1\right]$ which implies that $0 \leq f(x) \leq 1$.}
$(\bm \, \bn) \leq 1$.
                     \item \underline{$\m+ \n <1$:} Here we expand the log
terms keeping only $N$. We find, after summing over $k$:
       \begin{align}
         \log \, \tnum \, \y_{opt} & \approx h \, \log N + \ldots\;, \nonumber
\\
               \log \, \calh_{opt} & \approx h\, \log \ty - \frac{3}{2} \, h +
\ldots \;, \nonumber \\
       \log \, \tdim_N \, \y_{opt} & \approx (1- \n-\m) \,h\, \log N \approx h
\, \log \frac{N}{n} \;, \\
       \log \, \tdim_h \, \y_{opt} & \approx (\m+\d) \, h\, \log N \approx h
\, \log \frac{h}{n}\;.
       \end{align}
                 \end{itemize}
        \end{itemize}
\end{itemize}
%
%
%
\subsection{Summary}
%
%
In the following, we summarize what we got in the previous two subsections. This
will make it much easier to compare with the general discussion in section
\ref{probes-section}. To make contact with the expressions there, we introduce
the quantity:
 $$ \y_0 = \tmax \, \{d \,,\, n \} \;.$$
We will also borrow the classification of probes from there. Looking at the
different cases above, we find the following leading behavior of $\log
\,\tdim_N \, \y$ and $\log \, \tdim_h \, \y$:
\begin{itemize}
  \item \underline{Generic Probes:} In the case $\m=0$, these probes correspond
to $\d<1$ and include the cases $\d < \n$, $\d =\n$ and $\d >\n$. In the case
$\m \neq 0$, these probes correspond to $\d<1$ once again and include the cases
$(\n < \d \,,\, \m+ \d <1)$, $\n = \d$ and $(\d < \n \,,\, \n+\m <1)$. In all
these cases, we find that:
 \begin{equation}
  \log \, \tdim_N \, \y \approx h \, \log \frac{N}{\y_0} \quad , \quad \log \,
\tdim_h \, \y \approx h \, \log \frac{h}{\y_0} + a \, h \;,
\label{opt-family-gen-probe}
 \end{equation}
where $a$ is some constant whose precise value is not of interest to us.
  \item \underline{Linear Probes:} In the case $\m=0$, these probes include
$\d=1$ and $\n =1$ cases. In the case $\m \neq 0$, we have the cases $(\n < \d
\,,\, \m+ \d =1)$ and $(\d < \n \,,\, \n+\m =1)$. In all this cases we find:
 \begin{equation}
   \log \, \tdim_N \, \y \approx b \, h \;,\label{opt-family-lin-probe}
 \end{equation}
where $b$ is some constant whose value is not important to us.
  \item \underline{Long Probes:} In the case $\m=0$, these probes correspond to
$\d >1$. In the other case $\m \neq 0$, the probes correspond to the cases $ \d
>1$, $\d =1$, and $(\d<1 \,,\, \d+\m >1)$. In all of these cases, we find:
 \begin{equation}
   \log \, \tdim_N \, \y \approx n \, N \, \log \frac{d}{N} \;.
\label{opt-family-long-probe}
 \end{equation}
\end{itemize}
These results are in complete agreement with what we found in the general
discussion in subsection-\ref{probes-section}.
%
%
%
\section{Some useful properties of Kostka numbers}
\label{Kostka-numbers-app}
%
%
%
In this appendix, we will discuss some of the properties of Kostka numbers
$K_{\y \,,\, \b}$ that will be very useful in subsections
\ref{tensor-prod-h-ll-N} and \ref{tensor-prod-h-sim-gg-N}. In the following,
the filling $\b$ has no zero entries, see the end of subsection
\ref{tensor-prod-repackage} for more details.
%
%
\subsection{Kostka numbers and fillings}
%
The first property we are going to discuss has to do with the behavior of the
Kostka numbers under the reshuffling of the numbers $\b_i$'s that define a
filling $\b$. The claim is that the Kostka numbers are invariant under that, and
hence, we can take $\tilde{\b}$ the ordered counterpart of $\b$ as a
representative of these fillings i.e. start from a filling $\b = (\b_1 \,,\,
\b_2 \,,\, \ldots \,,\, \b_m)$, and construct the filling $\tilde{\b} =
(\tilde{\b}_1 \,,\, \tilde{\b}_2 \,,\, \ldots \,,\, \tilde{\b}_m)$ such that:
  $$ \tilde{\b}_1 \geq \tilde{\b}_2 \geq \ldots \geq \tilde{\b}_m \;.$$
To prove the claim above, it is enough to prove that we can construct a
one-to-one map between the SSYTx $\y$ with fillings $\b$ and $\bar{\b}$ that are
related by exchanging two numbers $\b_i$ and $\b_{i+1}$. i.e.
$$ \b = (\b_1 \,,\, \b_2 \,,\, \ldots \,,\, \b_i \,,\, \b_{i+1} \,,\, \ldots \,
, \, \b_m ) \;, \qquad \bar{\b} = (\b_1 \,,\, \b_2 \,,\, \ldots \,,\, \b_{i+1}
\,,\, \b_i \,,\, \ldots \, , \, \b_m ) \;. $$
Let us assume that we constructed all the SSYTx $\y$ with filling $\b$ and we
want to construct the ones associated to the filling $\bar{\b}$. This is done in
three steps:
\begin{itemize}
   \item[1.] First, we relabel by $(i+1)$ the boxes labeled by $i$, and vice
versa. Notice that since we are not touching the other labels in this step,
we need only to deal with the order of the labels $i$ and $(i+1)$.
   \item[2.] Next, we first deal with the order of the labels $i$ and $(i+1)$
in each column. If the label $i$ sits below the label $(i+1)$ we exchange
their position, otherwise we leave the column unchanged. Notice that in this
step if the column in not left unchanged then all we have done is switching the
labels of only two boxes in this column.
   \item[3.] Finally, we look at the order of labels in each row. Once again we
need to reorder the boxes with labels $i$ and $(i+1)$ only. This is done by
putting the boxes with label $i$ that are to the right of the boxes with labels
$(i+1)$ to their left. At the end of this step, we get a SSYT $\y$ with filling
$\bar{\b}$ for each SSYT $\y$ with filling $\b$.
\end{itemize}
Notice that we could have done the same by starting with the filling $\bar{\b}$,
then construct the SSYTx associated to the filling $\bar{\b}$. Hence, the two
Kostka numbers associated to $\b$ and $\bar{\b}$ are equal.

Take for example the fillings $\b= (1 \,,\, 1 \,,\, 2 \,,\, 1)$ and $\bar{\b} =
(1 \,,\, 2 \,,\, 1\,,\, 1)$ and the YD:

$$ \yng(3,2) \;. $$
\\
The only SSYTx associated to the filling $\b$ are:

$$ \young(123,34) \quad , \quad \young(133,24) \quad , \quad \young(124,33)
\quad .$$
\\
Let us follow the steps above to construct the SSYTx associated to $\bar{\b}$.
The first step gives:

$$\young(132,24) \quad , \quad \young(122,34) \quad , \quad \young(134,22)
\quad . $$
\\
The only YT that is affected by the second step is the last one, so we have:

$$\young(132,24) \quad , \quad \young(122,34) \quad , \quad \young(124,23)
\quad . $$
\\
Finally, the last step remedies the first YT which is the only non-SSYT. We
get at the end the following SSYTx associated to the filling $\bar{\b} = (1
\,,\, 2 \,,\, 1 \,,\, 1)$:

$$\young(123,24) \quad , \quad \young(122,34) \quad , \quad \young(124,23)
\quad . $$
\\
It is not hard to see that the SSYT above are the only ones possible for the
filling $\bar{\b}$. In the remaining of this appendix, whenever we talk about a
filling $\b$, we assume that it is completely ordered i.e. $\b_1 \geq \b_2 \geq
\ldots $.
%
%
\subsection{The non-zero Kostka numbers}
%
%
The question we want to address here is: when is the Kostka number non
vanishing? Since Kostka numbers $K_{\y \,,\, \b}$ are related to SSYTx $\y$
which are themselves related to decomposition of the tensor product $\y \otimes
B$, where $B$ is some arbitrary YD\footnote{We choose $B$ such that the number
of its columns is large enough.}, through the ordering rule defined in
subsection \ref{tensor-prod-rules} according to our discussion in subsection
\ref{tensor-prod-repackage}, we can use the implications of the ordering rule
 as a guiding tool.

The first implication of the ordering rule is that a box in row $i$ of the YD
$\y$ cannot be attached to a row $j$ of the YD $B$ if $i <j$, see end of section
\ref{tensor-prod-rules}. This can be easily understood in the case of SSYT and
has to do with the fact that the labels in the same column should be strictly
increasing from top to bottom, and labels in the same row should be weakly
increasing from left to right. In the language of SSYT, $i$ will be the row
number and $j$ will be the label. Due to the aforementioned conditions on the
labels in a SSYT, the labels in row $i$ should bigger or equal than $i$, which
implies the absence of the label $j$ in row $i$ if $j<i$.

The second implication of the ordering rule (see end of section
\ref{tensor-prod-rules}) can also be understood as a consequence of the
conditions of the labels of SSYTx. In terms of the tensor product
decomposition, it was shown that if the boxes added to a row $i$ of $B$ come
from different rows of $\y$ with $m$ being the highest one among them, than the
number of added boxes to the row $i$ is less or equal than $\y_m$ the length of
this row. In terms of the SSYT labeling, this situation corresponds to the
presence of the label $i$ in the row $m$ of $\y$. Due to the conditions on the
labeling of SSYT, any other box with label $i$ in a row below this row should
be to the left of this box. As a result the number of labels $i$ in the SSYT
$\y$ is less or equal $\y_m$ the length of the row $m$ of $\y$.

The combination of the previous two conditions, together with the fact that the
filling $\b$ is a totally ordered filling, implies the following condition for a
non-vanishing Kostka number $K_{\y \,,\, \b}$, see for example \cite{stanley}.
The Kostka number $K_{\y \,,\, \b}$ is nonzero if and only if both $\y$ and $\b$
are partitions of the same integer $h$, and moreover, $\y$ is larger than $\b$ in
the dominance order. The latter is defined as follows. Let $\y = (\y_1 \,,\,
\y_2 \,,\, \ldots \,,\, \y_n)$, $\b = (\b_1 \,,\, \b_2 \,,\, \ldots \,,\, \b_m)$
with $\y_1 \geq \y_2 \geq \ldots$ and $ \b_1 \geq \b_2 \geq \ldots $ be two
ordered tuples of integers. We say that $\y$ is larger than $\b$ in dominance
order, and we write $ \y \trianglerighteq \b $, if and only if for each integer
$k \geq 1$, the following is true:
$$ \forall \; k \geq 1 \;; \quad \sum_{i=1}^k \y_i \geq \sum_{i=1}^k \b_i \;,$$
where we fill in the non-existing integers $\y_k$ for $k >n$ by zeros. An
immediate consequence of this is that the number of entries in $\b$ should be
bigger or equal to the the number of rows of $\y$. Notice that this condition is
only true for $\b$ being totally ordered. If $\b$ is not ordered than the
statement above works in one directions only: $K_{\y \,,\, \b}$ nonzero then we
have the dominance order condition but not the way around.
%
%
\subsection{Maximizing Kostka numbers}
%
%
The final point in our investigation on the Kostka numbers is to find
conditions on the filling $\b$ such that we get a maximum value for $K_{\a
\,,\, \b}$. This is important as we are interested in leading order behavior of
the degeneracy of YD appearing in the decomposition of $\calo \otimes \y$, and
as we will see in subsection \ref{tensor-prod-h-sim-gg-N} and
section \ref{h-sim-gg-n-discussion} of
appendix-\ref{app-inclusion-yd-antisym-rules}, this degeneracy is
intimately connected to the maximum Kostka number in cases of interest to us.
Notice that we are fixing the shape $\y$ here and varying $\b$. Two claims
can be made here. First of all, $|\b|$ the ``length'' of the filling $\b$ should
be maximized. By length we mean the number of non-zero entries $\b_i \neq 0$.
Secondly, the entries $\b_i$ should be very close to the average $h/|\b_0|$
\footnote{An intuitive argument to why such a choice is special is as follows.
We argued before that the Kostka numbers are invariant under permutations of the
entries of the filling $\b$. The fixed point of such a permutation is when all
the entries are equal which reproduces precisely this choice.}. In the following,
$h$ will stand, as usual, for the number of boxes of the YD $\y$ whose
shape is also denoted by $\y$, whereas $N$ is, as usual, the flux of the
background geometry which is the same $N$ that appears in the group of our
interest U$(N)$.

The first claim is easy to understand. The condition on the columns ordering
implies that we cannot have the same label on more than one box in the same
column. Hence, to have more options in labeling we need to have more labels. As
a result, the bigger $|\b|$ the more options we have. In the case $N \leq h$,
the biggest $|\b|$ possible is $h$ which implies that $\b_i =1$. So, the
condition on $|\b|$ is strong enough to imply the second claim above. Things are
more complicated in the case $N < h$ since here we have at least one $\b_i$ that
is bigger than ``$1$'' for fillings associated with the largest possible value
of $|\b|=N$. In this case the second claim adds a non trivial condition on the
filling $\b$. The argument for the validity of the second claim in this case is
as follows.

First of all, we are going to work with the ordered filling $\tilde{\b}$. We
will also concentrate on the case $N < h$. Hence, we are starting with the
$N$-tuple $\tb = (\tb_1 \,,\, \tb_2 \,,\, \ldots \,,\, \tb_N)$ such that:
 $$ \tb_1 \geq \tb_2 \geq \ldots \geq \tb_N \;, \qquad \sum_{i=1}^N \tb_i = h
\;. $$
Let us first concentrate on $\tb_1$ and $\tb_2$. Let us fix the sum $\tb_1 +
\tb_2$ but allow each of them to vary. Notice that the boxes with label
$1$ are confined to the first row, whereas the ones with label $2$ have the
freedom to be located in either of the first tow rows. So, the bigger $\tb_2$
the bigger the Kostka number. We reach the maximum if $\tb_2 \approx
\tb_1$ since $\tb_2 \leq \tb_1$. Repeating the same argument for the labels
$3 \leq i \leq n$, where we use that boxes with label $j$ are confined to be in
the first $j$ rows, we conclude that to maximize the Kostka number we should
choose:
$$ \tb_1 \approx \tb_2 \approx \ldots \approx \tb_n \;.$$
To complete the argument we start from below. Once again fixing the sum $\tb_N
+ \tb_{N-1}$, but allowing both of them to vary. The position of the boxes with
label $N$ is at the end of the rows with the condition that there are no boxes
below them. For the boxes with label $(N-1)$, they can be either on top of the
boxes with label $N$, or to their left under the condition that if there are
boxes below, they should carry the label $N$. So, if we have in a row $k$
boxes with label $N$, we have at least $k$ possible boxes to carry the label
$(N-1)$. Hence the bigger $\tb_N$, the more options we have which leads to the
optimum situation $\tb_{N-1} \approx \tb_N$. We can repeat the same argument for
the other labels and reach a similar conclusion. Hence the second claim.

In the argument above, we cheated a little bit. Remember that there should be no
boxes below the ones with label $N$. So the number of such boxes i.e. boxes
with no box below in the YD, constitute an upper bound on $\tb_N$. So, the
claim above should be modified accordingly. However, as we will see in
subsection \ref{tensor-prod-h-sim-gg-N}, our case of interest corresponds to
$(h/N)$ being small enough, together with the type of YD we are dealing with, puts
us in a safe position.
%
%
%
\section{From Kostka numbers to actual degeneracies $d_k$}
\label{app-inclusion-yd-antisym-rules}
%
%
%
In the following, we will discuss the possible modifications to the conclusions
derived in subsections \ref{tensor-prod-h-ll-N} and \ref{tensor-prod-h-sim-gg-N}
for SSYTx $\y$ and their Kostka numbers once the YD and the
antisymmetry rules are included to make contact with the tensor decomposition
(\ref{tensor-prod-decomp-schem}). Following the arguments for the Kostka
numbers, the discussion depends on whether $h \ll N$, or $(h \sim N \,,\, h \gg
N)$.
%
%
\subsection{The case $h \ll N$} \label{h-ll-n-discussion}
%
%
We arrived in subsection \ref{tensor-prod-h-ll-N} at the conclusion that,
without the inclusion of the YD and the antisymmetry rule, we can take the YDs
$\vf_k^0$ as representatives of the the YDs $\vf_k$ that appear in the
decomposition of the tensor product $\calo \otimes \y$. Remember that the YDs
$\vf_k^0$ are the result of adding at most one box to each row of $\calo$. Their
degeneracy $d_k^0$ as well as their number $\caln_0$ have the leading terms
given in equation (\ref{tensor-prod-info-h-ll-N}), which we rewrite here for
convenience.
\begin{equation}
 \log \, d_k \approx h \, \log \frac{h}{\y_0} \;, \quad \log \caln_0 \approx
h \, \log \frac{N}{h} \;. \label{tensor-prod-info-h-ll-N-app}
\end{equation}
What happens when we take into account the YD and the antisymmetry rules? For
concreteness, let us restrict ourselves to the special case of the background YD
$\calo_0$ whose rows' lengths are given by:
$$ \calo_i = N-i \;; \qquad i=1 \,,\, 2 \,,\, \ldots \,,\, N-1 \;. $$
This corresponds to the limit shape YD of the superstar ensemble of YDs with
the same number of columns and rows, $N$. We will discuss the possible
modifications that we need to take into account for the general case at
the end.

Notice that for this background $\calo_0$, neither the YD rule nor the
antisymmetry rule modify the results above, since both are trivially satisfied.
Hence, there are $\caln_0$ YDs that are constructed by adding at most one box
to each row, each of them has the degeneracy $d_k^0$, which in total
reproduces the leading behavior\footnote{This is after taking the log of the
numbers under consideration.} of $\tdim_N \, \y$. We will not discuss the other
YDs $\vf_k$ since they will not play a role in what we are trying to do, see
subsection \ref{two-pt-funct-full-gen-h-ll-N} for more details.

To close the line of thoughts, we need to discuss what happens in the case of a
different background $\calo$. We claim that the worst scenario that can happen
is to replace $N$ by $\k \, N$, with $\k \leq 1$ in the formulas above. The idea
is to replace every set of consecutive equal length rows with just one row,
which brings our discussion close enough\footnote{In general, the shift in the
length of consecutive rows can be more than one box which does not change the
essence of the discussion of the case of $\calo_0$.} to the one for $\calo_0$.
The only thing we need to make sure about is that, after doing so, we are still
left with order $N$ rows. This is the case since we know that typical YDs of
the superstar ensemble have order $N$ corners, see the end of subsection
\ref{superstar-ensemble} for more details.

All in all, in the tensor product decomposition of $\calo \otimes \y$, where
$\y$ is a YD with number of boxes $h \ll N$, the number of representative YDs $\vf_k^0$ and
their degeneracy has the following leading behavior:
\begin{equation}
 \log \, d_k \approx h \, \log \frac{h}{\y_0} \;, \quad \log \caln \approx
h \, \log \frac{N}{h} \;,\label{tensor-prod-info-h-ll-N-gen}
\end{equation}
where, according to our conventions (appendix-\ref{conventions}), $\y_0 = \tmax
\{n \,,\, d\}$, $d$ is the number of columns of $\y$, and $n$ is the number
of its rows.
%
%
\subsection{The cases $h \sim N$ and $h \gg N$} \label{h-sim-gg-n-discussion}
%
%
We arrived in subsection \ref{tensor-prod-h-sim-gg-N} to the conclusion that,
in the case where $\y$ is a generic probe and $h \sim N$ or $h \gg N$, and
forgetting about the antisymmetry rule, there are special YDs $\vf_k^\ast$ in
the tensor product decomposition (\ref{tensor-prod-decomp-schem}) whose
degeneracy $d^\ast_k$ has a leading behavior given by equation
(\ref{tensor-prod-info-h-sim-gg-N}), which we rewrite here for convenience:
\begin{equation}
 \log \, d_k^\ast \approx \tdim_N \, \y \approx h \, \log \frac{N}{\y_0} \;.
\label{tensor-prod-info-h-sim-gg-n-app}
\end{equation}
These YDs are constructed by adding $h_i \approx (h/N)$ boxes to each
row of the YD $\calo$. Notice that $h_i \ll d$ and $h_i \ll n$ since $n$, $d
\ll N$ and $h \sim n\, d$. In this appendix, we want to study the implication of
taking into account the antisymmetry rule. To fix the notations below, we will
denote by $\b_0$ the filling of the SSYT $\y^\ast$ associated to one of the YDs
$\vf_k^\ast$. Hence:
$$ \b_0 = (h_1 \,,\, h_2 \,,\, \ldots \,,\, h_N) \;, \quad h_i \approx
\frac{h}{N} \;, \quad h_1 \geq h_2 \geq \ldots \geq h_N \;, $$
where $h_i$ is as usual the degeneracy of the label $i$. We have the following
leading behavior of the associated Kostka number:
$$ \log \, K_{\a \,,\, \b} \approx \tdim_N \, \y \;, $$
We further concentrate, as in the previous section, on the simple case where the
background YD $\calo_0$ whose rows' lengths are given by:
$$ \calo_i = N-i \;; \qquad i=1 \,,\, 2 \,,\, \ldots \,,\, N-1 \;. $$
We will come back to the possible modifications to the arguments below at the
end of this section.

What does the antisymmetry rule imply? First of all, the labels associated to
the first raw should be different for each box except for the first few ones
that carry the label $1$. So, for $i >1$ we cannot find more than one box in the
first row with this label. The same argument leads to the conclusion that in
row $k$, there should be at most one box with label $i > k$. So essentially,
taking into account the antisymmetry rule forces the labeling of the YD
$\y^\ast$ to be of the following form. The label $ 1 \leq i \leq n$ is
mostly confined to the row $i$. In other words there will be at most one box
with label $j$ in a row $i \neq j$. The remaining labels $i > n$ will appear at
most once in each row. To proceed further, we write schematically the different
contributions to the associated Kostka number as:
\begin{equation}
K_{\a \,,\, \b_0} = K^{(0)}_{\a \,,\, \b_0} + \sum_a K^{(a)}_{\a \,,\, \b_0}
\;, \label{Kostka-schem}
\end{equation}
where $K^{(0)}_{\a\,,\, \b_0}$ stands for the contribution giving rise to the
degeneracy we are after i.e. the labelings that satisfies the required
conditions discussed above, and the sum is over the remaining ``bad''
contributions. We will call the former the good labelings and the latter the
bad labelings. The meaning of the index $a$ in the sum will be clear below, but
at the moment it stands for some indexing of the bad labelings. Our aim in the
following, is to relate somehow the bad labelings to the good ones. But
before that, let us first deal with the special labels $\{1\,,\,2\,,\, \ldots
\,,\, n\}$. Remember that these labels are special in the sense that they are
the only ones that are allowed to occur more than once given that they are in
the row with the same number as the label. In other words, we can have more
than one box with label $i$ if and only if $1 \leq i \leq n$ and the associated
boxes belong to row $i$ in the tableau $\y$. Let us count the number of
labelings of $\y$ that violate this requirement. It is easy to see that this
number is smaller than:
$$ \log \, n_{wrong} = \sum_{i=1}^n \log \, C_{h_i + i - 1}^{h_i} \sim \left(
\frac{n}{N} \right) \, h \, \log N \ll h \, \log N \;.$$
Given that $\log \, K_{\a \,,\, \b_0} \sim h \, \log N$, we can safely restrict
ourselves to counting the number of labelings where the first $n$ labels are
fixed as above. Said differently, we will try in the following to
estimate the number of labeling $K_{\a \,,\, \b_0}$ such that the labels $1 \leq
i \leq n$ are mainly confined to be each in its corresponding row $i$. Our
discussion below will concern only the labels $n < i \leq N$.

A useful way to think about counting the labelings $K_{\a\,,\, \b_0}$ is as
follows. First construct all the labelings where each row of $\y$ has at most
one box with label $i$, for all the labels $ n<i\leq N $. These labelings were
called good labelings and their number $K^{(0)}_{\a\,,\, \b_0}$ is the one we
are after. For each labelings among these, we will move the boxes around in
order to increase the degeneracy of labels in the rows of $\y$. For each good
labeling $a$, $\tilde{K}^{(a)}_{\a \,,\, \b_0}$ the number of bad labelings
constructed so, is smaller than:
$$ \caln_a = \prod_{k=1}^N C_{2\, h_k-1}^{h_k} \approx 2^{2\, h} \;, $$
where we used that $\sum_k h_k = h$. It is easy to convince yourself that in
this way, we will be able to construct all the labeling of $\y$ associated to
$\b_0$. The reason is once again the inequality $h_i \ll n$. This is because,
starting from a bad labeling, one can find a lot of possible rows to accommodate
the degenerate labels in order to construct at the end a good labeling of $\y$.

Now identifying the label $a$ in the sum (\ref{Kostka-schem}) with the index of
good labels, together with the identification $\tilde{K}^{(a)}_{\a \,,\, \b_0}
= K^{(a)}_{\a \,,\, \b_0}$ and its upper bound above, we conclude that to
leading order:
\begin{equation}
\log \, K^{(0)}_{\a \,,\, \b_0} \approx \log \, K_{\a \,,\, \b_0} \approx
\log \, \tdim_N \, \y \sim h\, \log N\;. \label{tensor-prod-deg-estimate}
\end{equation}
This result looks universal despite the fact that we started with a specific
background tableau $\calo_0$. Actually one can easily generalize the arguments
above to other backgrounds, either belonging to the same ensemble as $\calo_0$,
or other ensembles. The key idea is the observation mentioned above regarding
the scale of $h_i$. Remember we have $h_i \ll n$, $d$. Let us discuss each case
on its own.

For other backgrounds in the ensemble of $\calo_0$, the complication arises
because in this case we do not have the simple picture of a jump by one box
between consecutive rows as in $\calo_0$. We could either have some rows with
the same length, or the difference in length of some consecutive rows will be
larger than one box. The first situation in principle reduces the number of
possible YDs in the tensor decomposition, however we can easily see that the
leading order does not change. An easy way to see this is to follow the steps
bellow to construct YDs in the decomposition of the tensor product $\calo
\otimes \y$, where $\calo$ is some YD in the same ensemble of $\calo_0$.
\begin{itemize}
  \item The effect of having sets of rows with equal length has at worst the
effect of reducing $N$ inside the $\log$ to a fraction of it, which does not
modify the leading order (see the end of the previous section). A way to think
about this is to imagine that we replaced our background YD $\calo$
with a new one $\overline{\calo}$ by replacing all the sets of rows of equal
length by a single row. By doing so, we reduce the range of possible labels to
just a fraction of $N$. But since we are not altering the inequalities $h_i \ll
n$, $d$, we can still use our original arguments used in the case of $\calo_0$.
  \item After deflating $\calo$ to $\overline{\calo}$, we first construct
all the good labelings associated to the latter YD. After that, we inflate back
$\overline{\calo}$ to  the original YD $\calo$. At the end of this operation we
will end up with diagrams that are not YDs. Essentially, the rows
right after each set of equal length rows of $\calo$ might be bigger than the
one above it. But we can redistribute the extra boxes on the empty rows since
they come from different rows in $\y$, and hence they do not violate the
antisymmetry rule. The ordering rule is also preserved because of the way
we construct the tensor product if we move the boxes in the following way. We
just cut the excess boxes and slide them up without changing their order. This
is explained in the following example:

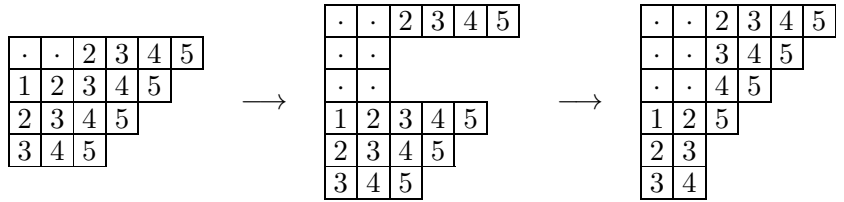
\begin{align*}
  \young(\cdot\cdot 2345,12345,2345,345) \quad \longrightarrow \quad
\young(\cdot\cdot 2345,\cdot\cdot,\cdot\cdot,12345,2345,345) \quad
\longrightarrow \quad \young(\cdot\cdot 2345,\cdot\cdot 345,\cdot\cdot
45,125,23,34) \;.
\end{align*}
\\
As a result, we do not change the leading order given in equation
(\ref{Kostka-schem}).
\end{itemize}
The second difference, the difference in length of consecutive rows is bigger
than one box, is beneficial as it allows to have more options because now we can
allow for more than one box with the same label $i >n$, for some $i$'s. But,
since the leading behavior of the degeneracy is already saturated by
(\ref{Kostka-schem}), we conclude once again that the leading term remains
intact.

All in all, dealing with a different background YD in the same ensemble of
$\calo_0$ leads to the same leading term of the largest degeneracy of YDs in
the decomposition of $\calo \otimes \y$ as in equation (\ref{Kostka-schem}). In
the case of other ensembles, the same story goes through as the only change
in this case can be reabsorbed by replacing $N$ by a fraction of it in the
worst case. This is because the background YDs we are dealing with are the ones
where the number of rows is $N$ and the number of columns $D$ is
of order $N$, and the number of boxes in fixed to be $(N \, D)/2$.
%
%
%
%
%
%
\bibliographystyle{jhep}
\bibliography{refslist}

\providecommand{\href}[2]{#2}\begingroup\raggedright\begin{thebibliography}{10}

\bibitem{Balasubramanian:2005mg}
V.~Balasubramanian, J.~de~Boer, V.~Jejjala, and J.~Simon, {\it {The Library of
  Babel: On the origin of gravitational thermodynamics}},  {\em JHEP} {\bf
  0512} (2005) 006, [\href{http://xxx.lanl.gov/abs/hep-th/0508023}{{\tt
  hep-th/0508023}}].

\bibitem{Bardeen:1973gs}
J.~M. Bardeen, B.~Carter, and S.~W. Hawking, {\it {The Four laws of black hole
  mechanics}},  {\em Commun. Math. Phys.} {\bf 31} (1973) 161--170.

\bibitem{Hawking:1974sw}
S.~W. Hawking, {\it {Particle Creation by Black Holes}},  {\em Commun. Math.
  Phys.} {\bf 43} (1975) 199--220.

\bibitem{Hawking:1976de}
S.~W. Hawking, {\it {Black Holes and Thermodynamics}},  {\em Phys. Rev.} {\bf
  D13} (1976) 191--197.

\bibitem{Lunin:2001jy}
O.~Lunin and S.~D. Mathur, {\it {AdS / CFT duality and the black hole
  information paradox}},  {\em Nucl.Phys.} {\bf B623} (2002) 342--394,
  [\href{http://xxx.lanl.gov/abs/hep-th/0109154}{{\tt hep-th/0109154}}].

\bibitem{Mathur:2002ie}
S.~D. Mathur, {\it {A proposal to resolve the black hole information paradox}},
   {\em Int. J. Mod. Phys.} {\bf D11} (2002) 1537--1540,
  [\href{http://xxx.lanl.gov/abs/hep-th/0205192}{{\tt hep-th/0205192}}].

\bibitem{Mathur:2005zp}
S.~D. Mathur, {\it {The Fuzzball proposal for black holes: An Elementary
  review}},  {\em Fortsch.Phys.} {\bf 53} (2005) 793--827,
  [\href{http://xxx.lanl.gov/abs/hep-th/0502050}{{\tt hep-th/0502050}}].

\bibitem{Mathur:2005ai}
S.~D. Mathur, {\it {The Quantum structure of black holes}},  {\em
  Class.Quant.Grav.} {\bf 23} (2006) R115,
  [\href{http://xxx.lanl.gov/abs/hep-th/0510180}{{\tt hep-th/0510180}}].

\bibitem{Bena:2007kg}
I.~Bena and N.~P. Warner, {\it {Black holes, black rings and their
  microstates}},  {\em Lect.Notes Phys.} {\bf 755} (2008) 1--92,
  [\href{http://xxx.lanl.gov/abs/hep-th/0701216}{{\tt hep-th/0701216}}].

\bibitem{Skenderis:2008qn}
K.~Skenderis and M.~Taylor, {\it {The fuzzball proposal for black holes}},
  {\em Phys.Rept.} {\bf 467} (2008) 117--171,
  [\href{http://xxx.lanl.gov/abs/0804.0552}{{\tt arXiv:0804.0552}}].

\bibitem{Mathur:2008nj}
S.~D. Mathur, {\it {Fuzzballs and the information paradox: A Summary and
  conjectures}},  \href{http://xxx.lanl.gov/abs/0810.4525}{{\tt
  arXiv:0810.4525}}.

\bibitem{Balasubramanian:2008da}
V.~Balasubramanian, J.~de~Boer, S.~El-Showk, and I.~Messamah, {\it {Black Holes
  as Effective Geometries}},  {\em Class.Quant.Grav.} {\bf 25} (2008) 214004,
  [\href{http://xxx.lanl.gov/abs/0811.0263}{{\tt arXiv:0811.0263}}].

\bibitem{Simon:2011zza}
J.~Simon, {\it {Extremal black holes, holography and coarse graining}},  {\em
  Int.J.Mod.Phys.} {\bf A26} (2011) 1903--1971,
  [\href{http://xxx.lanl.gov/abs/1106.0116}{{\tt arXiv:1106.0116}}].

\bibitem{deBoer:2009un}
J.~de~Boer, S.~El-Showk, I.~Messamah, and D.~Van~den Bleeken, {\it {A Bound on
  the entropy of supergravity?}},  {\em JHEP} {\bf 1002} (2010) 062,
  [\href{http://xxx.lanl.gov/abs/0906.0011}{{\tt arXiv:0906.0011}}].

\bibitem{Sen:2009bm}
A.~Sen, {\it {Two Charge System Revisited: Small Black Holes or Horizonless
  Solutions?}},  {\em JHEP} {\bf 1005} (2010) 097,
  [\href{http://xxx.lanl.gov/abs/0908.3402}{{\tt arXiv:0908.3402}}].

\bibitem{Simon:2009mf}
J.~Simon, {\it {Small Black holes versus horizonless solutions in AdS}},  {\em
  Phys.Rev.} {\bf D81} (2010) 024003,
  [\href{http://xxx.lanl.gov/abs/0910.3225}{{\tt arXiv:0910.3225}}].

\bibitem{Bena:2012hf}
I.~Bena, M.~Berkooz, J.~de~Boer, S.~El-Showk, and D.~Van~den Bleeken, {\it
  {Scaling BPS Solutions and pure-Higgs States}},
  \href{http://xxx.lanl.gov/abs/1205.5023}{{\tt arXiv:1205.5023}}.

\bibitem{Shepard:2005zc}
P.~G. Shepard, {\it {Black hole statistics from holography}},  {\em JHEP} {\bf
  0510} (2005) 072, [\href{http://xxx.lanl.gov/abs/hep-th/0507260}{{\tt
  hep-th/0507260}}].

\bibitem{Alday:2006nd}
L.~F. Alday, J.~de~Boer, and I.~Messamah, {\it The gravitational description of
  coarse grained microstates},  {\em JHEP} {\bf 12} (2006) 063,
  [\href{http://xxx.lanl.gov/abs/hep-th/0607222}{{\tt hep-th/0607222}}].

\bibitem{Myers:2001aq}
R.~C. Myers and O.~Tafjord, {\it {Superstars and giant gravitons}},  {\em JHEP}
  {\bf 0111} (2001) 009, [\href{http://xxx.lanl.gov/abs/hep-th/0109127}{{\tt
  hep-th/0109127}}].

\bibitem{deMelloKoch:2004ws}
R.~de~Mello~Koch and R.~Gwyn, {\it {Giant graviton correlators from dual SU(N)
  super Yang-Mills theory}},  {\em JHEP} {\bf 0411} (2004) 081,
  [\href{http://xxx.lanl.gov/abs/hep-th/0410236}{{\tt hep-th/0410236}}].

\bibitem{deMelloKoch:2005rq}
R.~de~Mello~Koch, A.~Jevicki, and S.~Ramgoolam, {\it {On exponential
  corrections to the 1/N expansion in two-dimensional Yang Mills theory}},
  {\em JHEP} {\bf 0508} (2005) 077,
  [\href{http://xxx.lanl.gov/abs/hep-th/0504115}{{\tt hep-th/0504115}}].

\bibitem{Corley:2001zk}
S.~Corley, A.~Jevicki, and S.~Ramgoolam, {\it {Exact correlators of giant
  gravitons from dual N=4 SYM theory}},  {\em Adv.Theor.Math.Phys.} {\bf 5}
  (2002) 809--839, [\href{http://xxx.lanl.gov/abs/hep-th/0111222}{{\tt
  hep-th/0111222}}].

\bibitem{Berenstein:2004kk}
D.~Berenstein, {\it {A Toy model for the AdS / CFT correspondence}},  {\em
  JHEP} {\bf 0407} (2004) 018,
  [\href{http://xxx.lanl.gov/abs/hep-th/0403110}{{\tt hep-th/0403110}}].

\bibitem{Lin:2004nb}
H.~Lin, O.~Lunin, and J.~M. Maldacena, {\it {Bubbling AdS space and 1/2 BPS
  geometries}},  {\em JHEP} {\bf 0410} (2004) 025,
  [\href{http://xxx.lanl.gov/abs/hep-th/0409174}{{\tt hep-th/0409174}}].

\bibitem{Maldacena:1997re}
J.~M. Maldacena, {\it {The Large N limit of superconformal field theories and
  supergravity}},  {\em Adv.Theor.Math.Phys.} {\bf 2} (1998) 231--252,
  [\href{http://xxx.lanl.gov/abs/hep-th/9711200}{{\tt hep-th/9711200}}].

\bibitem{Fulton}
W.~Fulton, {\em Young Tableaux: With Applications to Representation Theory and
  Geometry}, vol.~35 of {\em London Mathematical Society Student Texts}.
\newblock Cambridge University Press, 1996.

\bibitem{B-Sagan}
B.~Sagan, {\it {The ubiquitous Young tableau, Invariant theory and tableaux}},
  {\em IMA Vol. Math. Appl.} {\bf 19} (1990) 262--298.

\bibitem{Berenstein:2004hw}
D.~Berenstein, {\it {A Matrix model for a quantum Hall droplet with manifest
  particle-hole symmetry}},  {\em Phys.Rev.} {\bf D71} (2005) 085001,
  [\href{http://xxx.lanl.gov/abs/hep-th/0409115}{{\tt hep-th/0409115}}].

\bibitem{Ghodsi:2005ks}
A.~Ghodsi, A.~Mosaffa, O.~Saremi, and M.~Sheikh-Jabbari, {\it {LLL vs. LLM:
  Half BPS sector of N=4 SYM equals to quantum Hall system}},  {\em Nucl.Phys.}
  {\bf B729} (2005) 467--491,
  [\href{http://xxx.lanl.gov/abs/hep-th/0505129}{{\tt hep-th/0505129}}].

\bibitem{McGreevy:2000cw}
J.~McGreevy, L.~Susskind, and N.~Toumbas, {\it {Invasion of the giant gravitons
  from Anti-de Sitter space}},  {\em JHEP} {\bf 0006} (2000) 008,
  [\href{http://xxx.lanl.gov/abs/hep-th/0003075}{{\tt hep-th/0003075}}].

\bibitem{Hashimoto:2000zp}
A.~Hashimoto, S.~Hirano, and N.~Itzhaki, {\it {Large branes in AdS and their
  field theory dual}},  {\em JHEP} {\bf 0008} (2000) 051,
  [\href{http://xxx.lanl.gov/abs/hep-th/0008016}{{\tt hep-th/0008016}}].

\bibitem{Predrag}
C.~Predrag, {\em Group theory : birdtracks, Lie's, and exceptional groups}.
\newblock Princeton University Press, Princeton, NJ, 2008.

\bibitem{ettienne}
E.~Rassart, {\em Geometric approaches to computing Kostka numbers and
  Littlewood-Richardson coefficients}.
\newblock Thesis for the Ph.D. degree in Mathematics, Massachusetts Institute
  of Technology (MIT). 2004.

\bibitem{stanley}
R.~P. Stanley, {\em Enumerative Combinatorics. Volume 2}, vol.~62 of {\em
  Cambridge Studies in Advanced Mathematics}.
\newblock Cambridge University Press, Cambridge, UK, 1999.

\bibitem{Seiberg:1999xz}
N.~Seiberg and E.~Witten, {\it {The D1 / D5 system and singular CFT}},  {\em
  JHEP} {\bf 9904} (1999) 017,
  [\href{http://xxx.lanl.gov/abs/hep-th/9903224}{{\tt hep-th/9903224}}].

\bibitem{Maldacena:1998bw}
J.~M. Maldacena and A.~Strominger, {\it {AdS(3) black holes and a stringy
  exclusion principle}},  {\em JHEP} {\bf 9812} (1998) 005,
  [\href{http://xxx.lanl.gov/abs/hep-th/9804085}{{\tt hep-th/9804085}}].

\bibitem{Jevicki:1998rr}
A.~Jevicki and S.~Ramgoolam, {\it {Noncommutative gravity from the AdS / CFT
  correspondence}},  {\em JHEP} {\bf 9904} (1999) 032,
  [\href{http://xxx.lanl.gov/abs/hep-th/9902059}{{\tt hep-th/9902059}}].

\bibitem{Ho:1999bn}
P.-M. Ho, S.~Ramgoolam, and R.~Tatar, {\it {Quantum space-times and finite N
  effects in 4-D superYang-Mills theories}},  {\em Nucl.Phys.} {\bf B573}
  (2000) 364--376, [\href{http://xxx.lanl.gov/abs/hep-th/9907145}{{\tt
  hep-th/9907145}}].

\bibitem{math-funct}
M.~Abramowitz and I.~A. Stegun., {\em Handbook of mathematical functions with
  formulas, graphs, and mathematical tables}, vol.~55 of {\em Applied
  mathematics series}.
\newblock Dover Publications, New York, NY, 1964.

\end{thebibliography}\endgroup
%
%
%
\end{document}